\documentclass{aa}
\usepackage[dvipsnames,table]{xcolor}
\usepackage{chemformula}
\usepackage[version=4]{mhchem}
\usepackage{mathrsfs}
\usepackage{placeins}
\usepackage{tabularx}
\usepackage{graphicx}
\usepackage{txfonts}
\usepackage{subfig}
\usepackage{multirow}

\usepackage[colorlinks=true,
            linkcolor=red,
            urlcolor=blue,
            citecolor=blue]{hyperref}

\usepackage{arydshln}
\begin{document}

   \title{Large Interferometer For Exoplanets (LIFE):}

   \subtitle{III. Spectral resolution, wavelength range and sensitivity requirements based on
atmospheric retrieval analyses of an exo-Earth}

    \titlerunning{\emph{LIFE}: III. Spectral resolution, wavelength range and sensitivity requirements from Earth-twin retrievals}

    \authorrunning{Konrad et al.}

   \author{
   B.S. Konrad \inst{1,2} \fnmsep\thanks{Correspondence: \texttt{konradb@student.ethz.ch}}
   \and
   E. Alei \inst{1,2}
   \and
   D. Angerhausen \inst{1,2,3}
   \and
   \'O. Carri\'on-Gonz\'alez \inst{4}
   \and
   J.J. Fortney \inst{5}
   \and
   J.L. Grenfell \inst{6}
   \and
   D. Kitzmann \inst{7}
   \and
   P. Mollière \inst{8}
   \and
   S. Rugheimer \inst{9}
   \and
   F. Wunderlich \inst{6}
   \and
   S.P. Quanz \inst{1,2} \fnmsep\thanks{Correspondence: \texttt{sascha.quanz@phys.ethz.ch}}
   \and the
   \textit{LIFE} Collaboration \thanks{Webpage: \url{www.life-space-mission.com}}
   }

  \institute{
    ETH Zurich, Institute for Particle Physics \& Astrophysics, Wolfgang-Pauli-Str. 27, 8093 Zurich, Switzerland
    \and
    National Center of Competence in Research PlanetS (www.nccr-planets.ch)
    \and
    Blue Marble Space Institute of Science, Seattle, United States
    \and
    Zentrum für Astronomie und Astrophysik, Technische Universität Berlin, Hardenbergstrasse 36, D-10623 Berlin, Germany
    \and
    Department of Astronomy and Astrophysics, University of California, Santa Cruz, CA, USA 95064
    \and
    Department of Extrasolar Planets and Atmospheres (EPA), Institute of Planetary Research (PF), German Aerospace Center (DLR), Rutherfordstr. 2, 12489 Berlin, Germany
    \and
    University of Bern, Center for Space and Habitability, Gesellschaftsstrasse 6, 3012 Bern, Switzerland
    \and
    Max-Planck-Institut f\"ur Astronomie, Königstuhl 17, 69117 Heidelberg, Germany
    \and
    Dept of Physics, University of Oxford, Oxford, OX1 3PU, UK
}

   \date{Received -; accepted -} 

 
  \abstract
   {Temperate terrestrial exoplanets are likely to be common objects, but their discovery and characterization is very challenging because of the small intrinsic signal compared to that of their host star. Various concepts for optimized space missions to overcome these challenges are currently being studied. The \textit{LIFE} initiative focuses on the development of a space-based mid-infrared (MIR) nulling interferometer probing the thermal emission of a large sample of exoplanets.}
   {This study derives the minimum requirements for the signal-to-noise ratio (S/N), the spectral resolution (R), and the wavelength coverage for the \textit{LIFE} mission concept. Using an Earth-twin exoplanet as reference case, we quantify how well planetary/atmospheric properties can be derived from its MIR thermal emission spectrum as a function of wavelength range, S/N and R. }
   {We combine a cloud-free 1D atmospheric radiative transfer model, a noise model for observations with the \textit{LIFE} interferometer and the nested sampling algorithm for Bayesian parameter inference to retrieve for planetary/atmospheric properties. We simulate observations of an Earth-twin exoplanet orbiting a G2V star at $10$ pc from the Sun with different levels of exozodiacal dust emissions. We investigate a grid of wavelength ranges ($3-20\,\mu \mathrm{m}$, $4-18.5\,\mu \mathrm{m}$, and $6-17\,\mu \mathrm{m}$), S/Ns ($5$, $10$, $15$, and $20$ determined at a wavelength of $11.2\,\mu\mathrm{m}$) and Rs ($20$, $35$, $50$, and $100$).}
   {\ce{H2O}, \ce{CO2}, and \ce{O3} are detectable if S/N $\geq10$ (uncertainty $\leq\pm1.0$ dex). We find upper limits for \ce{N2O} (abundance $\lesssim10^{-3}$). \ce{CO}, \ce{N2}, and \ce{O2} are unconstrained in all cases.  The lower limits for a \ce{CH4} detection are R~=~50 and S/N~=~10. Our retrieval framework correctly determines the exoplanet's radius (uncertainty $\leq\pm10\%$), surface temperature (uncertainty $\leq\pm20\,\mathrm{K}$), and surface pressure (uncertainty $\leq\pm0.5$ dex) in all cloud-free retrieval analyses. Based on our current assumptions, the observation time required to reach the specified S/N for an Earth-twin at 10 pc when conservatively assuming a total instrument throughput of 5\% amounts to $\approx$6-7 weeks with 4$\times$2-m apertures.}
   {We provide first order estimates for the minimum technical requirements for \textit{LIFE} via the retrieval study of an Earth-twin exoplanet. We conclude that a minimum wavelength coverage of $4-18.5\,\mu$m, an R of 50, and an S/N of at least 10 is required. With the current assumptions, the atmospheric characterization of several Earth-like exoplanets at a distance of 10 pc and within a reasonable amount of observing time will require apertures $\ge$2 meters.} 


   \maketitle
%

\section{Introduction}

Since the detection of 51 Peg b, the first planetary companion to a solar-type star \citep{1995Natur.378..355M}, exoplanet research has become one of the pillars of modern astrophysics. With more than 4000 exoplanets currently known\footnote{\url{https://exoplanetarchive.ipac.caltech.edu}}, scientists have begun to uncover the vast diversity among exoplanet objects and extrasolar systems. 
A long-term goal is the discovery and the atmospheric characterization of a large sample of terrestrial exoplanets, with a specific focus on temperate objects.

In this context, a direct detection approach is essential in order to investigate the diversity of planetary atmospheres, assess the potential habitability of some objects, and look for so-called biosignatures in their atmospheres. Different concepts for large exoplanet imaging space-missions are currently being assessed, with 
\textit{LUVOIR} \citep{2017AAS...22940504P} and \textit{HabEx} \citep{2020arXiv200106683G}, which aim at directly measuring the reflected spectrum of exoplanets in the visible (VIS) and near-infrared (NIR) range, being prominent examples. The \textit{Large Interferometer For Exoplanets} (\textit{LIFE}) initiative\footnote{\url{www.life-space-mission.com}} follows a complementary approach by focusing on the prospects of a large, space-based mid-infrared (MIR) nulling interferometer which will observe the thermal emission spectrum and subsequently characterize the atmospheres of a large sample of (terrestrial) exoplanets \citep{Quanz:LIFE,Quanz:exoplanets_and_atmospheric_characterization,quanz2021}. The \textit{LIFE} initiative aims to combine various efforts to push towards an eventual launch of such a large, space-based MIR nulling interferometer. The work we present here aims to constrain some of the instrument requirements for \textit{LIFE} and is the third paper in a series. The first paper of the series \citep{quanz2021} quantifies the exoplanet detection performance of \textit{LIFE} and compares it with large single-aperture mission concepts for reflected light. The second paper \citep{dannert2022large} introduces the LIFE\textsc{sim} instrument simulator and the necessary signal extraction algorithms.

The choice of the MIR wavelength range for \textit{LIFE} pays dividends. More molecules have strong absorption bands in the MIR spectra of Earth-like planets which allows to better assess the atmospheric structure and composition \citep[e.g.,][]{desmarais2002,line2019} and the infrared appears to be less affected by the presence of hazes and clouds  \cite[see, e.g.,][]{2011A&A...531A..62K, rugheimer2013spectral,arney2018,lavvas2019,fauchez2019,komacek2020,wunderlich2021}, which is a major challenge at visible wavelengths \cite[see, e.g.,][for Jovian planets]{2016Natur.529...59S}. \footnote{However, clouds can also increase reflectivity and signal for molecules like molecular oxygen (\ce{O2}) that are well-mixed in the atmosphere of a terrestrial planet \citep{kawashima2019theoretical}.} Importantly, emission spectra allow us to constrain the planetary radius \citep[e.g.,][]{line2019}, which is degenerate with the planetary albedo in reflected-light measurements \citep{2020A&A...640A.136C}.
 
 
Finally, the MIR range includes a rich portfolio of biosignatures \citep[e.g.,][]{catling2018exoplanet}. Biosignatures in an exoplanet context are gases or features that can be detected at interplanetary distances and that are produced by life. Among the main biosignature gases there are \ce{O2} and its photochemical product ozone (\ce{O3}), methane (\ce{CH4}), nitrous oxide (\ce{N2O}), chloromethane (\ce{CH3Cl}), phosphine (\ce{PH3}), and dimethyl sulfide (\ce{C2H6S}, commonly known as DMS).  Many of these gases could also be generated by abiotic processes and therefore be false positives in the search for biosignatures \citep{Schwieterman2018,2018haex.bookE..71H}. However, the presence of multiple biosignature gases in the spectrum, along with other planetary context that points towards habitability, would increase the robustness of the detection of life on an exoplanet.  The most widely known set of multiple biosignatures is the so-called 'triple fingerprint' of carbon dioxide (\ch{CO2}), water vapor (\ch{H2O}) and ozone (\ch{O3}), well detectable in Earth's thermal emission spectrum and potentially detectable in terrestrial exoplanets \citep[see, e.g.,][]{2002A&A...388..985S}.
The simultaneous presence of reducing and oxidizing species in an atmosphere, such as \ce{O2} and \ce{O3} in combination with \ce{CH4}, is a strong biosignature, with no currently known false positives. These species would not be both present in large quantities in an atmosphere over long timescales without disequilibrium processes driven by the presence of life  \citep{lederberg1965signs, Lovelock:CH4-O2}. We refer the reader to \citet{Schwieterman2018} for a review about biosignatures and \citet{catling2018exoplanet} for a Bayesian framework for assessing the confidence level of a biosignature detection.

While there is ample scientific justification for choosing the MIR wavelength regime for detailed (atmospheric) investigations of terrestrial exoplanets, deriving a concept for a space-mission, as pursued by the \textit{LIFE} initiative, requires the derivation of fundamental mission and instrument requirements, including instrument sensitivity, wavelength range coverage and spectral resolution. 
Earlier steps in this direction were presented in \citet{2013A&A...551A.120V} for the former \textit{Darwin} mission concept \citep{1996Icar..123..249L}. 
In a more recent study, \cite{2018AJ....155..200F} used a Bayesian atmospheric retrieval approach to quantify the power of reflected light observations, as foreseen by \textit{HabEx} or \textit{LUVOIR}, as a function of instrument parameters. 

Here, we aim at providing minimum requirements for the parameters listed above for the \textit{LIFE} mission concept using an atmospheric retrieval framework. Such a framework allows us to derive quantitative estimates on the main atmospheric and planetary parameters from a simulated exoplanetary spectrum \citep[see, e.g.,][for recent reviews]{Madhusudhan:Atmospheric_Retrieval,deming2018,barstow_heng2020}. Using a Bayesian approach, the space of input parameters (e.g., atmospheric abundances) is explored iteratively to assess which combination of values best fits the simulated observations. Doing that for various combinations of  signal-to-noise ratio (S/N) of the emission spectrum, wavelength range and spectral resolution allows us to understand, how well the simulated planet can be characterized. In the present study, we use a cloud-free modern Earth-twin exoplanet as our reference case. As Earth is the only planet known to host life, we are particularly interested in assessing if and how well some of its main atmospheric constituents can be detected for various combinations of instrument parameters. We are aware that by ignoring clouds, we are somewhat simplifying the problem. However, we remind the reader that the main aim of our analysis is to get first estimates for sensitivity, wavelength range and spectral resolution requirements for \textit{LIFE}. Subsequent work will investigate other types of exoplanets and atmospheric compositions, providing additional constraints on some of the requirements.

In Sect.~\ref{Methods}, we introduce our retrieval framework. It combines a 1-D atmospheric forward model based on the \texttt{petitRADTRANS} radiative transfer code \citep{2019AA...627A..67M}, with the \textsc{LIFEsim} instrument simulator \citep[][]{dannert2022large}, that adds astrophysical noise terms to the simulated Earth spectrum, and a Nested Sampling approach \citep{Skilling:Nested_Sampling} for Bayesian parameter inference. We validate the retrieval framework in Sect.~\ref{Validation}. Then, we perform a series of retrievals of the theoretical emission spectrum of a cloud-free Earth-twin exoplanet as it would be observed by \textit{LIFE}, varying the wavelength range, the S/N, and the spectral resolution. Results are presented in Sect.~\ref{Results}. We discuss our results and compare our study with other works in Sect.~\ref{Discussion}. We summarize our main findings and conclusions in Sect.~\ref{Summary}.


\begin{figure*}
  \includegraphics[width=0.9\linewidth]{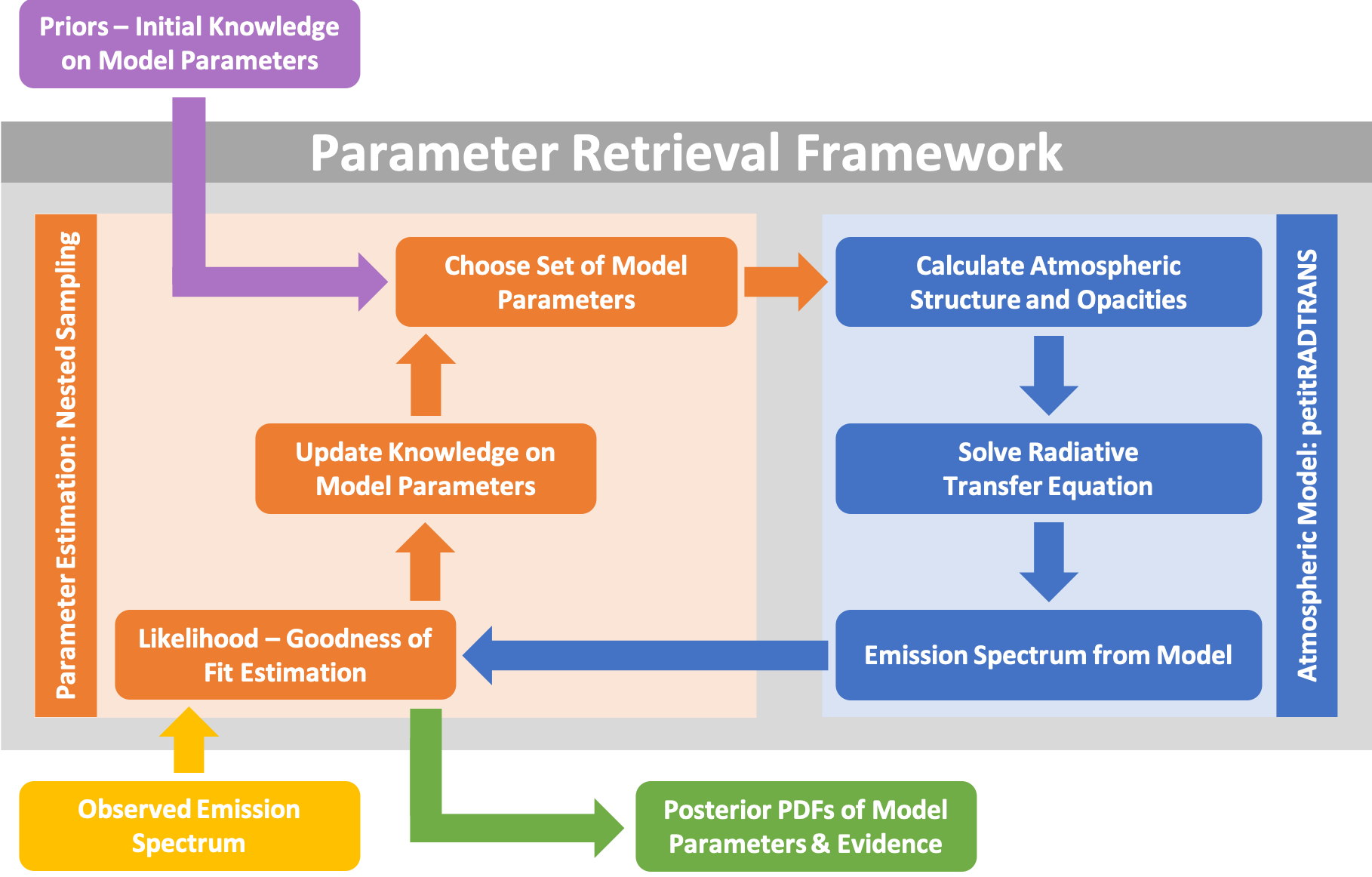}
  \caption{Schematic illustrating our atmospheric retrieval framework.}
  \label{fig:1}
\end{figure*}

\section{Methods}
\label{Methods}

Our atmospheric retrieval framework aims to infer the atmospheric pressure-temperature (P-T) structure and composition from simulated mock-observations of the MIR thermal emission spectrum of an Earth-twin planet. At its heart, the framework consists of two elements. Firstly, we need a parametric model for the atmosphere, which calculates the emergent light spectrum corresponding to a set of model parameters (Sect.~\ref{atmosphericmodel}). Secondly, a parameter estimation algorithm is required to optimize the model parameters, such that the spectrum produced by the atmospheric model best fits the simulated observational data (Sect.~\ref{bayesiantechnique}). These two elements are then combined to form our retrieval framework (Sect.~\ref{retrievalsetup}). An illustrative schematic summarizing our retrieval framework is given in Fig.~\ref{fig:1}.

\subsection{Atmospheric model}\label{atmosphericmodel}

We use the 1-D radiative transfer code \texttt{petitRADTRANS} \citep{2019AA...627A..67M}. To calculate the thermal emission spectrum of terrestrial exoplanets in the MIR wavelength range, \texttt{petitRADTRANS} passes a featureless black-body spectrum at the surface temperature through discrete atmospheric layers, which interact with the radiation field. We characterize each layer by its temperature, pressure, and the opacity sources present. 

\subsubsection{Atmospheric P-T structure}

Throughout this work, we parametrized the P-T structure of Earth's atmosphere using a 4\textsuperscript{th} order polynomial:
\begin{equation}
\centering
T(P)=\sum_{i=0}^4 a_i P^i.   
\end{equation}
Here, $P$ denotes the atmospheric pressure and $T$ the corresponding temperature. The parameters $a_i$ are the parameters describing the polynomial P-T model. We chose this polynomial P-T model over other P-T profile parametrizations \citep[e.g., models from][]{2009ApJ...707...24M,Guillot:PT_Profile,2019AA...627A..67M}, since it provides a comparable description of the atmospheric P-T structure using fewer model parameters (see Appendix \ref{P-T-choice} for further information).

\begin{figure*}
  \centering
  \includegraphics[width=1\textwidth]{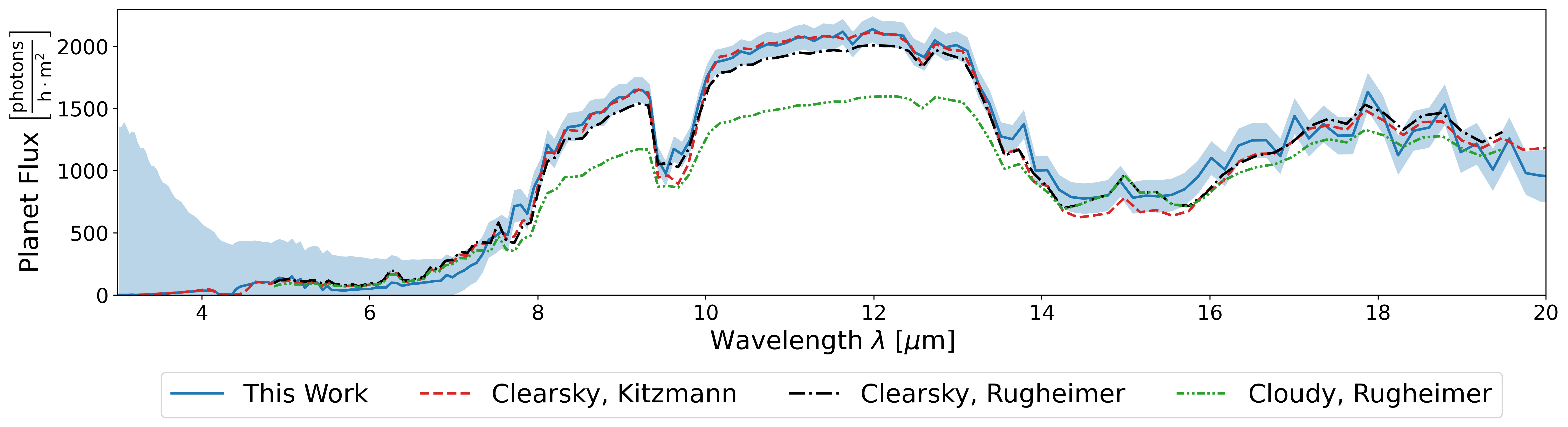}
\caption{Comparison of the Earth-twin MIR emission spectra calculated with various different models. We plot the photon flux received from an Earth-twin located 10 pc from the sun. The solid blue line is the MIR thermal emission calculated with \texttt{petitRADTRANS} using the settings discussed in Sect.~\ref{pR_setup}. The blue-shaded region indicates the most optimistic LIFE\textsc{sim} uncertainty (S/N~=~20) used in our retrievals. The red dashed line represents a cloud-fee Earth model which accounts for scattering by Daniel Kitzmann (private communications). The green and black dashed-dotted lines are the cloudy (60\% cloud coverage) and cloud-free modern Earth spectra from \citet{Rugheimer_2015} that account for scattering.}
\label{fig:2}
\end{figure*}

\subsubsection{Opacity sources in \texttt{petitRADTRANS}}

\texttt{petitRADTRANS} is capable of considering several different opacity sources that are potentially present in exoplanet atmospheres. In the computationally favorable low spectral resolution mode (R~=~1000), the opacities originating from different atmospheric gases are calculated via the correlated-k method \citep{Goody:correlated_k,Lacis:correlated-k,Fu:correlated_k}. Absorption lines from different molecules, collision-induced absorption (CIA), and atmospheric pressure broadening effects can be taken into account. Rayleigh and cloud scattering, as well as the scattering of direct radiation from the host star, can be considered. However, the scattering solution is achieved in an iterative way and the opacity handling is different \citep[see][for more information]{2020A&A...640A.131M}. These changes increase the spectrum calculation time by roughly an order of magnitude when scattering is included in the computation. 

In Fig.~\ref{fig:2} we compare our scattering- and cloud-free Earth-twin MIR emission spectrum, which assumes uniform chemical abundances throughout the atmosphere to other cloud-free and cloudy models \citep[Daniel Kitzmann, private communications;][]{Rugheimer_2015}. These models consider scattering as well as non-uniform abundances. We see that scattering contributions and non-uniform abundances only have a minor impact on Earth's spectrum in the MIR and the differences between the cloud-free spectra are of similar magnitude as the most optimistic LIFE\textsc{sim} noise estimate. This justifies our approach of excluding scattering from our retrieval routine. Thus, we neglect effects linked to the incident stellar radiation, which reduces the number of parameters retrieved and thereby the computing time.

We also neglect scattering and absorption by clouds. As can be seen from Fig.~\ref{fig:2}, this is a significant simplification of the problem. The presence of (opaque) clouds in an atmosphere will partially or fully obscure the view of the exoplanet's surface. Therefore, we expect systematic shifts in the surface temperature/pressure and potentially also in the retrieved planetary radius if clouds are present in an atmosphere. Additionally, a partial cloud coverage combined with vertically non-constant atmospheric abundances could lead to systematic offsets in the retrieved abundances. For example, if an atmospheric gas is only present below an optically thick cloud deck, it is likely not detectable via a MIR retrieval study. These potential effects require further attention in future works. However, as we are interested in first estimates for specific instrument requirements, focusing on a cloud-free atmosphere is justifiable. Complete details on the implementation are given in Sect.~\ref{pR_setup}.

\begin{figure*}
  \includegraphics[width=\linewidth]{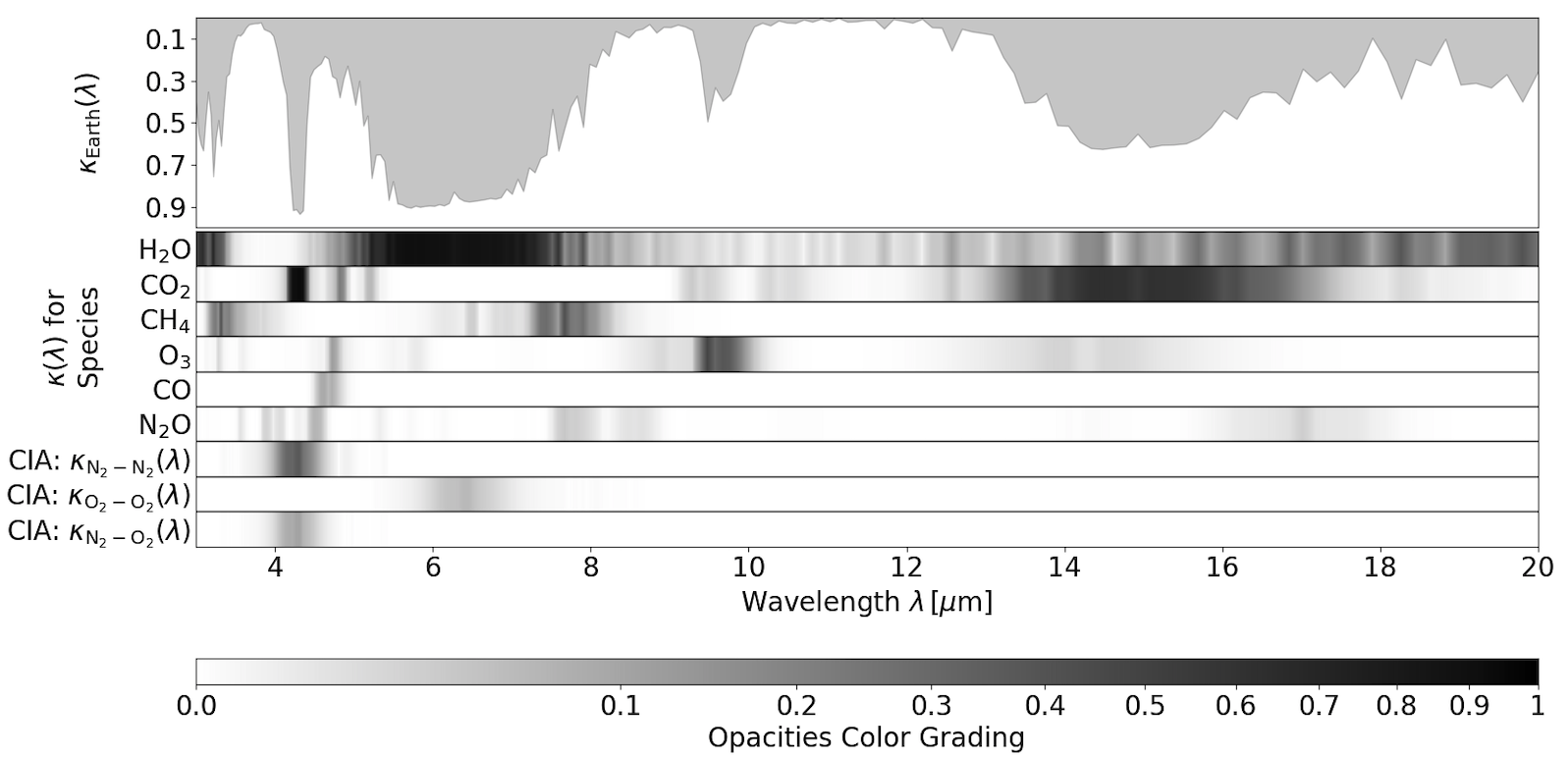}
\caption{\textit{Upper panel}: The opacity of a cloudless Earth-twin atmosphere as a function of wavelength. Gray shading shows the amount of light blocked by the atmosphere. \textit{Lower panel}: Contribution of the different molecules to the opacity of the Earth-twin atmosphere as a function of wavelength. Dark regions indicate a high opacity, as is indicated by the colorbar.}
  \label{fig:3}
\end{figure*}

\subsection{Parameter estimation}\label{bayesiantechnique}

Our retrieval study utilizes a Bayesian parameter inference tool to sample the posterior probability distributions of the atmospheric forward model parameters.

Bayesian Parameter inference methods are based on Bayes' theorem, which provides a method for estimating model parameters based on experimental data \citep[see, e.g.,][]{Trotta:Bayesian_Cosmology,primer2021}. Let us consider a model $\mathcal{M}$ described by a set of parameters $\Theta_\mathcal{M}$ and experimental data $\mathcal{D}$. Bayes' theorem states:
\begin{equation}
P(\Theta_\mathcal{M}|\mathcal{D},\mathcal{M})=\frac{P(\mathcal{D}|\Theta_\mathcal{M},\mathcal{M})P(\Theta_\mathcal{M}|\mathcal{M})}{P(\mathcal{D}|\mathcal{M})}.
\label{equ:baye_theorem_inference}
\end{equation}

$P(\Theta_\mathcal{M}|\mathcal{D},\mathcal{M})$ is called posterior (or posterior probability) and represents the probability of different sets of model parameters $\Theta_\mathcal{M}$ under the constraint of the experimental data $\mathcal{D}$ and model $\mathcal{M}$.

$P(\mathcal{D}|\Theta_\mathcal{M},\mathcal{M})$ provides a probabilistic measure of how well a specific set of parameters $\Theta_\mathcal{M}$ for the model $\mathcal{M}$ describes the data $\mathcal{D}$. We calculate this probability via a log-likelihood function $\ln\left(\mathscr{L}(\Theta_\mathcal{M})\right)$:

\begin{equation}
\ln(\mathscr{L}(\Theta_\mathcal{M}))=\sum_{i=1}^N\left(-\frac{1}{2}\ln{(2\pi\sigma_i^2)}-\frac{1}{2}\frac{(\mathcal{D}_i-\mu_i(\Theta_\mathcal{M}))^2}{\sigma_i^2}\right).
\label{equ:loglike}
\end{equation}
Our log-likelihood function assumes that each of the $N$ measured data points $\mathcal{D}_i$ behaves as a normally distributed quantity. The mean $\mu_i(\Theta_\mathcal{M})$ is a value predicted by the model $\mathcal{M}$ using parameters $\Theta_\mathcal{M}$ corresponding to $\mathcal{D}_i$. $\sigma_i$ is the measurement error on the data point $\mathcal{D}_i$.

$P(\Theta_\mathcal{M}|\mathcal{M})$ is the prior probability (or `prior') of the model parameters $\Theta_\mathcal{M}$ and represents the knowledge on the model parameters before taking the observational data into account.

$P(\mathcal{D}|\mathcal{M})$ is a normalization constant, which ensures that the posterior is normalized to unity and is frequently referred to as the Bayesian evidence $\mathcal{Z}_\mathcal{M}(\mathcal{D})$:
\begin{equation}
\mathcal{Z}_\mathcal{M}(\mathcal{D}) = P(\mathcal{D}|\mathcal{M})=\int \mathscr{L}(\Theta_\mathcal{M})P(\Theta_\mathcal{M} |\mathcal{M})d\Theta_\mathcal{M}.
\label{equ:Bayesian_Evidence}
\end{equation}
Additionally, the evidence $\mathcal{Z}_\mathcal{M}(\mathcal{D})$ enables the comparison of the performance of different models $\mathcal{M}_i$ to each other and to decide which model best describes the observed data $\mathcal{D}$. We can compare two models $\mathcal{M}_1$ and $\mathcal{M}_2$ by considering the Bayes' factor $K$

\begin{equation}\label{equ:bayes_factor}
    K = \frac{P(\mathcal{M}_1|\mathcal{D})}{P(\mathcal{M}_2|\mathcal{D})}=\frac{\mathcal{Z}_{\mathcal{M}_1}(\mathcal{D})}{\mathcal{Z}_{\mathcal{M}_2}(\mathcal{D})}\quad ,
\end{equation}
where we assumed that the prior probabilities $\mathcal{P}(\mathcal{M}_i)$ for both models are the same. An approach to interpreting the value of $K$ is via Jeffrey's scale \citep{Jeffreys:Theory_of_prob} given in Table~\ref{table:1}.

\begin{table}
\caption{Jeffrey's scale \citep{Jeffreys:Theory_of_prob}.}
\label{table:1}      
\centering                          
\begin{tabular}{c c c}        
\hline\hline                 
$\log_{10}\left(K\right)$ &Probability &Strength of Evidence\\    
\hline 
   $<0$     &$<0.5$         &Support for $\mathcal{M}_2$\\
   $0-0.5$  &$0.5-0.75$     &Very weak support for $\mathcal{M}_1$\\
   $0.5-1$  &$0.75-0.91$    &Substantial support for $\mathcal{M}_1$\\
   $1-2$    &$0.91-0.99$    &Strong support for $\mathcal{M}_1$\\
   $>2$     &$>0.99$        &Decisive support for $\mathcal{M}_1$\\ 
\hline 
\end{tabular}
\tablefoot{Scale for the interpretation of the Bayes' factor $K=\mathcal{Z}_{\mathcal{M}_1}(\mathcal{D})/\mathcal{Z}_{\mathcal{M}_2}(\mathcal{D})$. The scale is symmetrical, which means that negative values correspond to very weak/substantial/strong/decisive support for $\mathcal{M}_2$.}
\end{table}

Due to the model comparison capabilities of the evidence $\mathcal{Z}_\mathcal{M}(\mathcal{D})$, we choose the `nested sampling' algorithm \citep{Skilling:Nested_Sampling} over MCMC algorithms, such as the Metropolis-Hastings algorithm \citep{Hastings:MCMC,Metropolis:MCMC}, since it provides a direct estimate for the Bayesian evidence $\mathcal{Z}_\mathcal{M}(\mathcal{D})$. Furthermore, nested sampling is computationally less expensive and better at handling multimodal posterior distributions \citep{Skilling:Nested_Sampling}.

Specifically, we utilize the open-source \texttt{pyMultiNest} package \citep{Buchner:PyMultinest}, which makes the nested sampling implementation \texttt{MultiNest} \citep{Feroz:Multinest} accessible to the Python language. \texttt{MultiNest} is based on the original nested sampling algorithm \citep{Skilling:Nested_Sampling} and uses the `importance nested sampling' algorithm \citep{Feroz:Importance_Nested_Sampling} to obtain more accurate estimates of the Bayesian evidence $\mathcal{Z}_\mathcal{M}(\mathcal{D})$.

\subsection{Retrieval setup}\label{retrievalsetup}
\subsubsection{Forward model and noise terms}\label{pR_setup}

\begin{table}[]

\caption{Line lists used throughout this study.}           
\label{table:2}      
\centering                          
\begin{tabular}{lll}    
\hline\hline                 
Species & Database & References\\
\hline

\ce{CO2}    &ExoMol         &\citet{10.1093/mnras/staa1874}   \\
\ce{O3}     &HITRAN 2012    &\citet{2019AA...627A..67M};\\
            &               &\citet{2013JQSRT.130....4R} \\
\ce{CH4}    &ExoMol         &\citet{2021A...646A..21C};\\
            &               &\citet{10.1093/mnras/staa1874} \\
\ce{CO}     &HITEMP         &\citet{ROTHMAN20102139}   \\
\ce{H2O}    &HITEMP         &\citet{ROTHMAN20102139} \\
\ce{N2O}    &ExoMol         &\citet{2021A...646A..21C} \\
\hline 
\end{tabular}
\end{table}

We generate a correlated-k \texttt{petitRADTRANS} spectrum (R~=~1000) of an Earth-twin exoplanet using the parameter values listed in Table~\ref{table:3} (Input column) assuming an atmosphere consisting of 100 layers. We used the line lists from the ExoMol database \citep{TENNYSON201673} for \ce{CO2}, \ce{CH4}, and \ce{N2O} and from the HITRAN/HITEMP database \citep{10.1117/12.211919,ROTHMAN20102139} for \ce{O3}, \ce{CO}, and \ce{H2O}. We summarize the reference papers corresponding to the opacity line lists in Table~\ref{table:2}. For \ce{N2} and \ce{O2} we consider collision induced absorption (CIA) \citep[as discussed in][]{Schwieterman:N2_CIA}. All other atmospheric gases have distinct absorption features within the MIR range, even at low abundances (see Fig.~\ref{fig:3}). Furthermore, we assume constant abundances of all molecules vertically throughout the atmosphere.

We then resample the calculated MIR emission spectrum to the desired spectral resolution, keeping a constant resolution $\mathrm{R} = \lambda/\Delta\lambda$ across the spectrum. This results in wavelength bins of variable width, where the width at short wavelengths is smaller than at long wavelengths. For the spectral resampling, we use the \texttt{SpectRes} tool \citep{Carnall:SpectRes}, which allows for time efficient resolution reduction of MIR spectra whilst keeping the overall flux and energy conserved. We then use the radius $R_\mathrm{pl}$ and the distance from Earth $d_\mathrm{Earth}$ to scale the photon flux found with \texttt{petitRADTRANS} $F_\mathrm{pRT}$ to the flux \textit{LIFE} would detect ($F_{LIFE}$):

\begin{equation}
    F_{LIFE}=F_\mathrm{pRT}\frac{R_\mathrm{pl}^2}{d_\mathrm{Earth}^2}    
\end{equation} 

Throughout this work, we define the S/N of a spectrum as the S/N calculated in the reference bin at 11.2 micron. This wavelength was chosen as it lies close to the peak flux and it does not coincide with strong absorption lines. The S/N at all other wavelengths is determined by the noise model, which relates the S/N for the reference bin to the S/N in all other wavelength bins.

For the retrieval validation we perform in Sect.~\ref{Validation}, we only consider the photon noise of the planet spectrum.
For the grid retrievals performed in Sect.~\ref{Retrievalgrid}, we obtain noise estimates via the LIFE\textsc{sim} tool \citep[see][]{dannert2022large}. LIFE\textsc{sim} accounts for photon noise contributions from the planet's emission spectrum, stellar leakage as well as local- and exozodiacal dust emission. In our simulations, we assume that the noise does not impact the predicted flux values, but instead only adds uncertainties to the simulated spectral points. As discussed in \citet{2018AJ....155..200F}, randomization of the individual spectral points based on the S/N allows to simulate accurate observational instances. At the same time, retrieval studies on such instances will result in biased results, since the random placement of the small number of data points will impact the retrieval's performance. The ideal analysis would study many ($\gtrsim10$) data realizations for each considered spectrum and assess instrument performance by considering the posteriors found for these different noise instances. However, the number of cases (96) we consider and the computation time per case ($\approx1$ day on 90 CPUs) make such a study computationally unfeasible ($\gtrsim30$ months of cluster time). By not randomizing the individual spectral points, we eliminate the biases introduced by noise instances. However, we are aware that this approach will likely result in optimistic results. Namely, we expect an unrealistic centering of posteriors on the truths. Additionally, for molecules at the sensitivity limit (weak spectral features), we expect overly optimistic results \citep[see, e.g., ][]{2016Feng}. In Appendix \ref{RandNoise}, we perform retrievals on randomized noise instances to study these effects in more detail.

\subsubsection{Priors}
The assumed prior ranges for the polynomial P-T profile and the ground pressure shown in Table~\ref{table:3} are chosen such that a wide range of atmospheric structures (e.g., Venus or Mars atmosphere) are allowed (e.g., $P_0$ is considered over the range $10^{-4}$ to $1000$ bar as indicated by $\mathcal{U}(-4,3)$).

\begin{table*}
\renewcommand{\arraystretch}{1.5}

\caption{List of parameters used in the retrievals, their input values, prior distributions, and the validation results.}             
\label{table:3}      
\centering                          
\begin{tabular}{llllr}    
\hline\hline                 
Parameter &Description &Input &Prior & Validation \\    
\hline

$\sqrt[4]{a_4}$ &P-T Parameter (Degree 4)  & $1.14$   & $\mathcal{U}(0.5,1.8)$ &$1.12^{+0.05}_{-0.04}$\\
$a_3$           &P-T parameter (Degree 3)  & $23.12$  & $\mathcal{U}(0,100)$  & $22.65^{+1.74}_{-1.54}$\\
$a_2$           &P-T Parameter (Degree 2)  & $99.70$  & $\mathcal{U}(0,500)$   & $98.90^{+3.72}_{-3.44}$\\
$a_1$           &P-T Parameter  (Degree 1) & $146.63$ & $\mathcal{U}(0,500)$   & $146.07^{+4.75}_{-4.84}$\\
$a_0$           &P-T Parameter (Degree 0) & $285.22$ & $\mathcal{U}(0,1000)$  & $285.01^{+3.16}_{-3.08}$\\

$\log_{10}\left(P_0\left[\mathrm{bar}\right]\right)$&Surface Pressure  & $0.006$\tablefootmark{1}  & $\mathcal{U}(-4,3)$& $0.01^{+0.02}_{-0.02}$\\
$R_{\text{pl}}\,\left[R_\oplus\right]$&Planet Radius& $1.0$  & $\mathcal{G}(1.0,0.2)$ & $1.00^{+0.01}_{-0.01}$\\ 
$\log_{10}\left(M_{\text{pl}}\,\left[M_\oplus\right]\right)$&Planet Mass & $0.0$  & $\mathcal{G}(0.0,0.4)$& $-0.09^{+0.18}_{-0.22}$\\

$\log_{10}(\mathrm{N_2})$    &\ce{N2}  Mass Fraction  & $-0.107$\tablefootmark{1,2} & $\mathcal{U}(-3,0)$  & $-0.14^{+0.09}_{-0.11}$\\
$\log_{10}(\mathrm{O_2})$    & \ce{O2} Mass Fraction     & $-0.679$\tablefootmark{1,2} & $\mathcal{U}(-3,0)$  & $-0.75^{+0.18}_{-0.17}$\\
$\log_{10}(\mathrm{H_2O})$   & \ce{H2O} Mass Fraction           & $-3.000$ & $\mathcal{U}(-15,0)$ & $-3.09^{+0.18}_{-0.22}$\\ 
$\log_{10}(\mathrm{CO_2})$   & \ce{CO2} Mass Fraction    & $-3.387$\tablefootmark{1,2} & $\mathcal{U}(-15,0)$ & $-3.47^{+0.18}_{-0.22}$\\
$\log_{10}(\mathrm{CH_4})$   & \ce{CH4} Mass Fraction          & $-5.770$\tablefootmark{1,2} & $\mathcal{U}(-15,0)$ & $-5.85^{+0.18}_{-0.23}$\\
$\log_{10}(\mathrm{O_3})$    & \ce{O3} Mass Fraction           & $-6.523$\tablefootmark{2} & $\mathcal{U}(-15,0)$ & $-6.61^{+0.18}_{-0.23}$\\
$\log_{10}(\mathrm{CO})$     &\ce{CO}  Mass Fraction      & $-6.903$\tablefootmark{2} & $\mathcal{U}(-15,0)$& $-7.02^{+0.23}_{-0.28}$\\
$\log_{10}(\mathrm{N_2O})$   &\ce{N2O} Mass Fraction    & $-6.495$\tablefootmark{2} & $\mathcal{U}(-15,0)$& $-6.60^{+0.24}_{-0.27}$\\
\hline 
\end{tabular}
\tablefoot{The last column shows the output value with 1$\sigma$ uncertainties for each parameter obtained in the validation run (Sect.~\ref{Validation}). $\mathcal{U}(x,y)$ denotes a boxcar prior with a lower threshold $x$ and upper threshold $y$; $\mathcal{G}(\mu,\sigma)$ represents a Gaussian prior with mean $\mu$ and standard deviation $\sigma$. For $a_4$ we choose a prior on $\sqrt[4]{a_4}$, which allows us to sample small values more densely, typical of a 4\textsuperscript{th} order coefficient, and then take the 4\textsuperscript{th} power to obtain $a_4$.}
\tablebib{\tablefoottext{1}{NASA's planet factsheet: \url{https://nssdc.gsfc.nasa.gov/planetary/factsheet/earthfact.html}}
\tablefoottext{2}{Table A1 in \citet{doi:10.1089/ast.2015.1404}}}
\end{table*}

As demonstrated in \citet{dannert2022large}, the detection of a planet during the search phase of the \textit{LIFE} mission will already provide first estimates for the radius $R_{\text{pl}}$ of the object. Specifically, for small, rocky planets within the habitable zone, the authors showed that a detection in the search phase would provide an estimate $R_\mathrm{est}$ for the true planet radius $R_\mathrm{true}$ with an accuracy of $R_\mathrm{est}/R_\mathrm{true}=0.97^{\pm0.18}$. For our simulations, we therefore assume that a rough estimate of the radius is already known and we choose a Gaussian prior for this parameter (with 20\% uncertainty). A constraint on $R_{\text{pl}}$ can then be used to obtain a constraint on the planet's mass $M_{\text{pl}}$ via a statistical mass-radius relation \citep[see, e.g., ][]{Hatzes_2015,Wolfgang_2016,Zeng_2016,Kipping:Forecaster,otegi2020}. In our retrievals, we use  \texttt{Forecaster}\footnote{\url{https://github.com/chenjj2/forecaster}} \citep{Kipping:Forecaster}, a tool that allows us to set a prior on the planetary mass $M_{\text{pl}}$ from the known estimate for the radius $R_{\text{pl}}$. The tool relies on the statistical analysis of 316 objects (Solar System objects and exoplanets), for which well-constrained mass and radius estimates are available. It produces accurate predictions for a large variety of different objects, spanning from dwarf planets to late-type stars.

For the trace gases, we assume uniform priors between $-15$ and $0$ in $\mathrm{log}_{10}$ mass fraction. For the bulk constituents \ce{N2} and \ce{O2} we assume a uniform prior between $0$ and $-3$ in $\mathrm{log}_{10}$ mass fraction. This range gives us an increased sampling density in the high abundance regime, where we expect the sensitivity limit to be located for \ce{N2} and \ce{O2}. Furthermore, we use \ce{N2} as filling gas in our atmosphere, which ensures that $\sum(\mathrm{gas\,abundances})=1$.

\section{Validation}\label{Validation}

\begin{figure*}\centering
    \subfloat{\label{fig:4a}}
    \subfloat{\label{fig:4b}}
    \subfloat{\label{fig:4c}}
  \includegraphics[width=1\textwidth]{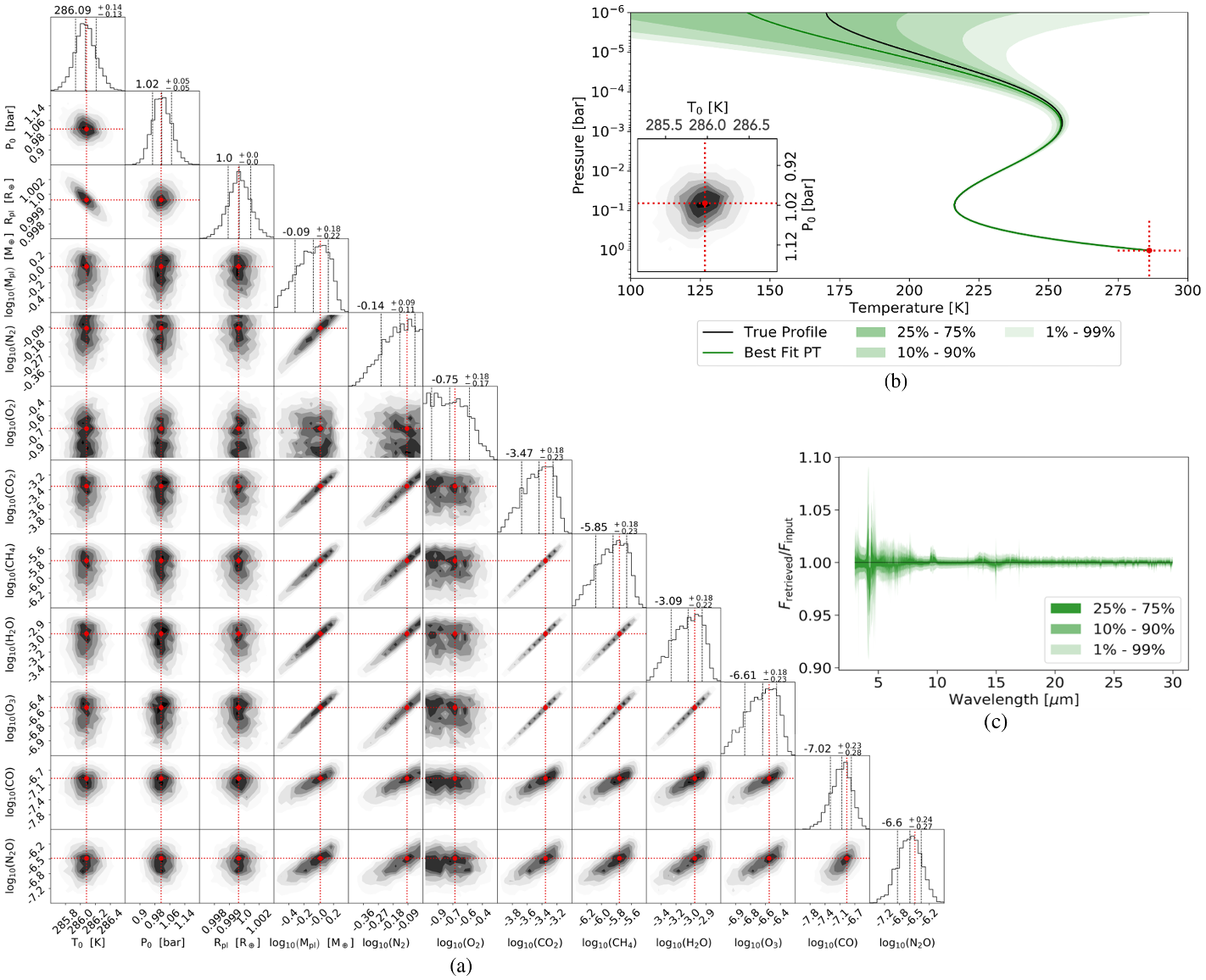}
  \caption{Summary of the validation run outlined in Sect.~\ref{Validation}. (a): Corner plot for the posterior distributions of the planetary surface temperature $T_0$, surface pressure $P_0$, radius $R_{\text{pl}}$, mass $M_{\text{pl}}$ and retrieved abundances of different molecules. The red lines indicate the values used to generate the input spectrum. Additionally, we plot the retrieved median and the $16^{\mathrm{th}}$ and $84^{\mathrm{th}}$ percentile as dashed lines in every posterior plot. (b): Retrieved P-T profile. The shaded green regions show the uncertainties on the retrieved profile. In the bottom left corner of the P-T profile plot, we display $P_0$ and $T_0$. The red cross marks the input values. (c): The retrieved emission spectrum $F_\mathrm{retrieved}$ relative to the input emission spectrum for the retrieval $F_\mathrm{input}$.}
  \label{fig:4}
\end{figure*}

Before running our retrieval framework for simulated \textit{LIFE} data, we validate its accuracy and performance. For the retrieval validation, we retrieve a full resolution (R~=~1000) Earth-twin MIR thermal emission spectrum covering the wavelength range $3-20\,\mu\mathrm{m}$. We generate the validation input spectrum using \texttt{petitRADTRANS} and the parameter values provided in Table~\ref{table:3}. We only consider photon noise from the spectrum itself and chose an S/N of $50$ at $11.2\,\mu\mathrm{m}$. Furthermore, we assume that the photon noise does not impact the simulated flux values, but instead only adds uncertainties to the spectral points.

In our retrievals, we run the \texttt{pyMultiNest} package using $700$ live points and a sampling efficiency of $0.3$ as suggested for parameter retrieval by the documentation\footnote{\url{https://johannesbuchner.github.io/PyMultiNest/pymultinest_run.html}}. We summarize the results in Fig.~\ref{fig:4} and Table~\ref{table:3} (last column).

The corner plot in Fig.~\ref{fig:4a} suggests that the exoplanet's radius $R_{\text{pl}}$ is retrieved to a very high precision with an uncertainty of roughly $0.001$\, $R_\oplus$, a significant improvement over the assumed prior distribution. Similarly, the retrieved posterior for the exoplanet mass $M_{\text{pl}}$ is more strongly constrained than the assumed prior distribution, with the standard deviation of the posterior ($0.2\cdot\mathrm{log_{10}(M_\oplus)}$) being significantly smaller than the standard deviation of the prior ($0.4\cdot\mathrm{log_{10}(M_\oplus)}$). The centering of the $R_{\text{pl}}$ and $M_{\text{pl}}$ posteriors on the true values and the lack of significant correlation between the two posteriors implies that the surface gravity $g_{pl}$ is estimated accurately. The surface pressure $P_0$ and surface temperature $T_0$ are both accurately retrieved to a very high precision, with an uncertainty of roughly 0.1 K for $T_0$ and 0.1 bar for $P_0$ (see Figs.~\ref{fig:4a} and \ref{fig:4b}). Further, we observe a correlation between the planetary radius $R_{\text{pl}}$ and the surface temperature $T_0$. This indicates that a higher $T_0$, which results in more emission per surface area, can be compensated by a smaller $R_{\text{pl}}$, which results in a smaller emitting area.

From the retrieved posterior distribution of \ce{N2} we see that our retrieval framework allows us to rule out low \ce{N2} abundances in Earth's atmosphere via the \ce{N2}$-$\ce{N2} CIA feature at 4$\mu$m.  However, the same \ce{N2}-\ce{N2} CIA feature is too weak to rule out very high \ce{N2} abundances. In contrast, the retrieval did not manage to find evidence for \ce{O2} in Earth's atmosphere. However, the retrieval managed to limit the \ce{O2} abundance to maximally $0.35$ in mass fraction.

The retrieved posterior distributions for the remaining molecules show a strong correlation with the $M_{\text{pl}}$ posterior and consequently with the surface gravity $g_{pl}$. This correlation is evident in the corner plot of Fig.~\ref{fig:4a} since both $M_{\text{pl}}$ and the molecular abundances of most retrieved trace gases exhibit a similarly shaped, non-Gaussian posterior distribution. This is a well known physical degeneracy and has been described in other studies \citep[see,  e.g.,][]{Molliere:Gravity_Abundance_Degeneracy,2018AJ....155..200F,Madhusudhan:Atmospheric_Retrieval,Quanz:exoplanets_and_atmospheric_characterization}; it is not related to a numerical artifact. The degeneracy appears since the same spectral feature can be explained by different combinations of gravity and atmospheric composition. This degeneracy originates from the mass appearing in the form of the surface gravity in the hydrostatic equilibrium. Therein, the surface gravity is degenerate with the mean molecular weight of the atmospheric species. Since we derive the mean molecular weight from the abundances of the trace gases, this connects the planet's mass to the trace gas abundances present in its atmosphere. Further evidence for this degeneracy can be found in Fig.~\ref{fig:4c}. Despite the degeneracy between the retrieved mass and molecule abundances, the relative difference between the input spectrum and the spectra corresponding to the retrieved parameters is small.

The posteriors of \ce{CO} and \ce{N2O} are broader, roughly Gaussian, and less correlated with $M_{\text{pl}}$. This indicates that the constraint imposed by the retrieval framework on the abundances of these species is not solely limited by the degeneracy with $M_{\text{pl}}$, but also by our retrieval's sensitivity for \ce{CO} and \ce{N2O}.

A method of dealing with these strong correlations is to consider relative instead of absolute abundances. This allows us to minimize the impact of systematic uncertainties that affect all retrieved trace gas abundances in the same way at the cost of losing information on the absolute abundances. Relative abundances of trace gases are of interest to us since they provide a probe to whether an atmosphere is in chemical disequilibrium, which could potentially be upheld by life. For example, we could consider the abundance of \ce{CH4} or \ce{N2O} relative to a strongly oxidizing species such as \ce{O2} (or \ce{O3}, which is a photochemical product of \ce{O2}) as first suggested in \citet{Lovelock:CH4-O2} and \citet{Lippincott:CH4-O2_N2O-O2}. These gases react rapidly with each other and therefore the simultaneous presence of both molecules is only possible if they are continually replenished at a high rate. On Earth, \ce{O2} is constantly produced via photosynthesis and there is a continuous flux of \ce{CH4} into the atmosphere due to biological methanogenesis and anthropogenic methane production. Similarly, the \ce{N2O} in Earth's atmosphere is continually replenished by a large range of microorganisms via incomplete denitrification. On Earth, these biological processes lead to \ce{CH4} and \ce{N2O} abundances that are many orders of magnitude larger than the chemical equilibrium. Another interesting ratio of atmospheric gases to consider is the ratio between \ce{CO} and \ce{CH4}. A large amount of \ce{CO} accompanied by a lack of significant \ce{CH4} could be interpreted as an anti-biosignature as suggested in \citet{Zahnle:CO}. For a more exhaustive discussion of potential biosignatures we refer the reader to e.g. \citet{Schwieterman2018}.

The retrieved atmospheric P-T structure is displayed in Fig.~\ref{fig:4b}. Our retrieval framework extracts the P-T structure of Earth's lower atmosphere accurately to very high precision. With decreasing pressure, the uncertainty on the retrieved P-T structure increases due to a lack of signatures from the low-pressure atmospheric layers ($\lesssim10^{-4}$ bar) in the exoplanet's thermal emission spectrum.

\section{Results}
\label{Results}

In the following, we analyze the performance of the retrieval suite in characterizing an Earth-twin planet orbiting a Sun-like star at 10 pc from our Earth.  We first estimate the fundamental detection limits of our retrieval suite for the trace gas abundances for different input spectra, as described in Sect.~\ref{sens_anal}. We then introduce the full grid of retrievals that was run in Sect.~\ref{Retrievalgrid}. The results are presented in Sects.~\ref{subsec:ret_pl_grid} and \ref{subsec:ret_abund}.

\subsection{Detection limit analysis}\label{sens_anal}

We define the fundamental detection limits of our retrieval suite as the lowest possible molecular abundances that are retrievable for the atmospheric gases considered. For abundances below the detection limit, the corresponding spectral features are lost in the observational noise. Specifically, we estimate the detection limits of the trace gases considered in our retrievals at different values for R (20, 50, 1000), S/N (10, 20, 50), and for the largest and smallest assumed bandwidth ($3-20\ \mu\mathrm{m}$, $6-17\ \mu\mathrm{m}$).

To do so, we generate spectra from the parameter values listed in Table~\ref{table:3}, but set the abundance of one trace gas to 0 for each generated spectrum. We then use these spectra as retrieval input assuming only photon noise of the planet to be present and retrieve the absent trace gas. We pass all other model parameters to the retrieval as knowns. 

\bgroup
\def\arraystretch{1.3}%

\begin{table*}
\caption{Results obtained in the retrieval sensitivity analysis.}          
\label{table:4}      
\centering          
\begin{tabular}{lcccccccc}     
\hline\hline       
R & S/N &Range $\left[\mu\mathrm{m}\right]$ &$\mathrm{log_{10}\left(H_2O\right)}$ &$\mathrm{log_{10}\left(CO_2\right)}$ &$\mathrm{log_{10}\left(O_3\right)}$ &$\mathrm{log_{10}\left(CH_4\right)}$ &$\mathrm{log_{10}\left(CO\right)}$ &$\mathrm{log_{10}\left(N_2O\right)}$\\ 
\hline                    
\multirow{4}{*}{$20$} &\multirow{2}{*}{$10$}    &$3-20$ &$-6.13^{\pm0.38}$                              &$-6.51^{\pm0.33}$  &-7.22$^{\pm0.38}$   &$-6.04^{\pm0.51}$                         &$-4.14^{\pm1.32}$  &$-3.39^{\pm1.23}$   \\  
                    &                           &$6-17$ &$-6.11^{\pm0.54}$   &$-6.61^{\pm0.52}$  &$-7.04^{\pm0.30}$   &$-5.91^{\pm0.57}$ &UC                 &$-3.09^{\pm1.04}$   \\\cdashline{3-9}
                    &\multirow{2}{*}{$20$}      &$3-20$ &$-6.68^{\pm0.45}$   &$-7.08^{\pm0.48}$  &$-7.55^{\pm0.27}$   &$-6.68^{\pm0.53}$ &$-5.28^{\pm0.63}$  &$-4.55^{\pm1.05}$   \\
                    &                           &$6-17$ &$-6.63^{\pm0.40}$   &$-7.05^{\pm0.44}$  &$-7.47^{\pm0.37}$   &$-6.67^{\pm0.56}$ &UC                 &$-4.30^{\pm1.02}$   \\\cdashline{2-9}
\multirow{4}{*}{$50$} &\multirow{2}{*}{$10$}    &$3-20$ &$-6.60^{\pm0.49}$                              &$-7.08^{\pm0.52}$  &$-7.45^{\pm0.31}$   &$-6.49^{\pm0.42}$                         &$-5.26^{\pm0.67}$  &$-4.33^{\pm1.07}$   \\  
                    &                           &$6-17$ &$-6.47^{\pm0.45}$   &$-7.05^{\pm0.46}$  &$-7.46^{\pm0.28}$   &$-6.56^{\pm0.55}$   &UC                 &$-3.86^{\pm0.98}$   \\\cdashline{3-9}
                    &\multirow{2}{*}{$20$}      &$3-20$ &$-6.99^{\pm0.34}$   &$-7.48^{\pm0.34}$  &$-7.75^{\pm0.22}$   &$-7.09^{\pm0.45}$   &$-5.82^{\pm0.36}$  &$-5.28^{\pm0.75}$   \\
                    &                           &$6-17$ &$-6.98^{\pm0.34}$   &$-7.50^{\pm0.41}$  &$-7.84^{\pm0.36}$   &$-7.08^{\pm0.33}$   &UC                 &$-5.10^{\pm0.87}$   \\\cdashline{2-9}
$1000$&$50$&$3-20$&$-9.29^{\pm0.25}$&$-9.27^{\pm0.27}$&$-9.02^{\pm0.27}$&$-8.81^{\pm0.29}$&$-7.88^{\pm0.29}$&$-8.35^{\pm0.35}$\\
\hline 
\multicolumn{3}{l}{Terrestrial abundances:}&$-3.00$&$-3.39$&$-5.77$&$-6.52$&$-6.90$&$-6.50$\\
\multicolumn{3}{l}{\textit{LIFE} retrieval expectation:}&\textcolor{Green}{$\boldsymbol{\checkmark}$}&\textcolor{Green}{$\boldsymbol{\checkmark}$}&\textcolor{Green}{$\boldsymbol{\checkmark}$}&\textcolor{BurntOrange}{$\boldsymbol{\sim}$}&\textcolor{Red}{$\boldsymbol{\times}$}&\textcolor{Red}{$\boldsymbol{\times}$}\\\hline
\end{tabular}
\tablefoot{All abundances are given in mass fractions in $\log_{10}$. We provide the threshold abundances for a detection at different R and S/N configurations, which we define as the half maximum  of the retrieved soft step posterior distribution. Additionally, we provide the 16\textsuperscript{th} and 84\textsuperscript{th} percentile of the logistic function as denoted by the '$\pm$'. 'UC' stands for unconstrained and signifies that the retrieval does not manage to constrain the species of interest. In the last two rows, we display the assumed Earth-twin abundances and whether we expect such an abundance to be retrievable in low R and S/N cases, respectively: '\textcolor{Green}{$\boldsymbol{\checkmark}$}' $=$ retrievable; '\textcolor{BurntOrange}{$\boldsymbol{\sim}$}' $=$ case dependant; '\textcolor{Red}{$\boldsymbol{\times}$}' $=$ not retrievable.}
\end{table*}

\egroup

By construction, the retrieval routine should rule out any abundance of the missing gas down to the detection limit abundance at which the spectral features can no longer be distinguished from the noise. The posterior distribution should therefore be a flat distribution (all values are equally probable) for abundances smaller than the threshold value. For values greater than the detection limit, the probability should be close to null. The threshold value can therefore be interpreted as a detection limit for the gas abundance.

The resulting posterior distribution for the trace gas can be approximated by the logistic function (a soft step function):
\begin{equation}\label{equ:logistic_f}
    f(x) = \frac{c}{1+e^{a\cdot x+b}}.
\end{equation}
Here, $x$ is the concentration of the trace gas for which the detectability is being considered. The constants a, b, and c are unique for each posterior. In Table~\ref{table:4}, we provide the abundance corresponding to the half maximum and the $16$\textsuperscript{th} and $84$\textsuperscript{th} percentile of the fitted logistic function for all tests we ran. The half maximum together with the percentiles provides an estimate for the detection limit of our retrieval suite.

The results we obtain for the R~=~1000 test case (see Table~\ref{table:4}) predict that all atmospheric trace gases should be detectable in such a retrieval setup. This is in agreement with the retrieval validation presented in Sect.~\ref{Validation}.

For the cases with R~$\leq$~50, we predict \ce{H2O}, \ce{CO2} and \ce{O3} to be easily detectable in an Earth-like atmosphere, since the true abundances are more than one order of magnitude larger than the retrieval's estimated detection limit. For \ce{CH4}, the true abundance is comparable to the retrieval's detection limit. Thus, we expect the performance for \ce{CH4} to depend strongly on the R and S/N of the input spectrum. The true \ce{CO} and \ce{N2O} abundances lie at least one order of magnitude below the estimated detection limits and are therefore irretrievable. The upper limit of \ce{CO} exhibits a strong dependence on the wavelength range considered because the only \ce{CO} feature in the MIR is located at $\sim4.7\,\mu\mathrm{m}$ (see Fig.~\ref{fig:3}). Excluding the wavelength range $3-6\ \mu\mathrm{m}$ from the analysis makes the \ce{CO} abundance impossible to constrain.

This test study has provided best-case detection limits for the abundance in Earth-like atmospheres. However, retrieval of all parameters will lead to an overall increase in these detection limits. Additionally, adding additional astrophysical noise terms will also negatively impact the detection limits for the trace gases.

\subsection{Retrieval grid}\label{Retrievalgrid}

We choose to consider the following grid of wavelength ranges, R values and S/N values in our final grid of retrieval studies:
\begin{itemize}
    \item Wavelength ranges: $3-20\,\mu\mathrm{m}$, $4-18.5\,\mu\mathrm{m}$, and $6-17\,\mu\mathrm{m}$
    \item Spectral resolutions: R~=~20, 35, 50, and 100.
    \item Signal-to-noise ratios: S/N~=~5, 10, 15, and 20 fixed at the wavelength bin centered at $11.2\,\mu\mathrm{m}$.
\end{itemize}
The short end of the wavelength range tests,  whether the CO band at $\sim$4.7 $\mu$m can be retrieved and whether including the $\sim$3.3 $\mu$m band of CH$_4$ helps the detection of this molecule. At the long wavelength end, the main question is how much of the extended water feature should be included in the analysis. The choice for the spectral resolution range was motivated by earlier studies \citep[e.g.,][]{desmarais2002}.

For the reference planet, we again assume the thermal emission spectrum of an Earth-twin in orbit around a G$2$ star located at a distance of $10$ pc from Earth. We consider two different observational cases with slightly different noise properties and observational setup:

\begin{itemize}
    \item Nominal case: (1) the \textit{LIFE} baselines (physical distance between the four mirrors) were optimized for the detection of habitable zone planets at a wavelength of $\lambda = 15\,\mu\mathrm{m}$ \citep[cf.][]{quanz2021}; (2) the level of exozodiacal dust emission corresponds to 3-times the level of the local zodiacal light. 
    \item Optimized case: (1) the \textit{LIFE} baselines were optimized for the detection of habitable zone planets at the short wavelength end; (2) the level of exozodiacal dust emission corresponds to 0.5-times the level of the local zodiacal light\footnote{According to the brightness distribution of exozodi disks used in recent \emph{LIFE} detection yield estimate \citep{quanz2021}, which is based on the results from the HOSTS survey \citep{ertel2020}, $\lesssim$20\% of the disks show such low emission.}.
\end{itemize}

\begin{figure}[t]
\centering
\includegraphics[width=0.49\textwidth]{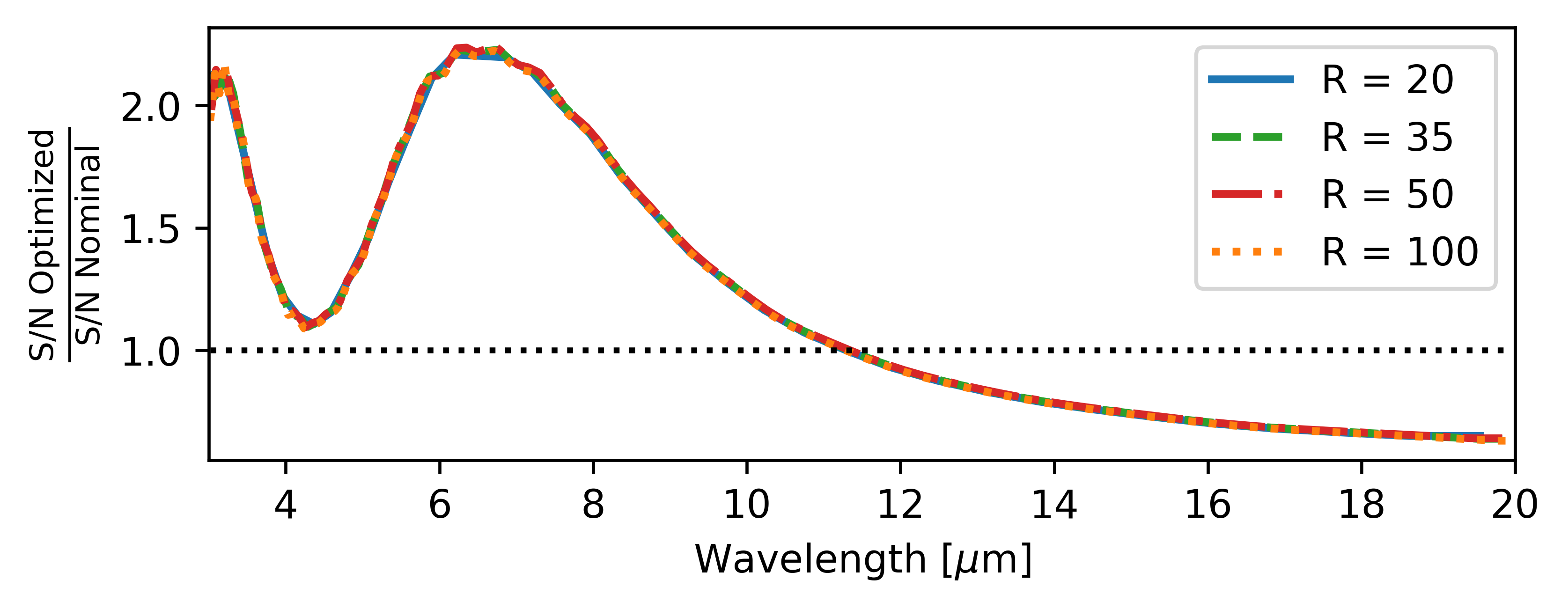}
\caption{Ratio between the wavelength-dependent S/N of the optimized case and the nominal case. This ratio is independent of the overall S/N and the R of the Spectrum.}
\label{fig:5}
\end{figure}

Figure~\ref{fig:5} visualizes the difference between the nominal and the optimized S/N case by plotting the ratio between the two S/N instances. The full noise terms (including stellar leakage, local zodi and exozodi emission, and photon noise from the planet) were computed with \textsc{LIFEsim} \citep[see][]{dannert2022large}. In total, the grid comprised 96 retrieval analyses. 

\begin{figure*}[t]
\centering
\includegraphics[width=\textwidth]{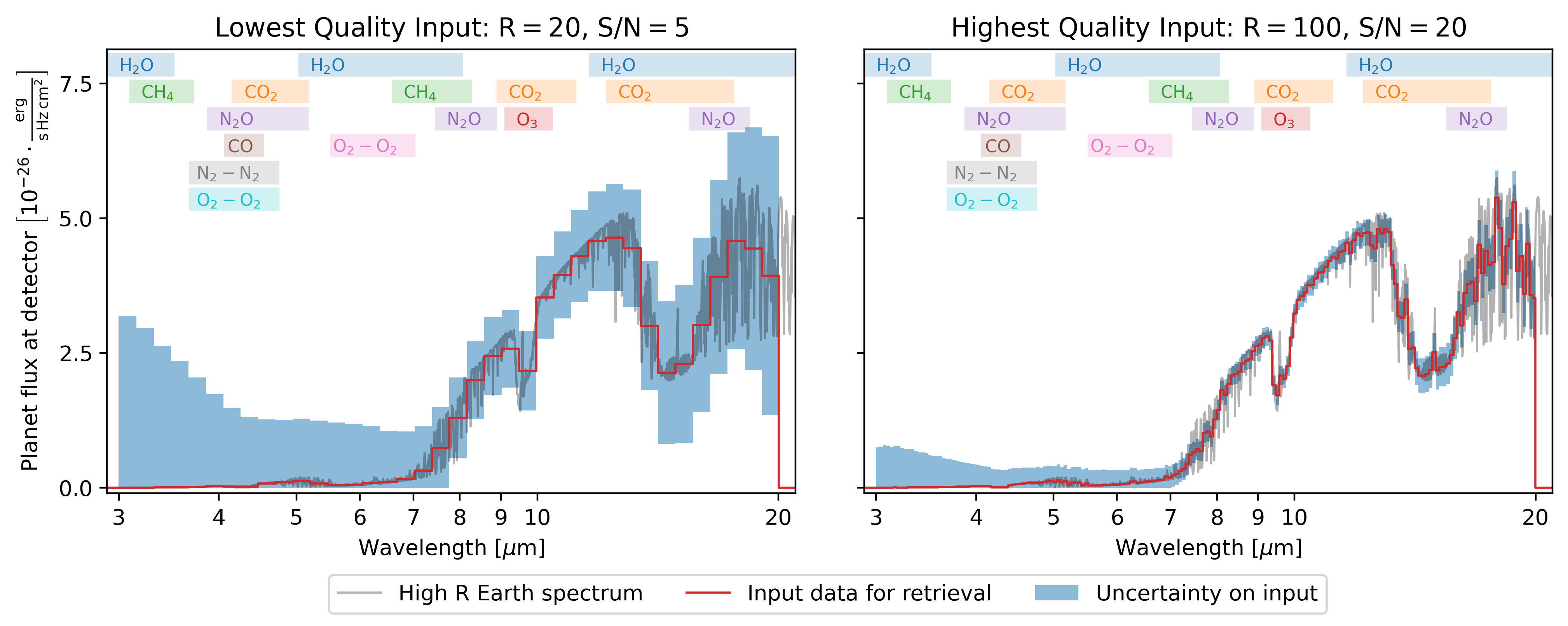}
\caption{Examples of input spectra used in the grid retrievals for the nominal case (left: the lowest quality input with R~=~20, S/N~=~5; right: the highest quality input with R~=~100, S/N~=~20). In gray, we provide the full resolution \texttt{petitRADTRANS} Earth spectrum. The red step function represents the wavelength-binning of the input data. Further, the blue shaded region represents the uncertainty for the corresponding bin. We also mark the absorption features of the considered atmospheric gases.}
\label{fig:6}
\end{figure*}

Figure~\ref{fig:6} visualizes the highest (R~=~100, S/N~=~20) and lowest (R~=~20, S/N~=~5) quality input spectra for the nominal case. For every grid point specified above, we run a retrieval assuming the prior distributions listed in Table~\ref{table:3}. Furthermore, we use the same \texttt{pyMultiNest} settings applied in the retrieval validation run (Sect.~\ref{Validation}).

\begin{figure}[t]
\centering
\subfloat{\includegraphics[width=.49\textwidth]{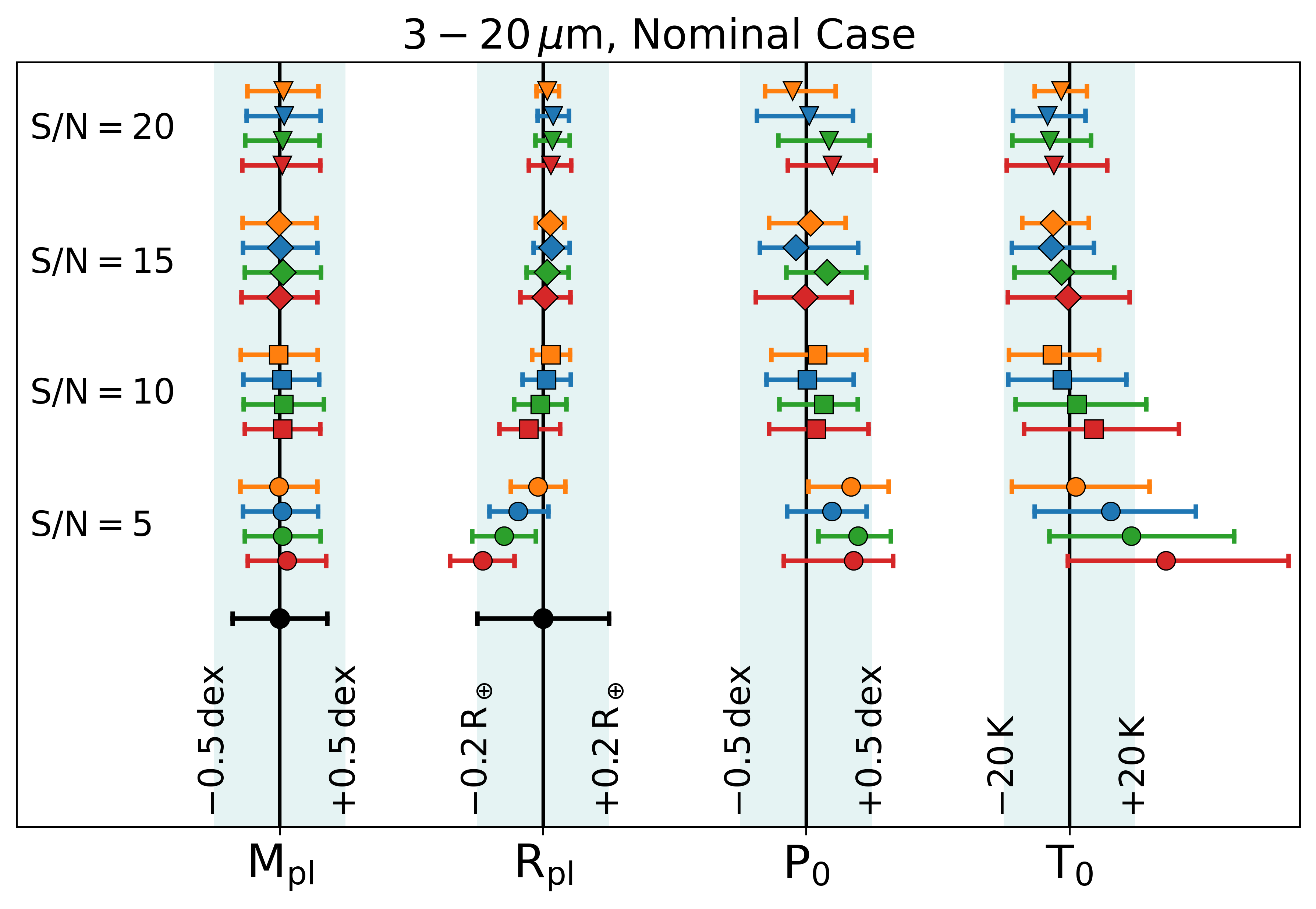}}
\,
\subfloat{\includegraphics[width=.40\textwidth]{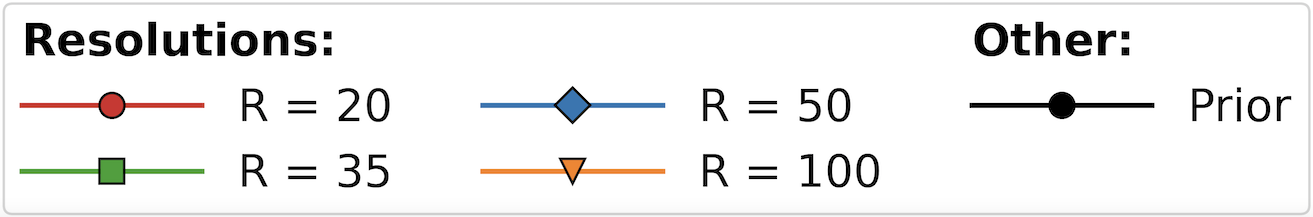}}
\caption{Retrieved exoplanet parameters for the different grid points for the wavelength range $3-20\,\mu\mathrm{m}$ in the nominal case. $M_{\text{pl}}$ is the mass, $R_{\text{pl}}$ the radius, $P_0$ the surface pressure and $T_0$ the surface temperature of the exoplanet. The error bars denote the $68\%$ confidence intervals. For $M_{\text{pl}}$ and $R_{\text{pl}}$, we also plot the assumed prior distributions. For $T_0$ and $P_0$, we assumed flat, broad priors. The vertical lines mark the true parameter values.}
\label{fig:7}
\end{figure}

\begin{figure*}[t]
\centering
\subfloat{\includegraphics[width=.49\textwidth]{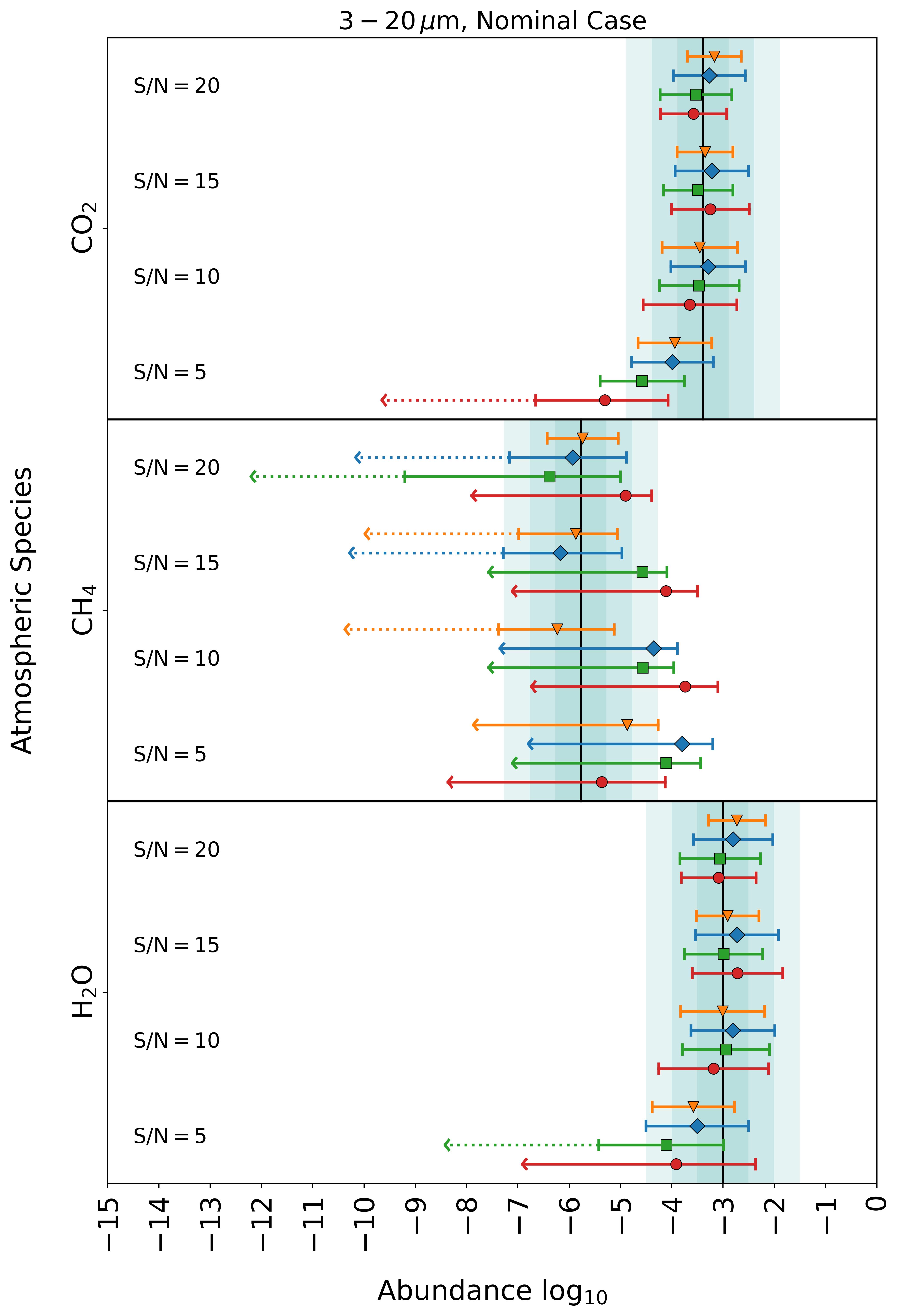}}
\,
\subfloat{\includegraphics[width=.49\textwidth]{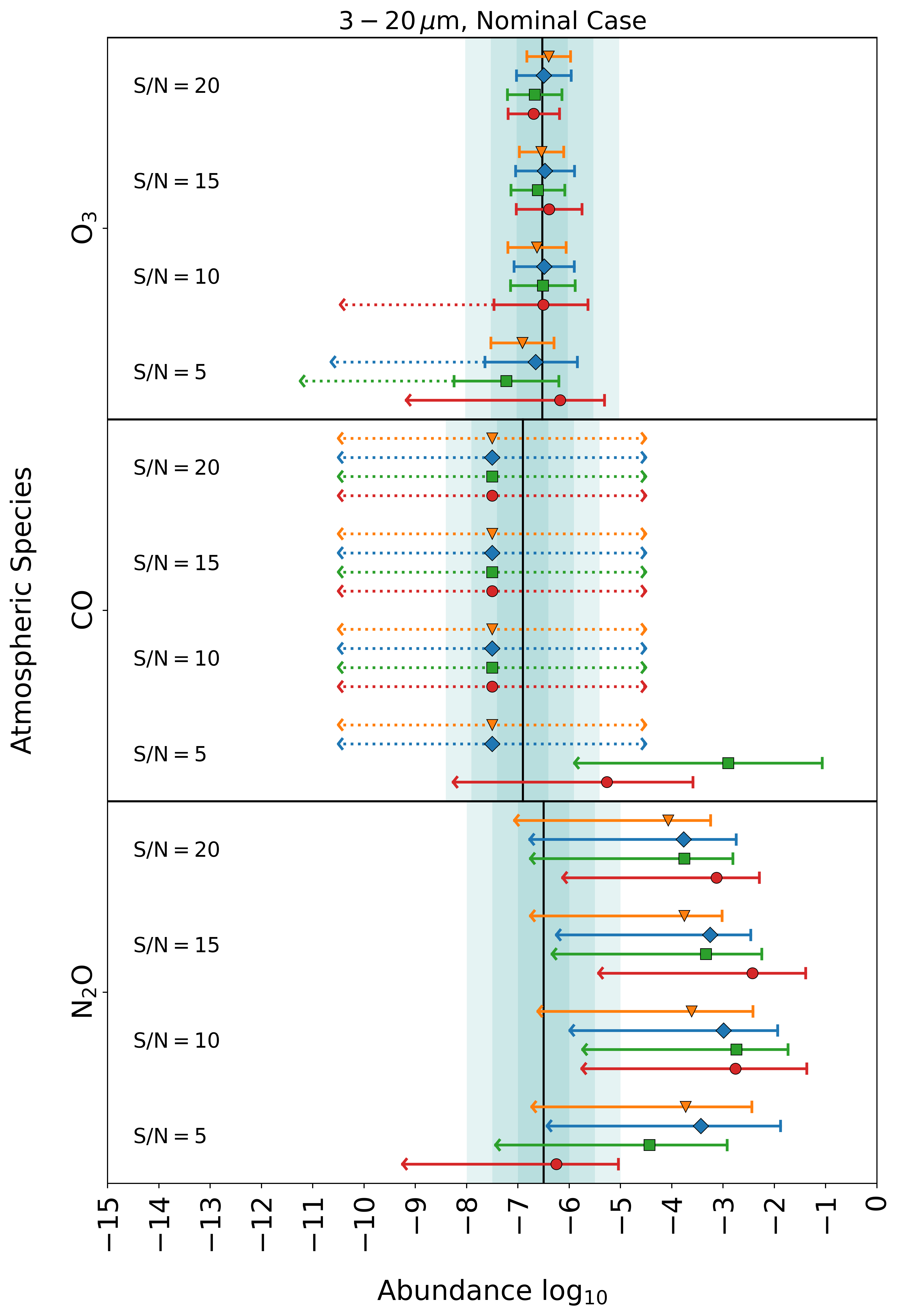}}
\,
\subfloat{\includegraphics[width=.75\textwidth]{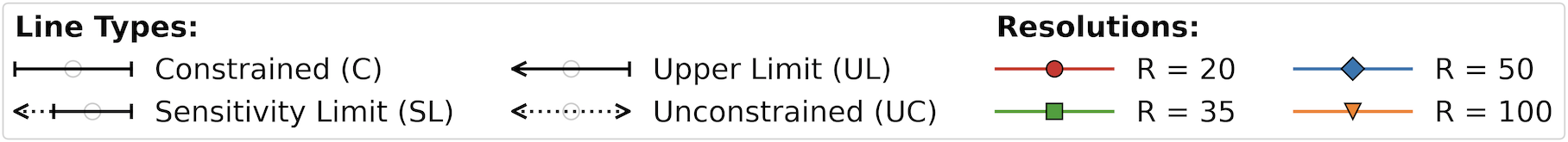}}
\caption{Retrieved mass mixing abundances of the different trace gases present in Earth's atmosphere for an input spectrum wavelength range of $3-20\,\mu\mathrm{m}$ in the nominal case. The vertical lines mark the true abundances whereas the shaded regions mark the $\pm0.5\,\mathrm{dex}$, $\pm1.0\,\mathrm{dex}$, and $\pm1.5\,\mathrm{dex}$ regions.}
\label{fig:8}
\end{figure*}

Taking the $3-20\,\mu$m nominal case input spectra as an example, we plot the posteriors of the planetary parameters (Fig.~\ref{fig:7}), as well as those for the absolute (Fig.~\ref{fig:8}) and the relative (Fig.~\ref{fig:9}) abundances. We will use the retrieved posterior probability distributions to underline trends with respect to the wavelength range, R and S/N. For the molecular abundances, we differentiate between four different classes of posterior distributions:
\begin{itemize}
    \item Constrained (C): The true atmospheric abundance lies above the retrieval's detection limit. The posterior is described satisfactorily by a Gaussian. Thus, both significantly higher and lower abundances are ruled out.
    \item Sensitivity limit (SL): The true atmospheric abundance is comparable to the detection limit of the retrieval. We observe a distinct peak in the posterior. However, lower abundances cannot be fully ruled out. The posterior is best described by the convolution of a logistic function with a Gaussian.
    \item Upper Limit (UL): The true atmospheric abundance lies below the retrieval’s detection limit and could be zero. The posterior is best described by a logistic function discussed in Sect.~\ref{sens_anal}. Abundances above the detection limit can be excluded, all abundances below the limit are equally likely.
    \item Unconstrained (UC): No information about the atmospheric abundance can be retrieved. The retrieved posterior is best described by a constant function.
\end{itemize}
For more detailed information about our posterior classification, see Appendix~\ref{PostClass}.


Figure~\ref{fig:10} summarizes the type of the retrieved posterior distribution for the molecular abundances for all considered R, S/N, wavelength ranges and the two different cases. Tables containing the median as well as the 16\textsuperscript{th} and 84\textsuperscript{th} percentile of the retrieved posteriors for all retrieval runs are provided in the Appendix~\ref{app:res} (Tables~\ref{table:3-20}-\ref{table:6-17_opt}).

\subsection{Retrieved planetary parameters}
\label{subsec:ret_pl_grid}

In Fig.~\ref{fig:7} we provide the retrieval results for the exoplanet parameters $M_{\text{pl}}$, $R_{\text{pl}}$, $P_0$, and $T_0$ ($3-20\,\mu$m input spectra in the nominal case). Our retrieval framework estimates all planetary parameters correctly.

For the planetary mass $M_{\text{pl}}$, the retrieved posterior is centered on the true value and roughly corresponds to the assumed prior distribution. This result indicates that our retrieval framework cannot extract further information from the input spectrum.

In contrast to $M_{\text{pl}}$, we manage to strongly constrain the exoplanet's radius $R_{\text{pl}}$ with respect to the assumed prior distribution. For all S/N~$\geq10$, we retrieve an accurate estimate for $R_{\text{pl}}$ ($\Delta R_{\text{pl}}\leq\pm10\%$).

For all S/N~$\geq10$, our retrievals yield strong constraints for the surface pressure $P_0$. The retrieved value lies within maximally $\pm0.5$ dex (a factor of $\approx3$) depending on the R and S/N combination. Similarly, the surface temperature $T_0$ (calculated from the retrieved P-T profile parameters and $P_0$) is accurately estimated by our retrieval framework for all S/N~$\geq10$. The retrieved values are centered on the correct value for most cases and the $1\,\sigma$ uncertainties are mostly smaller than $\pm20\,K$.

For an S/N of $5$, the retrieved parameters $T_0$ and $P_0$ exhibit significant offsets with respect to the input value. Furthermore, we find similar offsets for $R_{\text{pl}}$. These deviations indicate that an S/N of $5$ is too low for the accurate characterization of an Earth-twin exoplanet's atmospheric structure, as we discuss further in Sect.~\ref{subsec:ret_abund}. The same is true for all other wavelength ranges and both the nominal and the optimized cases considered in our work (see Appendix~\ref{app:res}).

\subsection{Retrieved abundances}
\label{subsec:ret_abund}

The retrieved molecular abundances of the atmospheric trace gases are provided in Fig.~\ref{fig:8} ($3-20\,\mu$m input spectra for the nominal case). Our results show the following:
\begin{itemize}

    \item The abundances of \ce{CO2}, \ce{H2O} and \ce{O3} are accurately retrieved to a precision of $\leq\pm1\,\mathrm{dex}$ for all cases where the S/N is $\geq10$.
    
    \item \ce{CH4} only becomes retrievable for high combinations of R and S/N (for S/N~=~10 at R~$\geq100$, for S/N~=~15 at R~$\geq50$ and for S/N~=~20 a R~$\geq35$). For other combinations of R and S/N, we only retrieve upper limits on the \ce{CH4} abundance.
    
    \item The \ce{CO} abundance is not constrained by our retrievals for any S/N $\geq10$. At an S/N of 5 we retrieve an upper limit on the abundance of \ce{CO}. This upper limit is a result of the poor overall retrieval performance in the S/N~=~5 case as we will motivate below. 
    
    \item We do not retrieve the \ce{N2O} abundance, but find an upper limit for the maximal possible abundance. The position of this upper limit decreases significantly with increasing R and S/N of the input spectrum.
    
    \item For all considered R and S/N combinations, the input spectra do not contain sufficient information to constrain the abundances of the bulk constituents \ce{N2} and \ce{O2} (not shown). We retrieve UC-type posteriors and cannot provide any constraint for \ce{N2} and \ce{O2} in our retrievals.
    
    \item Adding up the retrieved abundances, we find that there is at least one additional atmospheric gas that has no MIR signature probable by \textit{LIFE} and constitutes $\approx99\%$ of Earth's atmospheric mass and has no significant absorption feature in the MIR. We can exclude \ce{H}/\ce{He} dominated atmospheres due to the small retrieved radius and \ce{CO2}, \ce{CH4}, or \ce{H2O} dominated atmospheres due to our retrieval results. These findings already put strong constraints on the bulk atmospheric composition.
    
\end{itemize}
Overall, these findings agree well with the predictions obtained in the sensitivity analysis (Sect.~\ref{sens_anal}), demonstrating that the simplified assumptions used in that analysis are reasonable.

\begin{figure*}[t]
\centering
\subfloat{\includegraphics[width=.49\textwidth]{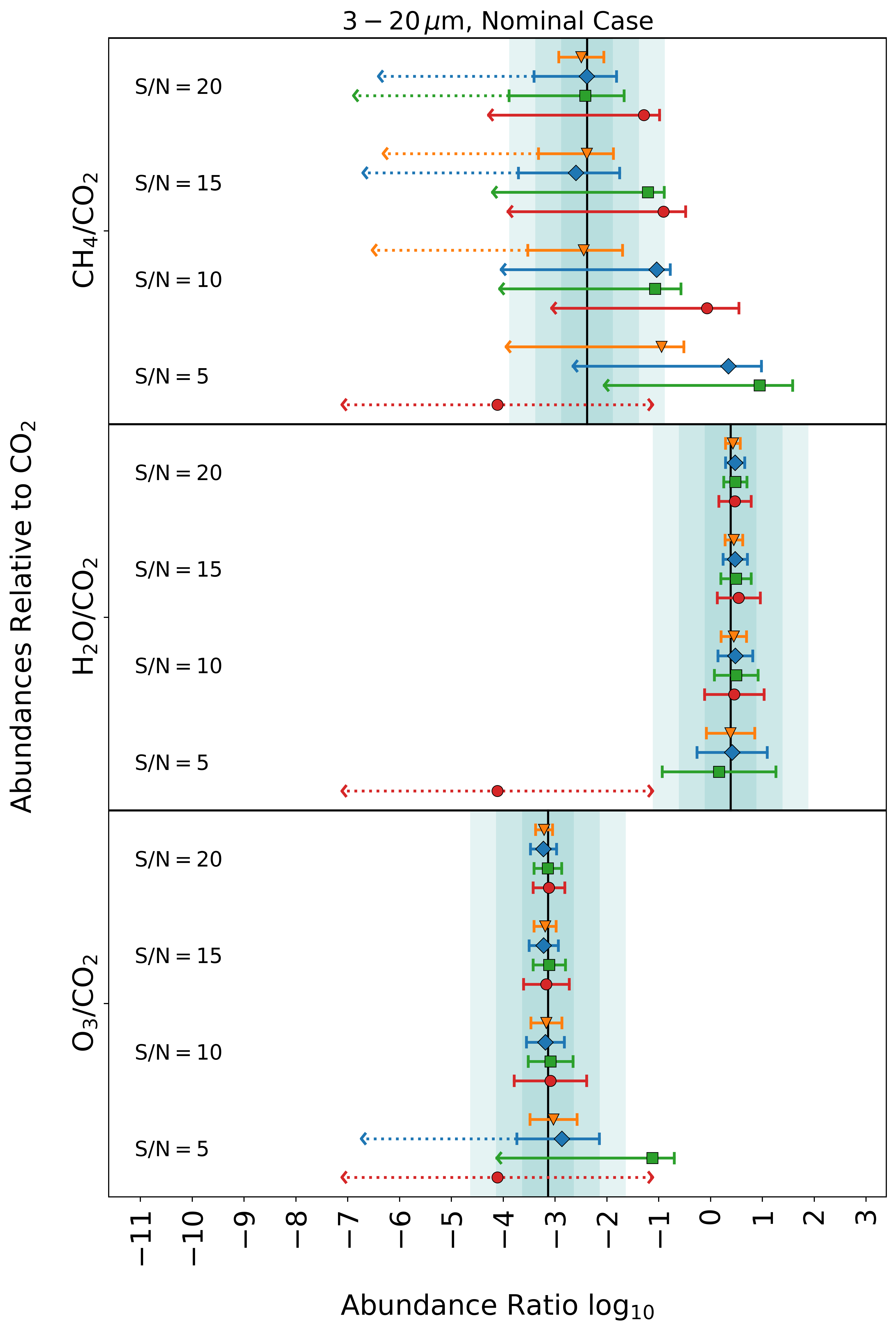}}
\,
\subfloat{\includegraphics[width=.49\textwidth]{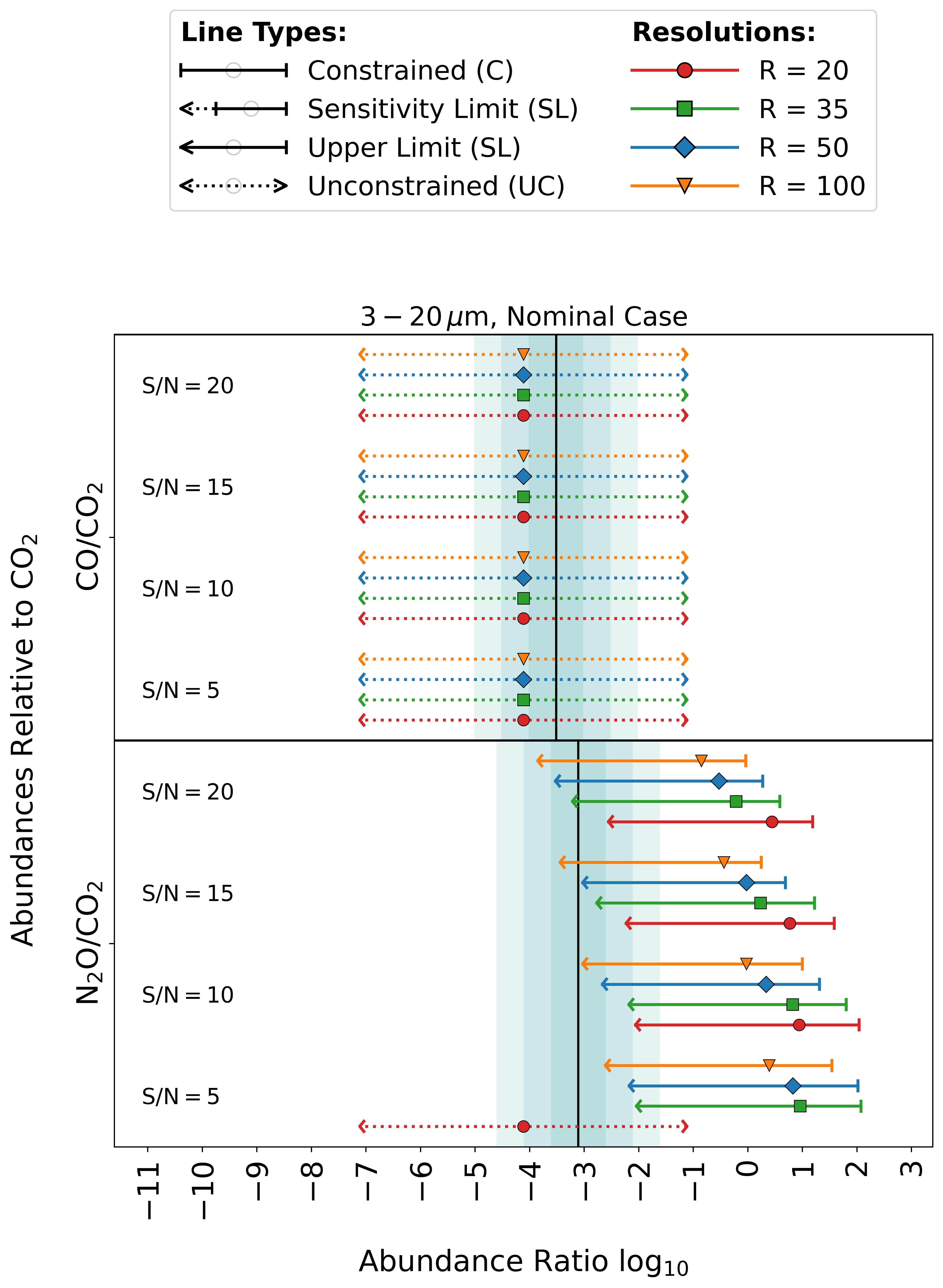}}

\caption{As for Fig.~\ref{fig:8} but for retrieved abundances relative to \ce{CO2}.}
\label{fig:9}
\end{figure*}

Input spectra with an S/N of $5$ do not contain sufficient information to yield accurate retrieval predictions for the absolute abundances of the considered trace gases. This is in accordance with our findings for the planetary parameters. For \ce{CO2}, \ce{H2O} and \ce{O3}, we tend to underestimate the true atmospheric abundances. Similarly, the upper limits on the abundances of \ce{N2O} retrieved at S/N~=~5 are lower than at S/N~$\geq10$ and for \ce{CO} we retrieve upper limits, which are no longer found at higher S/N. The underestimation of abundances at S/N~=~5 is coupled with the overestimation of the surface pressure $P_0$ and temperature $T_0$, which are again compensated by an underestimation of the radius $R_\mathrm{pl}$. A higher $P_0$ leads to a higher atmospheric mass and therefore to more absorbing material between the planet surface and the observer, which will lead to deeper absorption features at constant molecular abundances; the same line in a MIR emission spectrum would then be produced by a smaller abundance of the atmospheric species. Hence, in this case the retrieved abundances lie below the true ones. By considering the relative abundances we can reduce these offsets (Fig.~\ref{fig:9} shows the trace gas posteriors relative to the \ce{CO2} posterior for the nominal case, $3-20\,\mu$m input spectrum). This indicates that the offsets share the same properties for all atmospheric gases, indicating that they are caused by degeneracies between parameters (pressure-abundance and gravity-abundance degeneracies). These degeneracies are larger in the small S/N and R cases, since the constraints posed by the input spectrum are smaller. We further observe that the true values still lie within the posterior range (note that the plots only show the 16\textsuperscript{th}, 50\textsuperscript{th}, and 84\textsuperscript{th} percentiles). If the degeneracies are asymmetric (e.g. lower surface pressures and temperatures are easier to exclude than high ones), this would result in the observed asymmetric positioning of the posterior distribution around the truth, causing the observed offsets. The offsets diminish at higher R and S/N because the retrieval input provides stronger constraints and thus manages to reduce/break these degeneracies.

For S/N~$\geq$~10, the relative abundances of \ce{H2O}, \ce{O3} and \ce{CH4} (if retrieved) are centered on the true values and the corresponding uncertainties are significantly smaller than for the absolute abundances ($\leq\pm0.5\,\mathrm{dex}$). The reduction in uncertainty is due to the elimination of the gravity-abundance degeneracy, since this degeneracy affects all retrieved abundances comparably. Likewise, at S/N~=~5, the offsets we observed for the retrieved absolute abundances  are strongly diminished when considering relative abundances. This allows us to find accurate estimates for the relative abundances of \ce{CO2}, \ce{H2O} and \ce{O3} despite the inaccurately retrieved absolute abundances. 

These findings demonstrate that considering relative abundances can significantly diminish the effects of degeneracies between trace gas abundances and other atmospheric parameters. This occurs at the cost of losing information on the absolute abundances. However, the relative abundances of trace gases can still contain vital information on planetary conditions and provide potential biosignatures \citep[see Sect.~\ref{Validation} and, e.g.,][]{Lovelock:CH4-O2,Lippincott:CH4-O2_N2O-O2,Meadows2018}.

\begin{figure*}[t]
\centering
\subfloat[][$3-20\,\mu\mathrm{m}$, nominal case]{\includegraphics[width=.321\textwidth]{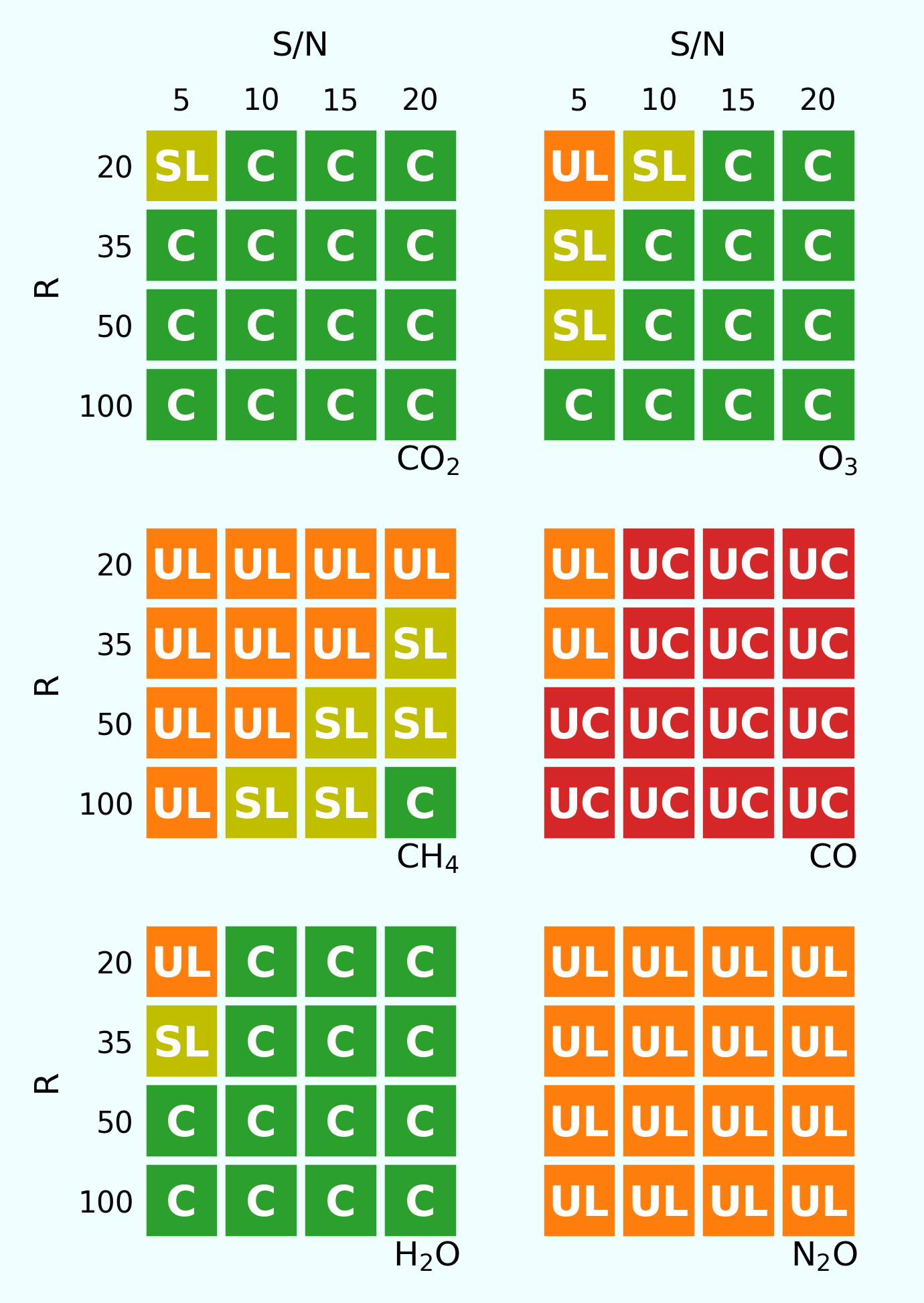}}
\,
\subfloat[][$4-18.5\,\mu\mathrm{m}$, nominal case]{\includegraphics[width=.321\textwidth]{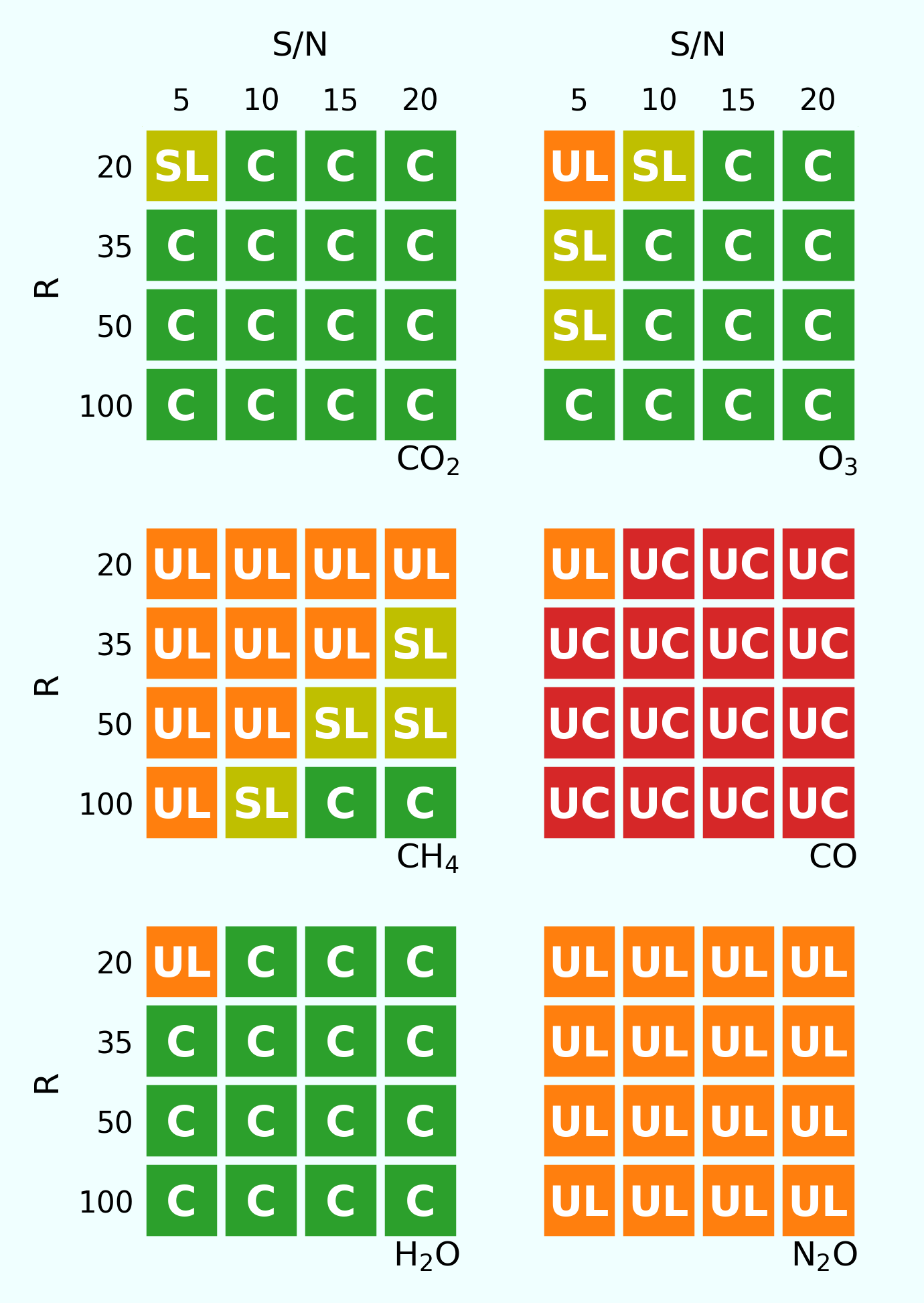}}
\,
\subfloat[][$6-17\,\mu\mathrm{m}$, nominal case]{\includegraphics[width=.321\textwidth]{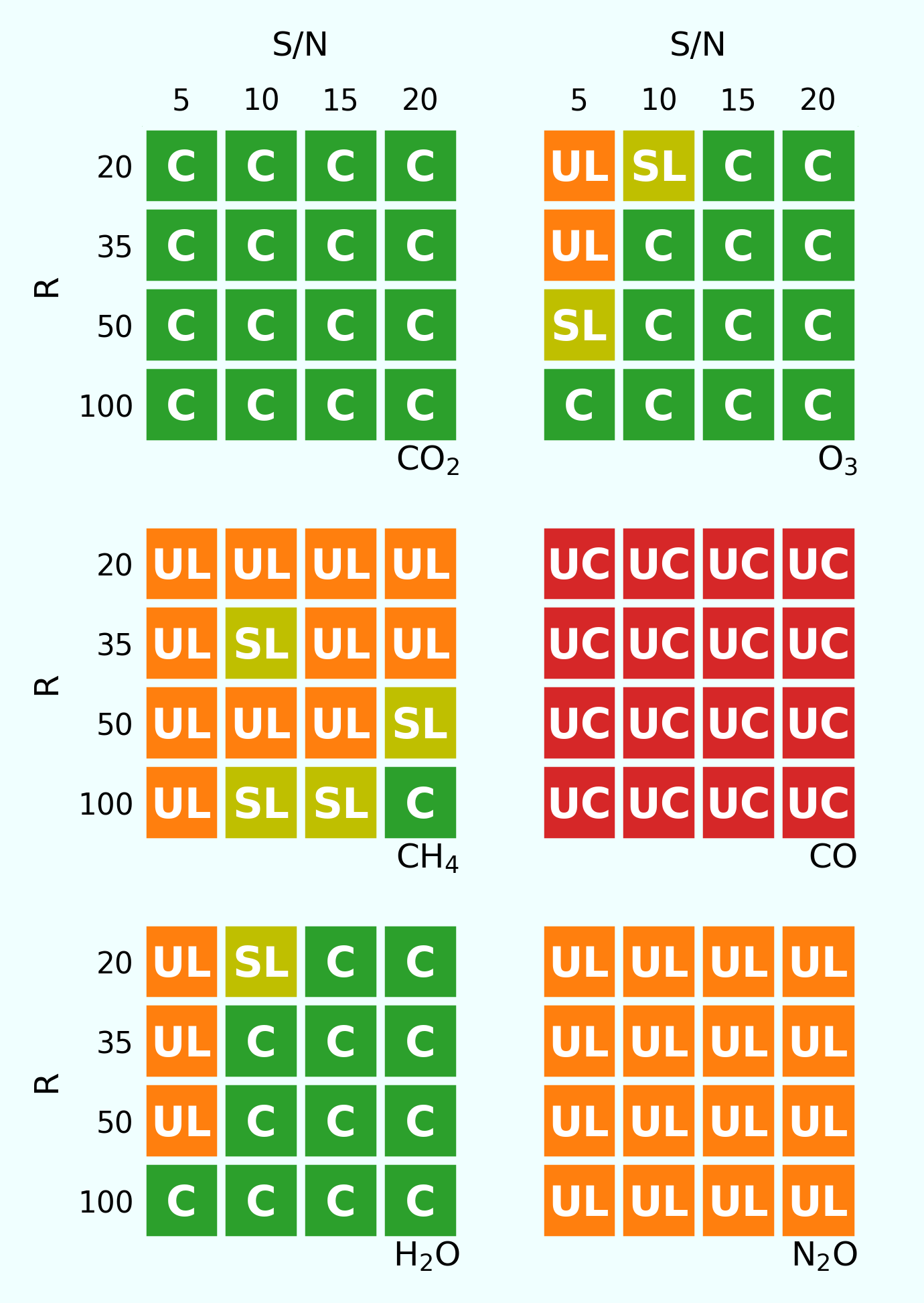}}
\,
\subfloat[][$3-20\,\mu\mathrm{m}$, optimized case]{\includegraphics[width=.321\textwidth]{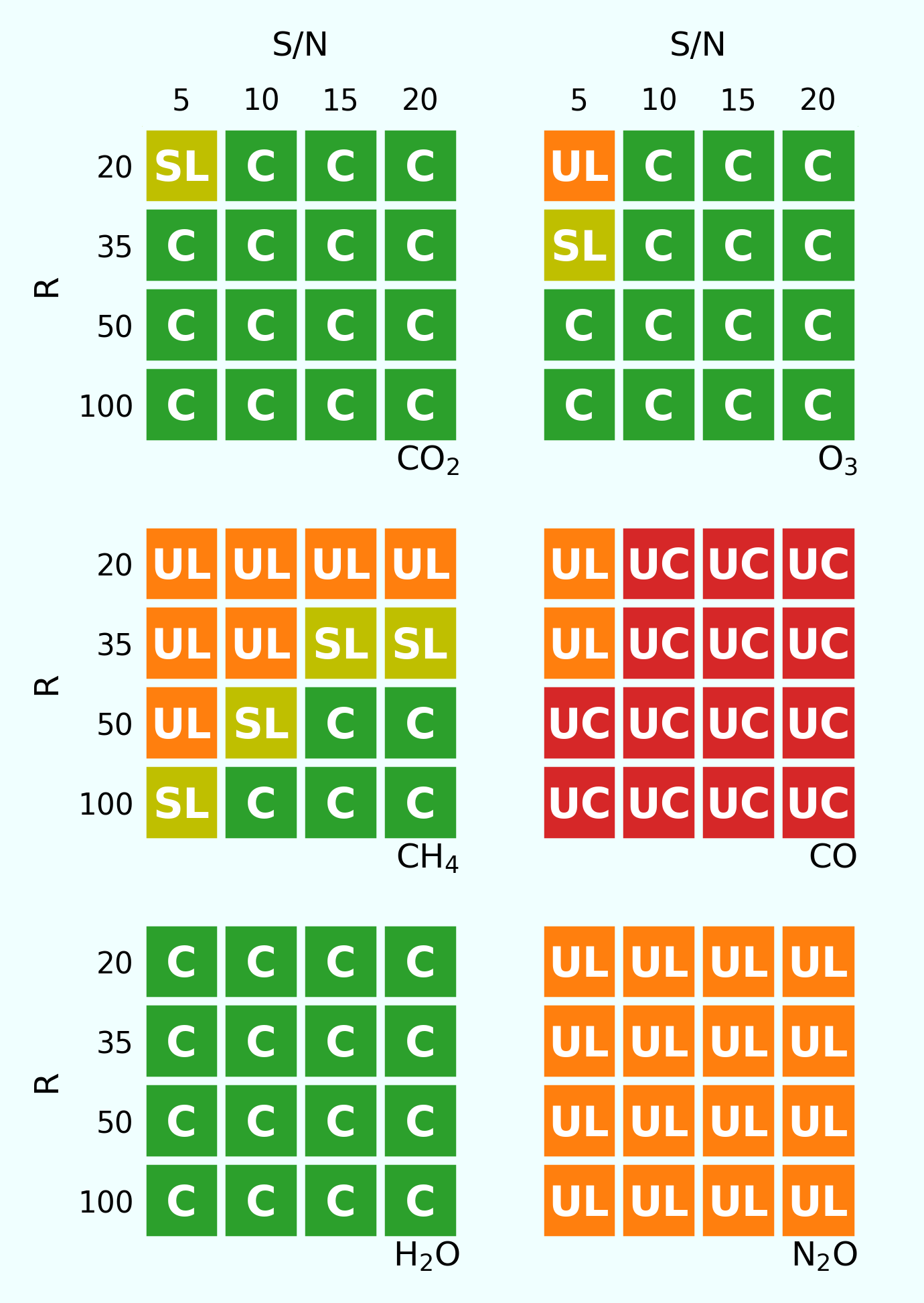}}
\,
\subfloat[][$4-18.5\,\mu\mathrm{m}$, optimized case]{\includegraphics[width=.321\textwidth]{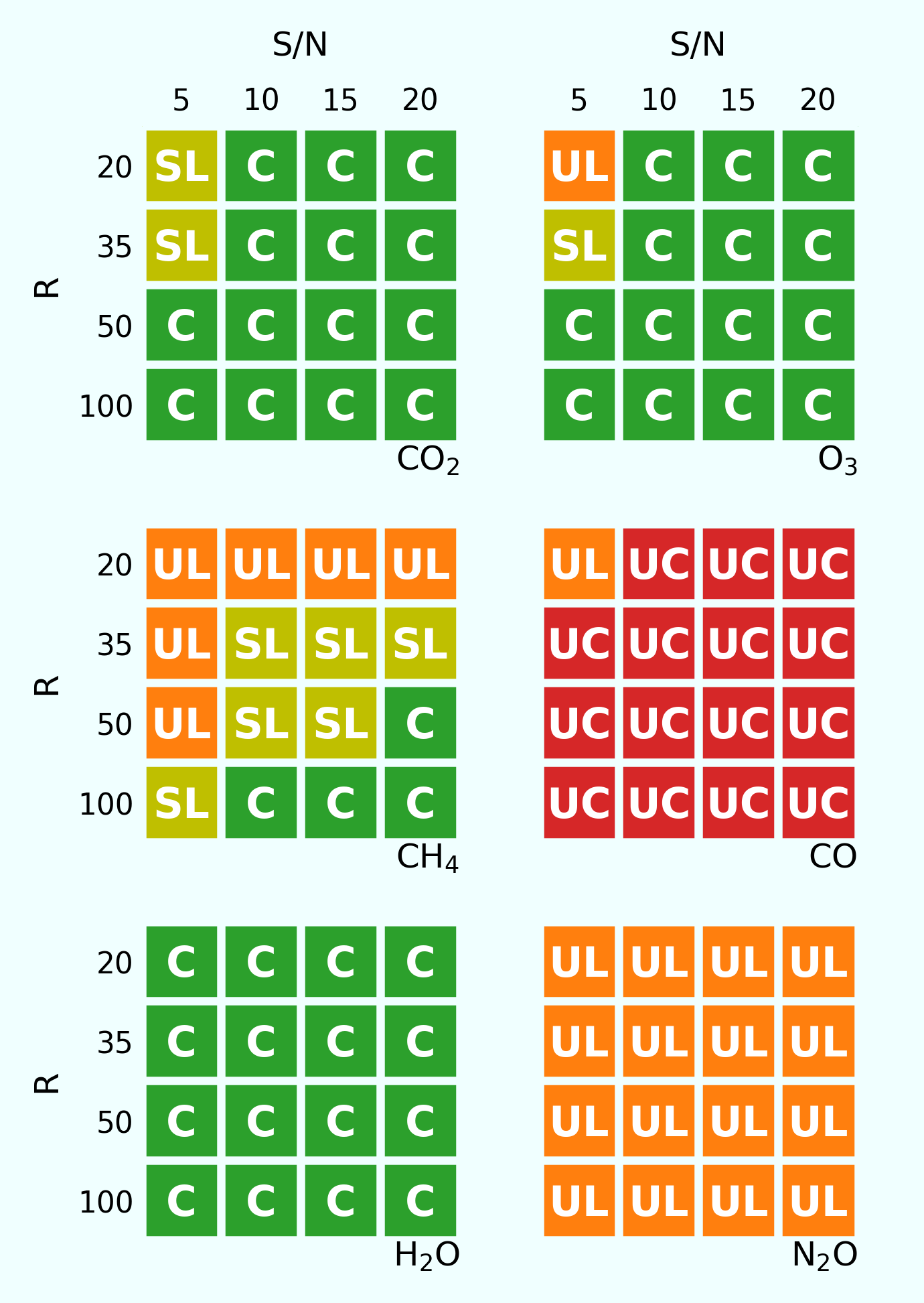}}
\,
\subfloat[][$6-17\,\mu\mathrm{m}$, optimized case]{\includegraphics[width=.321\textwidth]{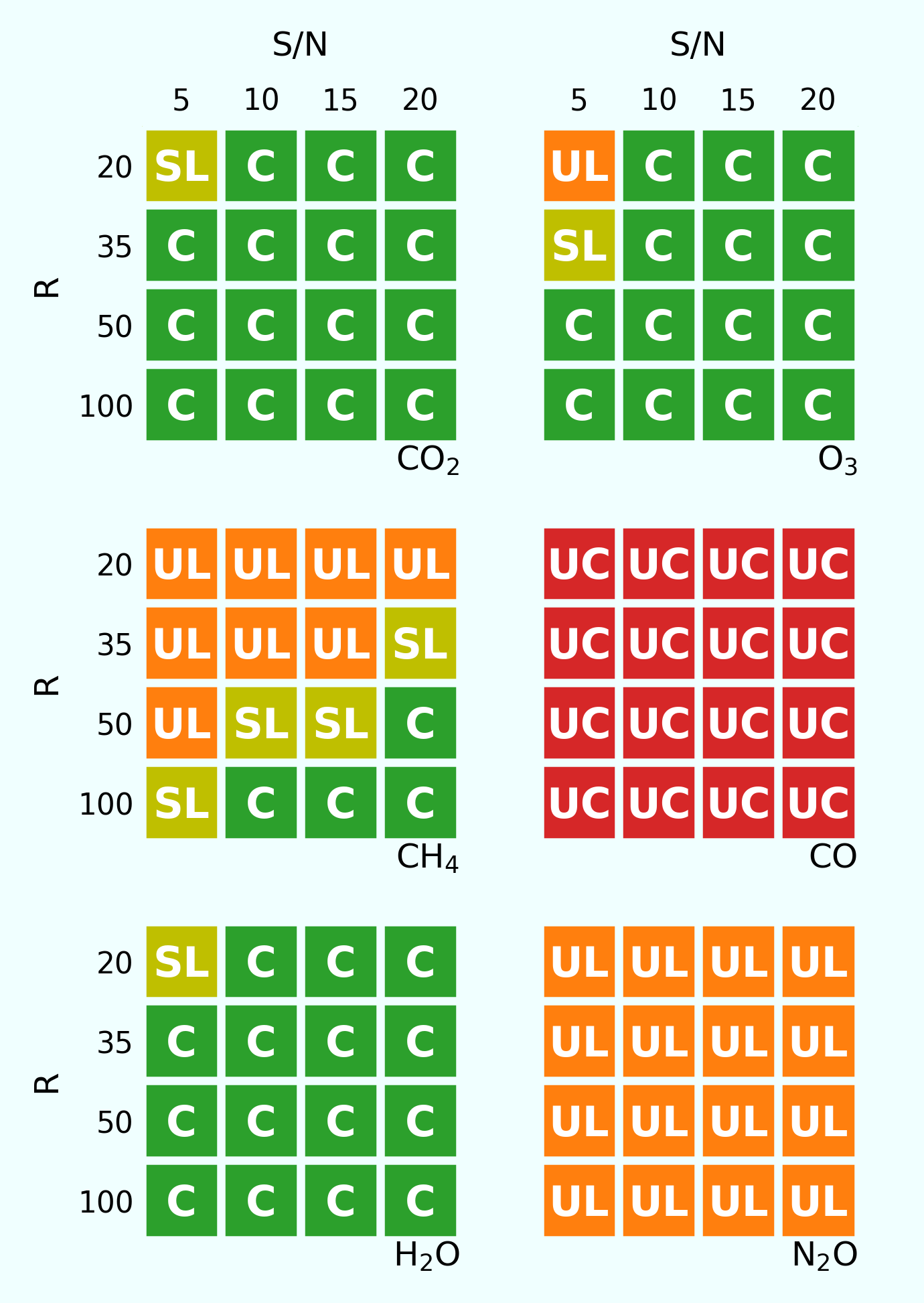}}
\,
\subfloat{\includegraphics[width=.4\textwidth]{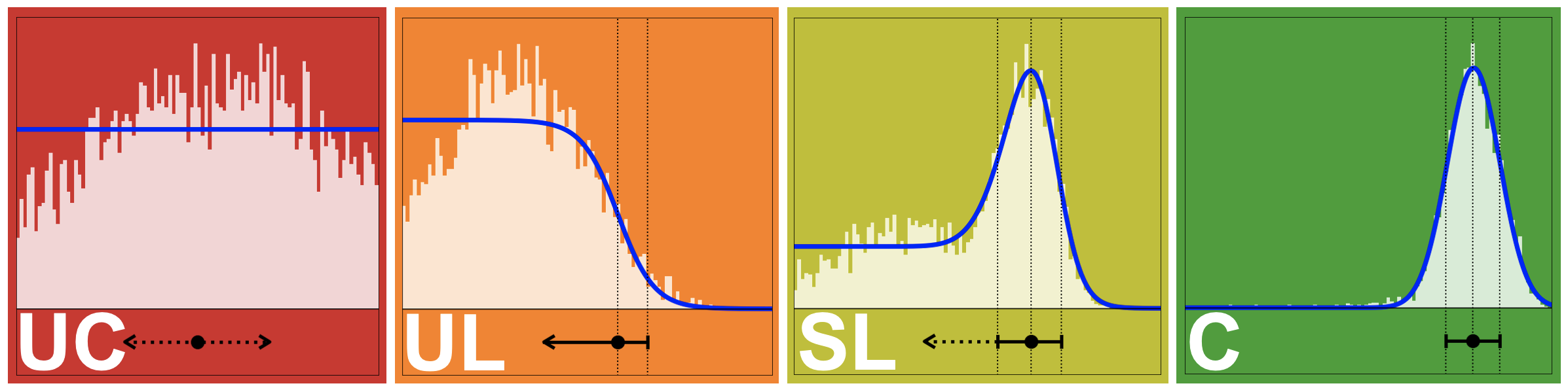}}
\caption{Wavelength-dependent posterior types retrieved for the different trace gases in the nominal case (a)-(c) and the optimized case (d)-(f). The lowermost panel gives the color coding for the different posterior types. The abbreviations used for the different posteriors are introduced in Sect.~\ref{Retrievalgrid}. (a),(d): $3-20\,\mu\mathrm{m}$, (b),(e): $4-18.5\,\mu\mathrm{m}$, (c),(f): $6-17\,\mu\mathrm{m}$.}
\label{fig:10}
\end{figure*}

\section{Discussion}
\label{Discussion}

After having provided an overview of the results we obtained for the nominal case (with an input spectrum covering $3-20\,\mu\mathrm{m}$), we now compare the retrieval results for different wavelength ranges in the nominal and optimized case. Thereby, we investigate the minimal requirements that the \textit{LIFE} mission needs to fulfill to characterize the atmospheric structure of an Earth-twin exoplanet and detect biosignature gases in the emission spectrum. In Sects.~\ref{subsubsec:choice_of_wlen} and~\ref{sec:Baselines} we derive requirements for the wavelength range, the spectral resolution R, and the S/N. We then quantify the integration times required to reach the desired S/N for a given R, conservatively assuming a total instrument throughput of 5\% \citep[cf.][]{dannert2022large}. and discuss the implications for the \textit{LIFE} design in Sect.~\ref{integration_time}. In Sect.~\ref{subsec:comp_other_studies}, we compare our work to similar retrieval studies in the literature to underline the unique scientific potential of \textit{LIFE} for atmospheric characterization. The limitations inherent to our work and possible future investigations follow in Sect.~\ref{subsec:limitations}.

\subsection{Wavelength range requirement }\label{subsubsec:choice_of_wlen}
\label{sec:Wavelength_Range}

Figure~\ref{fig:10} provides a concise summary of the retrieved posterior types for the trace gas abundances for varying wavelength range, R and S/N for both the nominal and optimized case. We are primarily interested in large systematic differences in the overall retrieval performance between the different wavelength ranges. Small differences (e.g., in the nominal case for \ce{H2O} at R~=~35, S/N~=~5 or for \ce{CH4} at R~=~35, S/N~=~10) are expected to disappear when averaging over multiple retrieval runs. In the following discussion, we focus on the results obtained for the nominal case (first row in Fig.~\ref{fig:10}).

We observe that for the trace gases \ce{CO2}, \ce{O3}, \ce{CO}, and \ce{N2O}, considering a broader wavelength range input spectrum will result in slightly smaller uncertainties on the retrieved parameters. This is expected, since more information is passed to the retrieval framework. However, there is no significant difference in performance between the considered wavelength ranges.

For the trace gases \ce{H2O} and \ce{CH4} the $6-17\,\mu\mathrm{m}$ case exhibits a considerably lower sensitivity than the $3-20\,\mu$m and $4-18.5\,\mu$m cases. An upper limit at $17\,\mu\mathrm{m}$ excludes the strong \ce{H2O} absorption features at $>17\,\mu\mathrm{m}$ (see Fig.~\ref{fig:3}). The \ce{H2O} lines between 8 and $17\,\mu\mathrm{m}$ are either weak or overlap with stronger absorption features of other molecules. Therefore, at low R and S/N, these features cannot constrain the \ce{H2O} abundance satisfactorily. Below $\approx8\,\mu\mathrm{m}$, we observe a strong \ce{H2O} band which overlaps with the \ce{CH4} feature at $7.7\,\mu\mathrm{m}$. In addition, at these wavelengths the S/N in the simulated spectra decreases drastically, further reducing the constraints on \ce{H2O}. Thus, when considering a wavelength range from $6-17\,\mu\mathrm{m}$, both the \ce{H2O} and \ce{CH4} abundance are estimated via their overlapping absorption feature at $\approx8\,\mu\mathrm{m}$. At low R and S/N this leads to larger uncertainties. If we consider a larger wavelength range, the long-wavelength tail of the emission spectrum provides more robust constraints on the \ce{H2O} abundance, which directly leads to an improvement for the \ce{CH4} estimates as well. We therefore can not recommend limiting the wavelength to $6-17\,\mu$m for \textit{LIFE} due to its negative impact on the sensitivity for \ce{H2O} and \ce{CH4}.

On the other hand, choosing the largest wavelength range considered in our study ($3-20\,\mu$m) provides only a negligible improvement over the $4-18.5\,\mu$m wavelength range. From this, we conclude that \textit{LIFE} should opt for a long wavelength cut-off of at least $18.5\,\mu\mathrm{m}$. For the short wavelength limit, we suggest a value of $4\,\mu\mathrm{m}$. This boundary would include the \ce{CO} feature at $4.67\,\mu\mathrm{m}$, as well as the \ce{N2} collision-induced absorption line at $4.3\,\mu\mathrm{m}$. In the Earth-twin retrievals we present in this work, the abundances of \ce{CO} and \ce{N2} are not constrainable for any of the considered cases that include these spectral features. However, making these spectral features accessible to \textit{LIFE} would enable us to potentially constrain the abundances of these important molecules in non Earth-twin atmospheres. More tests to explore this idea are foreseen for future retrieval studies.

For the retrievals performed for the optimized case, we reach similar conclusions and our recommended wavelength range remains unchanged. However, the interfering effects described above are less pronounced due to an improved S/N at short wavelengths (see Fig.~\ref{fig:10} and the tables in Appendix~\ref{app:res}).

\subsection{R and S/N requirements}
\label{sec:Baselines}

For most trace gases, we observe no significant difference between the nominal and optimized case for all S/N $\geq10$ as can be seen from Fig.~\ref{fig:10}. For an S/N of $5$, the optimized case yields better results for the retrieved abundances. However, due to the generally poor performance at this noise level (as previously seen), an S/N of $5$ is not sufficient to characterize the atmospheric structure and composition of an Earth-twin exoplanet satisfactorily.

Generally, we observe that \ce{CO2}, \ce{H2O} and \ce{O3} are easily retrievable (within $\pm 1$ dex) for an Earth-like atmosphere for all S/N $\geq10$. This finding is in accordance with the results presented in \citet{cockell2009b}. However, other studies suggest that for a clear \ce{O3} detection a higher R or S/N are necessary \citep[see, e.g., ][]{von_Paris_2013,leger_astrobio}. In contrast, \ce{CO} is not recoverable for any of the considered cases due to the large astrophysical noise at short wavelengths, which indicates that detecting \ce{CO} from the MIR emission spectrum of an Earth-twin exoplanet around a Sun-like star is extremely challenging and would require very high S/N and higher spectral resolution. Similarly, the \ce{N2O} abundance present in Earth's atmosphere is too low to be detected in all cases considered. However, we retrieve upper abundance limits, which indicates that high atmospheric concentrations of \ce{N2O} ($\gtrsim10^{-3}$) would likely be detectable.

In contrast, the retrieval results for \ce{CH4} depend strongly on the considered R and S/N for both the nominal and optimized case (see Fig.~\ref{fig:10}). Generally, our retrieval results for \ce{CH4} improve as we consider higher R and S/N. For both baseline configurations, we retrieve a threshold above which the retrieval framework manages to accurately estimate the \ce{CH4} abundance. This threshold manifests itself as a diagonal in the R-S/N space, which indicates that, e.g., a lower S/N can be compensated with a higher R without impacting the retrieval results significantly. The main difference is that R is intrinsic to the \textit{LIFE} design (dependent on the spectrograph specifications), whereas the S/N depends on the design (aperture size) and the integration time.

Generally, we find that in the optimized case the accuracy of the retrieval for \ce{CH4} is improved. Firstly, \textit{LIFE} is configured to optimize the signal at short wavelengths, which directly leads to an improvement in the S/N of the $\sim 7.7\,\mu$m \ce{CH4} feature. Secondly, the reduced noise contribution from exozodiacal dust also improves the S/N at short wavelengths. The resulting S/N enhancement at short wavelengths can be seen in Fig.~\ref{fig:5}. These two factors lead to an increase in the retrieval's overall performance. We observe that, in the optimized case for the $3-20\,\mu$m range, \ce{CH4} is detectable (at least an SL-type posterior) in an Earth-twin atmosphere with the following combinations:

\begin{itemize}
    \item S/N~=~15 for R~=~35,
    \item S/N~=~10 for R~=~50,
    \item S/N~=~5 for R~=~100.
\end{itemize}

R~=~20 is too low to allow for a meaningful constraint on the \ce{CH4} abundance for all considered cases. But if for technical reasons one would like to keep R as low as possible, we are left with R~=~35 and R~=~50. The higher resolution case potentially allows for the detection of a C-type posterior when going to higher S/N. In the nominal case, the S/N has to be increased to obtain comparable results at the same R (S/N~=~20 for R~=~35, or S/N~=~15 for R~=~50).

\begin{table}
\caption{Log-evidence for retrievals with and without $\mathrm{CH_4}$.}

\begin{tabular}{ccc:ccc}
\hline\hline
\multicolumn{3}{c:}{S/N}                   & $10$              & $15$              & $20$  \\\hline
\multirow{6}{*}{\rotatebox[origin=c]{90}{$\begin{array}{c}\text{Nominal}\\\text{Case}\end{array}$}}&\multirow{3}{*}{\rotatebox[origin=c]{90}{${\mathrm{R}=35}$}} & \begin{tabular}[c]{@{}c@{}}$\mathrm{ln(\mathcal{Z}_\mathrm{CH_4})}$\end{tabular}                                                    & $-23.6^{\pm0.2}$  & $-27.0^{\pm0.2}$  & $-28.8^{\pm0.2}$                   \\
&& \begin{tabular}[c]{@{}c@{}}$\mathrm{ln(\mathcal{Z}_\mathrm{-CH_4})}$\end{tabular}                                                                       & $-23.7^{\pm0.2}$                    & $-27.2^{\pm0.2}$                   & $-28.8^{\pm0.2}$                   \\
                        \cdashline{3-6}

&&\begin{tabular}[c]{@{}c@{}}$\mathrm{log_{10}\left(K\right)}$\end{tabular} & $0.0^{\pm0.1}$                      & $0.1^{\pm0.1}$                     & $0.0^{\pm0.1}$                      \\
                        \cline{2-6}
&\multirow{3}{*}{\rotatebox[origin=c]{90}{${\mathrm{R}=50}$}} & \begin{tabular}[c]{@{}c@{}}$\mathrm{ln(\mathcal{Z}_\mathrm{CH_4})}$\end{tabular}                                                                          & $-25.1^{\pm0.2}$                    & $-27.8^{\pm0.2}$                   & $-29.8^{\pm0.2}$              \\
&& \begin{tabular}[c]{@{}c@{}}$\mathrm{ln(\mathcal{Z}_\mathrm{-CH_4})}$\end{tabular}                                                                     & $-25.4^{\pm0.2}$                    & $-28.4^{\pm0.2}$                   & $-30.6^{\pm0.2}$             \\\cdashline{3-6}
&                        &\begin{tabular}[c]{@{}c@{}}$\mathrm{log_{10}\left(K\right)}$\end{tabular} & $0.1^{\pm0.1}$                      & $0.3^{\pm0.1}$                     & $0.3^{\pm0.1}$          \\
                        \hline

\multirow{6}{*}{\rotatebox[origin=c]{90}{$\begin{array}{c}\mathrm{Optimized}\\\mathrm{Case}\end{array}$}}&\multirow{3}{*}{\rotatebox[origin=c]{90}{${\mathrm{R}=35}$}} & \begin{tabular}[c]{@{}c@{}}$\mathrm{ln(\mathcal{Z}_\mathrm{CH_4})}$\end{tabular}                                                                                             & $-24.8^{\pm0.2}$               & $-27.9^{\pm0.2}$               & $-30.2^{\pm0.2}$              \\
&                        & \begin{tabular}[c]{@{}c@{}}$\mathrm{ln(\mathcal{Z}_\mathrm{-CH_4})}$\end{tabular}                                                                                        & $-24.9^{\pm0.2}$               & $-28.2^{\pm0.2}$               & $-30.1^{\pm0.2}$              \\
                        \cdashline{3-6}

&&\begin{tabular}[c]{@{}c@{}}$\mathrm{log_{10}\left(K\right)}$\end{tabular}              & $0.0^{\pm0.1}$                 & $0.1^{\pm0.1}$                 & $0.0^{\pm0.1}$                \\
                        \cline{2-6}
&\multirow{3}{*}{\rotatebox[origin=c]{90}{${\mathrm{R}=50}$}} & \begin{tabular}[c]{@{}c@{}}$\mathrm{ln(\mathcal{Z}_\mathrm{CH_4})}$\end{tabular}                                                                              & $-26.7^{\pm0.2}$               & $-29.6^{\pm0.2}$               & $-31.5^{\pm0.2}$              \\
&                        & \begin{tabular}[c]{@{}c@{}}$\mathrm{ln(\mathcal{Z}_\mathrm{-CH_4})}$\end{tabular}                                                                              & $-28.8^{\pm0.2}$               & $-31.7^{\pm0.2}$               & $-35.1^{\pm0.2}$              \\\cdashline{3-6}
&                        &\begin{tabular}[c]{@{}c@{}}$\mathrm{log_{10}\left(K\right)}$\end{tabular}                    & $0.8^{\pm0.1}$                 & $0.9^{\pm0.1}$                 & $1.6^{\pm0.1}$      \\
                        \hline
                         
\end{tabular}

\label{table:5}
\tablefoot{With $\mathrm{CH_4}$: $\mathrm{ln}(\mathcal{Z}_\mathrm{CH_4})$; Without $\mathrm{CH_4}$: $\mathrm{ln}(\mathcal{Z}_\mathrm{-CH_4})$. The results are for input spectra covering the $3-20\,\mu\mathrm{m}$ wavelength range. The last row for a given R lists $\mathrm{log}_{10}\left(K\right)$ of the Bayes' factor $K$, which we calculate via: $\mathrm{log}_{10}\left(K\right)=\left(\ln\left(\mathcal{Z}_\mathrm{CH_4}\right)-\ln\left({\mathcal{Z}_\mathrm{-CH_4}}\right)\right) /\ln\left(10\right)$ (derived from Eq. \ref{equ:bayes_factor}). We compare the performance of the models with and without $\mathrm{CH_4}$ to each other using Jeffrey's scale \citep{Jeffreys:Theory_of_prob} (see Table \ref{table:1}). The '${\pm}$' indicates the $68\%$ confidence interval $\mathrm{log}_{10}\left(K\right)$.}
\end{table}

Since the detection of \ce{CH4} depends strongly  on the combination of R and S/N, it is important to evaluate the significance of the \ce{CH4} detection in the cases R~$=\{35,\,50\}$ and S/N~$=\{10,\,15,\,20\}$. For every R-S/N pair, we run retrievals with and without \ce{CH4} being included in the forward model. We then compare the log-evidences corresponding to the retrievals for both scenarios via the Bayes' factor $K$. The results are summarized in Table~\ref{table:5}. 

The value of $K$, which corresponds to the ratio calculated from the Bayesian evidence of the model that considers \ce{CH4} divided by the evidence of the methane-free model, is a measure of which model is better at describing the observed data (see Table~\ref{table:1}). A positive $\mathrm{log_{10}\left(K\right)}$ indicates that the model including \ce{CH4} (model evidence is denoted $\mathcal{Z}_\mathrm{CH_4}$) describes the observed emission spectrum better than the model without \ce{CH4} (model evidence is denoted $\mathcal{Z}_\mathrm{-CH_4}$). In contrast, negative values favor the \ce{CH4}-free atmospheric model.

For all cases where R~=~35, the Bayes' factor $\mathrm{log_{10}} \left(K\right) \approx0.0$ (within 1-$\sigma$) indicating that there is no difference between the models with and without \ce{CH4}. Therefore, despite retrieving SL-type posteriors for S/N $\geq15$, the retrieval framework does not provide unambiguous evidence that \ce{CH4} is indeed present in the observed atmosphere. This finding underlines the nature of SL-type posteriors, where the long tail towards low abundances indicates that the retrieval can do without \ce{CH4}. For R~=~50, we generally observe larger $\mathrm{log_{10}} \left(K\right)$. In the nominal case, we find weak support for the model including \ce{CH4}. In the optimized case, we find substantial to strong evidence for the presence of \ce{CH4} as is indicated by $0.5<\mathrm{log_{10}} \left(K\right)<2.0$.

These findings suggest that \textit{LIFE} requires a minimal R of $50$ to confidently rule out the \ce{CH4}-free atmospheric model in retrievals of Earth-twin MIR emission spectra.

\subsection{Observation time estimates}
\label{integration_time}

It is crucial to derive estimates for the integration time required to reach a certain S/N. The integration time does, however, not only depend on R and S/N, but also on the aperture diameter of the \textit{LIFE} collector spacecraft. Hence, our analyses can provide some first order requirements for the aperture size and the total instrument throughput. We summarize the integration time estimates for the above-mentioned combinations of R and S/N in Table~\ref{table:6}, which are based on the conservative assumption of a total instrument throughput of 5\% \citep[cf.][]{dannert2022large}. We give a more exhaustive list containing observation time estimates for all retrieved input spectra in Appendix~\ref{app:integration}.

\begin{table}
\caption{Required observation time in days.}            
\label{table:6}      
\centering                          
\begin{tabular}{c c:c c c: c c c}        
\hline\hline                 
&&\multicolumn{6}{c}{Observation Time [days]}\\
& &\multicolumn{3}{c:}{Nominal Case} &\multicolumn{3}{c}{Optimized Case}\\
R&S/N &1 m&2 m&3.5 m&1 m&2 m&3.5 m\\
\hline 
$35$    &$15$  &$771$   &$74$   &$15$   &$922$  &$64$   &$9$  \\

$50$    &$10$  &$490$   &$47$   &$10$   &$589$  &$41$   &$6$  \\
\hline 
\end{tabular}
\tablefoot{Observation times for an Earth-twin at 10 pc for the two combinations of R and S/N discussed in Sect.~\ref{sec:Baselines} and three different aperture diameters (1m, 2m, and 3.5m). We conservatively assume a total instrument throughput of 5\% \citep[cf.][]{dannert2022large}.}
\end{table}

We find that with $4\times1$m apertures \textit{LIFE} will not be able to characterize terrestrial exoplanet atmospheres at a distance of 10 pc because unrealistic integration times $>1$ year would be required.
While \citet{bryson2021} used the results from NASA's \textit{Kepler} mission to estimate with 95\% confidence that the nearest terrestrial exoplanet orbiting with the habitable zone around a G or K dwarfs is only $\approx$ 6 pc away, they also estimate that there are only $\approx$ 4 such objects within 10 pc. Hence, a $4\times1$m configuration will not allow us to probe a somewhat sizable sample of temperate terrestrial exoplanets.

For the other two setups ($4\times2$m and $4\times3.5$m), the observation times are more feasible and would allow for the characterization of up to a few tens of terrestrial exoplanet atmospheres within \textit{LIFE}'s characterization phase. Specifically, for both the nominal and optimized case and assuming the $4\times2$m setup, having R~=~50 and S/N~=~10 would require less time (47 days for the nominal case, 41 days for the baseline optimized case) compared to R~=~35 and S/N~=~15 (74 and 64 days, respectively). This again underlines that a resolution of R~=~50 is more suitable. The saved integration time could be used to characterize the atmospheres of additional exoplanets or to measure higher S/N spectra for the most promising objects. Similar conclusions hold for the $4\times3.5$m case. However, the required observation times required would be significantly smaller.

\subsection{Comparison to other studies}\label{subsec:comp_other_studies}

In order to demonstrate the validity and understand the full implications of our retrieval results, we compare the results to previous studies in the literature. In Sect.~\ref{MIR}, we compare our work to previously published work on MIR thermal emission retrievals. In Sect.~\ref{NIR/VIS}, we consider findings from a comparable retrieval study for the NIR/VIS wavelength range and demonstrate the unique scientific potential of MIR observations with \textit{LIFE}.

\subsubsection{MIR thermal emission studies}
\label{MIR}

In the white paper by \citet{Quanz:LIFE}, a similar retrieval study was performed for an Earth-twin exoplanet. The study assumed R~=~100, S/N~=~20, covered a larger wavelength range ($3-30\,\mu\mathrm{m}$), and considered only photon noise of the planet on the input spectrum, neglecting any additional noise terms. We compare their findings to our results for the $3-20\,\mu$m wavelength range, R~=~100, S/N~=~20 for the nominal case.

Our study reaches a comparable accuracy ($\pm0.5$ dex) for the atmospheric trace gases \ce{CO2}, \ce{H2O}, \ce{O3}, and \ce{CH4}. This is achieved despite our usage of the more realistic LIFE\textsc{sim} noise model, which features additional noise sources that lead to considerably larger errors especially at short wavelengths ($\lambda\lesssim8\,\mu\mathrm{m}$), where the LIFE\textsc{sim} noise is dominated by the contributions from stellar leakage. The performance of our retrieval suite is likely a result of our prior assumption for the exoplanet mass, which was not made in the \citet{Quanz:LIFE} study. The width of the retrieved abundance posteriors is limited by the exoplanet's mass posterior due to the degeneracy between trace gas abundances and the surface gravity. The Gaussian prior we assumed for the exoplanet's mass limits the range of allowed masses. Thereby, also the surface gravity is constrained, limiting the range of possible abundances. We note, however, that our mass and radius prior are informed by empirical measurements.  

In contrast to \citet{Quanz:LIFE}, we do not succeed in constraining the \ce{CO} and \ce{N2O} abundances in our retrievals. Both atmospheric trace gases have their main absorption features at wavelengths $\lambda\lesssim8\,\mu$m (see Fig.~\ref{fig:3}), where the observational LIFE\textsc{sim}-noise dominates over the absorption features of interest. Furthermore, the retrievals performed in \citet{Quanz:LIFE} find significantly stronger constraints for the shape of the P-T profile and the planetary parameters $\mathrm{P_0}$ and $\mathrm{T_0}$. This is likely a combined result of the more heavily constrained P-T profile model and the more optimistic noise estimates they used in their retrieval analysis.

Finally, the Earth-twin's radius is retrieved to an extremely high precision in both studies. This underlines the scientific potential of observations probing the thermal emission of planets in the MIR wavelength range. Determining the radius in NIR/VIS reflected light spectra is not robust \citep[e.g.,][]{2018AJ....155..200F} due to a degeneracy between the planet's albedo and its radius (a large surface area and low albedo can lead to the same flux as a smaller area and higher albedo).

A similar study, albeit for lower resolutions (R~=~5, 20) and a less complete noise model, has been performed by \cite{von_Paris_2013} for the former \textit{Darwin} mission concept. They found, that at R~=~20, a MIR nulling interferometer would be capable of constraining the surface conditions ($\mathrm{P}_0$ to $\pm$~0.5dex, $T_0$ to $\pm$~10K) and planetary parameters ($\mathrm{R_{pl}}$ to $\pm$~10\%, $\mathrm{M_{pl}}$ to $\pm$~0.3dex) of a cloud-free exoplanet, which agrees well with the results presented here. Further, they manage to retrieve comparable constraints for the abundances of \ce{CO2} and \ce{O3} at 1$\sigma$ confidence levels. However, they also discuss, that for a 5$\sigma$ detection of \ce{CO2} and \ce{O3}, higher resolutions are mandatory. Similar results are presented in \cite{leger_astrobio}, where a resolution limit of R~=~40 is derived for a robust detection of \ce{CO2}, \ce{H2O} and \ce{O3} in the atmospheres of Earth-similar planets around M- and K-Type stars.

\begin{figure*}[t]
\centering
\includegraphics[width=\textwidth]{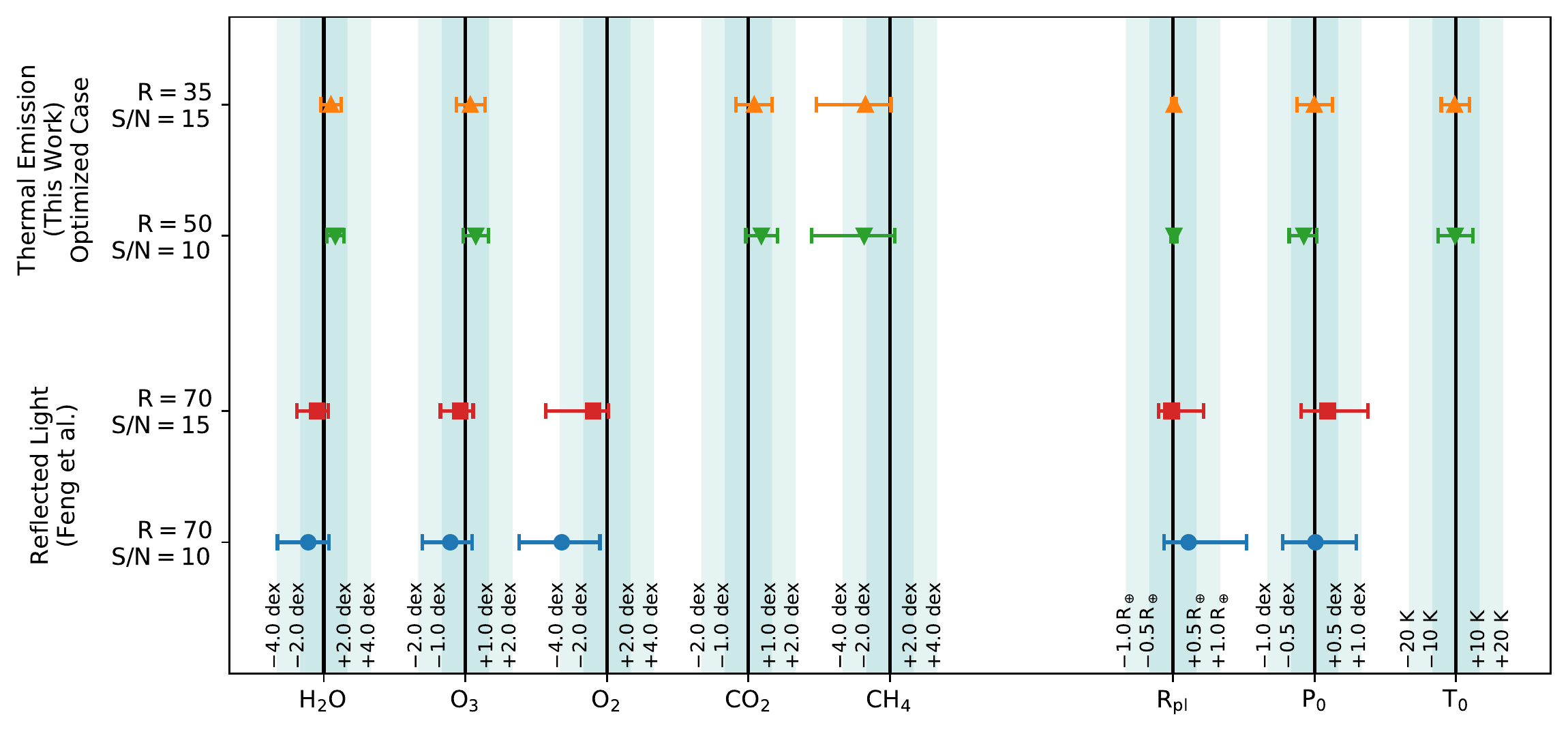}\label{fig:low_res_comparison}
\caption{Performance comparison between different retrieval studies. The error bars correspond to the $68\%$ confidence intervals of the retrieved posterior distributions. The emitted light study from this work for (R~=~50, S/N~=~10 (green downward triangle) and R~=~35, S/N~=~15 (orange upward triangle), $3-20\,\mu\mathrm{m}$ in the optimized case); the reflected light study by \citet{2018AJ....155..200F} (R~=~70, S/N~=~10 (blue dot), 15 (red square), $0.4-1.0\,\mu\mathrm{m}$).}
\label{fig:11}
\end{figure*}

\subsubsection{NIR/VIS reflected light studies}
\label{NIR/VIS}

\citet{2018AJ....155..200F} presented results for a similar retrieval study using reflected light input spectra in the wavelength range $0.4-1.0\,\mu\mathrm{m}$. These results are similar to the findings presented in \citet{Brandt13278}. We consider their results for the R~=~70, S/N~=~\{10, 15\} cases and compare them with our findings for the R~=~50, S/N~=~10 and R~=~35, S/N~=~15 optimized case. In Fig.~\ref{fig:11}, we show the results for all parameters of interest that are retrievable with at least one of the two approaches.

This comparison demonstrates that NIR/VIS and MIR wavelength studies yield partially complementary results. The MIR range enables us to search for signatures of the important trace gases \ce{CO2} and \ce{CH4}. Both these gases are not accessible through reflected NIR/VIS light observations at Earth's mixing ratios. Additionally, studying the MIR thermal emission spectra of exoplanets enables us to pose significant constraints on the important planetary parameters $T_0$, $P_0$ and $R_{\text{pl}}$. These three parameters are not easily accessible via reflected light observations in the NIR/VIS wavelength range \citep{Quanz:LIFE,2018AJ....155..200F, 2020A&A...640A.136C}.

However, studies in the NIR/VIS wavelength range provide a direct probe for the \ce{O2} abundance, whereas MIR observations can only probe the \ce{O3} abundance, a photochemical byproduct of \ce{O2} in our atmosphere. Additionally, the NIR/VIS wavelength range may  allow us to characterize the surface composition of an exoplanet \citep[e.g.,][]{Brandt13278} via the wavelength-dependent surface scattering albedo. Such NIR/VIS observations could potentially allow for a detection of liquid surface water via the ocean glint as suggested in \citet{2010ApJ...721L..67R} or a detection of the vegetation red edge, which is an increased reflectivity in the NIR due to photosynthetic life and therefore a surface biosignature \citep[see, e.g.,][]{Seager_2005,Schwieterman:Surface_Temporal_Biosignatures}. A planet accessible to both techniques would be a prime target for atmospheric characterization

\subsection{Limitations and future work}
\label{subsec:limitations}
Even though we have achieved our main goal of deriving first-order quantitative  requirements for \textit{LIFE}, the results have to be interpreted with care since there are fundamental limitations inherent to our approach:
\begin{itemize}
    \item The input spectra for our retrievals are generated using a $1$D radiative transfer model and a fully mixed atmosphere, which is a  simplification of reality. Additionally, we did not account for the effect of partial, full, or varying cloud coverage on the MIR Earth-twin emission spectrum. Retrieval analyses based on more complex forward models are foreseen in the future to investigate the effects of these simplifying assumptions.
    
    \item We do not retrieve for additional molecules that are not present in the input spectrum. However, when analyzing spectra from observations, we do not know what species are present in the atmosphere. Not retrieving for additional molecules that are not present might lead to overly confident estimates for \textit{LIFE}'s technical requirements. We provide a first test for the robustness of our results with respect to additional species in Appendix \ref{RandNoise} and plan further investigations for future work.
    
    \item The input spectrum is static and represents the ``average'' emission spectrum of the Earth, but the real emission spectrum varies over time (day vs. night and summer vs. winter) and it also depends on the viewing geometry (pole-on vs. equator-on) \citep[e.g.,][]{mettler2020}. 
    
    \item The presence of a moon can have an influence on the integrated thermal emission spectrum of the planet-moon system, in particular if the moon is as large as the Earth's moon and features day side temperatures higher than that of the planet \citep{robinson2011}. The quantitative impact a moon has on the retrieval results will be investigated in future work.
    
    \item We used \texttt{petitRADTRANS} both to generate the simulated input spectra and as atmospheric forward model for the retrieval framework. As demonstrated in \citet{Barstow:Retrieval_comparison}, retrieving the same input spectrum with different forward models can lead to inconsistencies between the retrieved parameter values due to differences in the forward models. Similar problems will likely arise when retrieving model parameters from an experimentally measured spectrum, since the forward model does not capture the full physics (and/or chemistry) of the observed atmosphere. Our results could therefore be overly-optimistic. Estimating the magnitude of this bias is the subject of future work.
    
    \item We added the LIFE\textsc{sim}-noise as uncertainty to the theoretically simulated theoretical flux without randomizing the value of the individual spectral points. This may lead to over-optimistic retrieval results, in particular for the small S/N cases. This is especially true for \ce{CH4}, whose Earth's abundance is close to the detection limit of \textit{LIFE}. A discussion on the potential impact of this simplification is provided in Appendix \ref{RandNoise}.
    
    \item Currently, LIFE\textsc{sim} only features dominant astrophysical noise terms \citep[][]{dannert2022large}. However, systematic instrumental effects will also impact the observations, even though, ideally, the instrument will only contribute to the noise, but not dominate the noise budget. Still, the required integration times should be considered a lower limit until the optical, thermal and detector designs of \textit{LIFE} have further matured and LIFE\textsc{sim} is updated accordingly.
    
    \item Finally, the study of an Earth-twin exoplanet is clearly a simplification. The known diversity of terrestrial exoplanets demonstrates that future studies will have to look beyond the Earth-twin case and consider a wider range of worlds. In order to obtain more rigorous constraints on the requirements for \textit{LIFE}, we will perform similar retrieval studies on a  variety of different exoplanet types.

\end{itemize}

\section{Summary and conclusions}\label{Summary}

Using an atmospheric retrieval framework, we have derived the minimal requirements for the spectral resolution R, the wavelength coverage and the S/N that need to be met by a space-based MIR nulling interferometer like \textit{LIFE} to characterize the atmospheres of Earth-twin exoplanets orbiting nearby solar-type stars. 

In our atmospheric model, we describe the atmospheric P-T structure of an Earth-twin using a 4\textsuperscript{th} order polynomial with  surface-pressure $P_0$ and surface temperature $T_0$. We assume constant, modern Earth abundances of \ce{N2}, \ce{O2}, \ce{CO2}, \ce{H2O}, \ce{O3}, \ce{CH4} and \ce{N2O} throughout the vertical extent of the atmosphere; clouds are not included. We generate the thermal emission spectrum corresponding to this Earth-twin atmosphere using the 1D radiative transfer model \texttt{petitRADTRANS} \citep{2019AA...627A..67M}, assuming an exoplanet mass $M_{\text{pl}}= 1 M_{\oplus}$ and a radius $R_{\text{pl}}=1 R_{\oplus}$. We used the LIFE\textsc{sim} tool \citep{dannert2022large} to estimate the wavelength-dependent observational noise (incl. noise from stellar leakage and from local and exozodiacal dust emission) for different wavelength ranges and combinations of R and S/N.

We created a Bayesian retrieval framework coupling  \texttt{petitRADTRANS} and \texttt{pyMultiNest} \citep{Buchner:PyMultinest} (python access of the \texttt{FORTRAN} \texttt{MultiNest} \citep{Feroz:Multinest} implementation of the Nested Sampling algorithm \citep{Skilling:Nested_Sampling}) to extract information about the atmospheric structure and composition of the Earth-twin exoplanet from the input spectra. These retrievals were performed considering different wavelength ranges ($3-20\,\mu$m, $4-18.5\,\mu$m, $6-17\,\mu$m), spectral resolutions R (20, 35, 50, 100) and signal-to-noise ratios S/N (5, 10, 15, 20) at a wavelength of 11.2~$\mu$m (the corresponding noise at other wavelengths is derived via LIFE\textsc{sim}). 

The performed retrieval analyses suggest that MIR observations with \textit{LIFE} at an S/N $\geq10$ can robustly constrain the radius (uncertainty $\leq\pm 10\%$), surface pressure $P_0$ (uncertainty $\leq\pm 0.5$ dex) and surface temperature $T_0$ (uncertainty $\leq\pm 20$ K) of an Earth-twin exoplanet. These parameters cannot be probed accurately via reflected light observations of an Earth-twin at NIR/VIS wavelengths. Furthermore, we predict \ce{CO2}, \ce{H2O} and \ce{O3} to be detectable by \textit{LIFE} (error $\leq\pm1.0$ dex) given an input spectrum with an S/N $\geq10$ for all the considered wavelength ranges and R. In contrast, the potential biosignature \ce{N2O} and the potential anti-biosignature \ce{CO} are not detectable for any of the considered \textit{LIFE} configurations.  For \ce{N2O} we find an upper limit on the abundance, indicating that high abundances ($\gtrsim10^{-3}$ in mass fraction) are potentially detectable in MIR observations. For \ce{CO}, we do not retrieve any constraint.

Concerning the potential biosignature \ce{CH4}, our retrieval results strongly depend on the properties of the input spectrum. If we aim to detect \ce{CH4} at Earth-like abundances ($\approx10^{-6}$ in mass fraction) in exoplanet atmospheres, we estimate a minimal requirement of R~=~50 and S/N~=~10. Furthermore, we observe a performance drop for \ce{CH4} for the $6-17\,\mu\mathrm{m}$ wavelength range, which is a result of cutting off the \ce{H2O} absorption bands at wavelengths $\geq 17\,\mu$m. Including \ce{H2O} bands at wavelengths $>17\,\mu$m improves the accuracy of the \ce{H2O} abundance estimate, which in turn helps to disentangle contributions from species overlapping at shorter wavelengths.  Between the $3-20\,\mu\mathrm{m}$ and $4-18.5\,\mu\mathrm{m}$ wavelength range, we do not observe significant differences in the performance of the retrieval framework. Therefore, a wavelength coverage of at least $4-18.5\,\mu\mathrm{m}$ is desirable.

Turning the S/N requirement into estimates for integration times \citep[conservatively assuming a total instrument throughput of 5\%, cf.][]{dannert2022large} and considering recent estimates for the number of temperate terrestrial exoplanets around Solar-type stars within 10 pc, we find that \textit{LIFE} should feature at least $4\times$2 m apertures to be able to investigate a somewhat sizeable sample of these objects. With $4\times$1 m apertures, the integration time to study an Earth-twin located at 10 pc will be prohibitively long ($>$1 year). 

By comparing our results with those obtained in similar studies focusing on reflected light observations of Earth-twin exoplanets \citep[e.g.,][]{2018AJ....155..200F}, we find that both approaches complement each other. However, MIR emission spectra of terrestrial exoplanets can provide access to surface conditions and accurately constrain the radii of the objects, parameters that are challenging for reflected light observations. In addition, we have shown that MIR spectra will allow us to probe for the simultaneous presence of \ce{O3} and \ce{CH4}, a strong combinatory biosignature with no currently known  false positives, underlining again the large and unique opportunity that MIR observations provide.

Overall, our results suggest that pursuing a concept for a space-based MIR nulling interferometer, like \textit{LIFE}, in addition to the proposed NIR/VIS mission concepts, currently under consideration by NASA, will be a key element for the future of exoplanet characterization.  In particular, the combination of results from both approaches would vastly expand our knowledge about worlds outside the Solar System.

\begin{acknowledgements}
    This  work  has  been carried  out within  the  framework  of the National Center of Competence in Research PlanetS supported by the Swiss National  Science  Foundation. S.P.Q. and E.A. acknowledge the financial support from the SNSF. This work benefited from the 2019 Exoplanet Summer Program in the Other Worlds Laboratory (OWL) at the University of California, Santa Cruz, a program funded by the Heising-Simons Foundation. Further, we would like to thank Michael Line for useful discussions and analyses from which the project benefited. P.M. acknowledges support from the European Research Council under the European Union’s Horizon 2020 research and innovation program under grant agreement No. 832428. J.L.G. thanks ISSI Team 464 for useful discussions.
    \\
    \\
    \emph{Author contributions.} BSK carried out the analyses, created the figures and wrote the bulk part of the manuscript. SPQ initiated the project. SPQ and EA guided the project and wrote part of the manuscript. All authors discussed the results and commented on the manuscript.
\end{acknowledgements}

%
%
\bibliographystyle{aa}
\bibliography{bibliography}

\begin{appendix}
\section{Selection of P-T profile model}\label{P-T-choice}

In the following, we analyze the performance of different parametric models in describing the atmospheric P-T structure of different terrestrial planets in our Solar System and assess their applicability in our retrieval framework. 

\subsection{Considered P-T models}
\label{P-T-options}

For our P-T model selection, we consider the following four P-T models:

\begin{itemize}
    \item Polynomial P-T parametrization: The P-T structure of an atmosphere is modelled via a n\textsuperscript{th} order polynomial:

    \begin{equation}
    T(P) = \sum_{i=0}^n a_iP^i
    \label{equ:poly_pt}
    \end{equation}
    
    The $n+1$ constants $a_i$ are the model parameters. In our P-T model selection, we consider polynomials up to order $n=9$.
    
    \item P-T parametrization proposed by \citet{2009ApJ...707...24M}: This is a P-T model for terrestrial planets. It is based on the P-T profiles of atmospheres of rocky Solar System objects and $1$D self-consistent exoplanet P-T profiles generated via model atmosphere calculations. It is defined by the following equations:
    
    \begin{align}
    \label{equ:ms_eq1}
    P_0<P<P_1:\qquad P&=P_0e^{\alpha_1(T-T_0)^{\beta_1}}\\
    \label{equ:ms_eq2}
    P_1<P<P_3:\qquad P&=P_2e^{\alpha_2(T-T_2)^{\beta_2}}\\
    \label{equ:ms_eq3}
    P>P_3:\qquad T&=T_3.
    \end{align}
    
    The middle pressure layer (Eq. \ref{equ:ms_eq2}) allows for thermal inversion, the deepest layer (Eq. \ref{equ:ms_eq3}) is set to be isothermal due to its large optical depth. Parameters $\beta_1$ and $\beta_2$ are set to $0.5$. $P_0$ is the pressure at the top of the atmosphere, which we fix to $10^{-6}$ bars. The remaining tunable parameters are: $T_0$, $\alpha_1$, $\alpha_2$, $P_1$, $P_2$, $P_3$, $T_2$ and $T_3$.  By requiring continuity between the three layers, the number of model parameters can be reduced to six by setting:

    \begin{align}
    \label{equ:ms_eq4}
    T_3 &=  T_2+\left(\frac{\log(P_3/P_2)}{\alpha_2}\right)^2 \\
    \label{equ:ms_eq5}
    T_2 &=  T_0+\left(\frac{\log(P_1/P_0)}{\alpha_1}\right)^2-\left(\frac{\log(P_3/P_2)}{\alpha_2}\right)^2.
    \end{align}
    
    \item P-T parametrization proposed by \citet{Guillot:PT_Profile}: This model calculates the P-T structure of an atmosphere assuming a radiative equilibrium for each atmospheric layer. The model is described by the following equation:
    \begin{equation}
        \begin{split}
            T_{Guillot}(P) &=\frac{3T_{int}^4}{4} \left(\frac{2}{3} +\frac{\kappa_{IR}}{g_{pl}}P\right) \\ &+\frac{3T_{equ}^4}{4}\left[\frac{2}{3}+\frac{1}{\gamma\sqrt{3}}+\left(\frac{\gamma}{\sqrt{3}}-\frac{1}{\gamma\sqrt{3}}\right)e^{-\gamma\sqrt{3}\frac{\kappa_{IR}}{g_{pl}}P}\right].
        \end{split}
        \label{equ:Guillot_PT}
    \end{equation}
    The factor $\left(\kappa_{IR}P/g_{pl}\right)$ is the atmosphere's optical depth $\tau$. The model assumes that the IR opacity can be approximated by a constant $\kappa_{IR}$. Further, the opacity in the visible, $\kappa_{VIS}$, is assumed to be constant and linked to $\kappa_{IR}$ via $\gamma$: $\kappa_{VIS} = \gamma\kappa_{IR}$. $g_{pl}$ is the planet's surface gravity; $T_{int}$ the planet's internal temperature (remaining heat from planet's gravitational collapse or energy from radioactive decay of elements). $T_{equ}$ is the planet's equilibrium temperature, which is the temperature the planet would have, if it were a black-body heated only by the radiation coming from its host star:
    \begin{equation}
        T_{equ} = T_*\sqrt{\frac{R_*}{2d}}\left(1-A_B\right)^{1/4}.
        \label{equ:equilibrium_temperature}
    \end{equation}
    $T_*$ is the host star's effective temperature, $R_*$ its radius, $d$ the separation between the planet and its host and $A_B$ the planet's Bond albedo.

    Overall, the \citet{Guillot:PT_Profile} P-T model requires four model parameters ($T_{int}$, $T_{equ}$, $\kappa_{IR}$ and $\gamma$; $g_{pl}$ is already a parameter of the retrieval).
    
    \item P-T parametrization proposed by \citet{2019AA...627A..67M}: The \citet{Guillot:PT_Profile} model does not allow for a non-isothermal structure in the upper atmosphere. The small modification proposed by \citet{2019AA...627A..67M} allows for a non-isothermal upper atmosphere:
    \begin{equation}
        \label{equ:Mod_Guillot}
        T(P)=T_\mathrm{Guillot}(P)\cdot\left(1-\frac{\alpha}{1+P/P_{\mathrm{Trans}}}\right)
    \end{equation}
    $T_\mathrm{Guillot}(P)$ is the \citet{Guillot:PT_Profile} P-T model (Eq. (\ref{equ:Guillot_PT})). This modification adds two parameters, $\alpha$ and $P_\mathrm{Trans}$, to the \citet{Guillot:PT_Profile} model, resulting in six model parameters.
    \end{itemize}

\begin{figure*}[]
  \centering
  \includegraphics[width=0.99\textwidth]{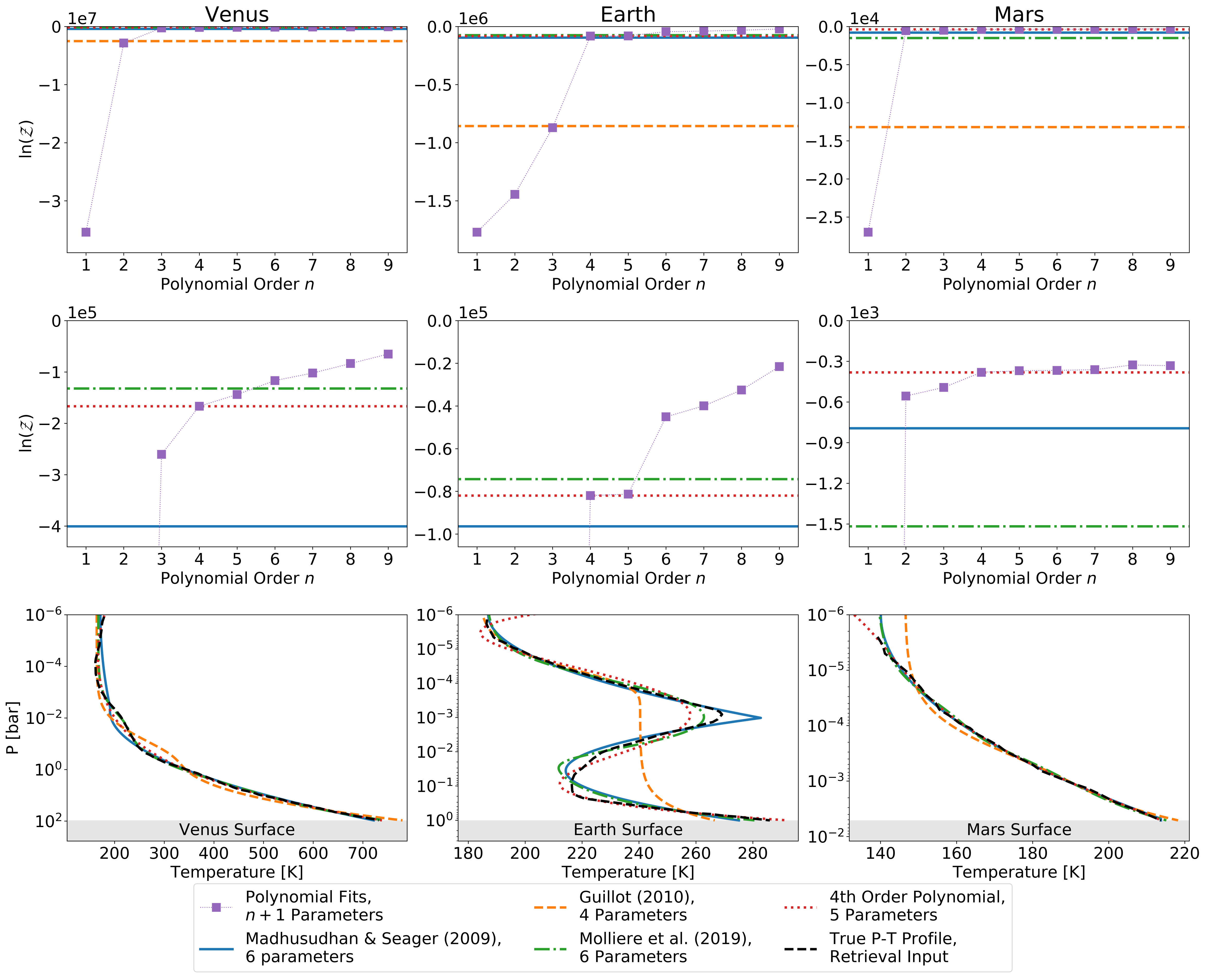}
\caption{Results from the P-T model retrievals. The first row displays the log-evidence $\mathrm{ln}(\mathcal{Z})$ of the different models. The second row shows a zoomed in view of the high log-evidence region of the first row to allow for better comparison. The bottom row displays the retrieved best fit P-T profiles for the $4$\textsuperscript{th} order polynomial, the \citet{2009ApJ...707...24M} model and the two Guillot models \citep{Guillot:PT_Profile,2019AA...627A..67M}. The data for the true P-T profiles is taken from Fig.~1 in \citet{Mueller-Wodarg:SS_PT_Profiles}.}
\label{fig:PT_Retrieval_Results}
\end{figure*}

\subsection{Choice of P-T model via retrievals}

\begin{table}
\caption{Priors used in the P-T profile retrievals.}
\label{tab:Priors_PT_Retrieval}
\centering
\begin{tabular}{ccc}
\hline\hline
Model                  &Model Parameter                  &Prior                \\ \hline
Polynomial                      & $a_i$                                     & $\mathcal{U}(-1000,1000)$     \\ \hdashline
Madhusudhan                     & $\mathrm{log}_{10}\left(P_i\right)$                & $\mathcal{U}(-10,4)$          \\
\& Seager                       & $T_0$                                     & $\mathcal{U}(0,1000)$         \\
                                & $\alpha_1, \alpha_2$                      & $\mathcal{U}(0,100)$          \\ \hdashline
Guillot                   & $T_{equ}$, $T_{int}$                      & $\mathcal{U}(0,1000)$         \\
                                & $\kappa_{IR}$                             & $\mathcal{U}(0,1)$            \\
                                & $\gamma$                                  & $\mathcal{U}(0,100)$          \\
                                \hdashline 
Addition for                    & $\mathrm{log}_{10}\left(P_{Trans}\right)$          & $\mathcal{U}(-10,4)$          \\
Modified Guillot                & $\alpha$                                  & $\mathcal{U}(-1,1)$           \\ \hline

\end{tabular}
\tablefoot{$\mathcal{U}(x,y)$: boxcar prior with lower limit $x$ and upper limit $y$.}
\end{table}

We use a Bayesian retrieval framework to assess the performance of the different P-T models in describing the P-T structure of the Solar System planets Venus, Earth and Mars. We include Venus and Mars to prevent the selected P-T model from being biased toward Earth-like P-T profiles. In the retrievals, the P-T models introduced in Appendix \ref{P-T-options} are used as forward models. Further, we use the \texttt{MultiNest} algorithm \citep{Feroz:Multinest} for parameter estimation, via \texttt{pyMultiNest} \citep{Buchner:PyMultinest}. We use $400$ live points and a sampling efficiency of $0.8$ as suggested for model comparison by the \texttt{pyMultiNest} documentation. The prior distributions assumed for the different model parameters are summarized in Table~\ref{tab:Priors_PT_Retrieval}. The P-T model we choose should provide an optimal combination of the following properties:
\begin{itemize}
\item Maximal model evidence $\mathcal{Z_M}$: A large $\mathcal{Z_M}$ implies a good fit of the P-T model (see Sect.~\ref{bayesiantechnique}).
\item Minimal parameter number: Since additional model parameters increase the computational cost significantly, we aim to keep the parameter number as low as possible. \end{itemize}

In the first two rows of Fig.~\ref{fig:PT_Retrieval_Results},
we plot the log-evidence $\mathrm{ln}(\mathcal{Z})$ corresponding to the retrieval results for the considered polynomial P-T models. Additionally, we indicate the $\mathrm{ln}(\mathcal{Z})$ corresponding to the non-polynomial P-T models as horizontal lines. For all considered atmospheres, we observe a continuous increase in $\mathrm{ln}(\mathcal{Z})$ with rising order of the polynomial P-T model. This indicates that higher-order polynomials provide a better fit to the atmosphere's P-T profile. However, we observe a distinct flattening in the increase of $\mathrm{ln}(\mathcal{Z})$ with increasing polynomial order for considered atmospheres. This flattening occurs at a $3$\textsuperscript{rd} order polynomial for Venus, a $4$\textsuperscript{th} order polynomial for Earth and a $2$\textsuperscript{nd} order polynomial for Mars. Increasing the polynomial order beyond these thresholds does not lead to a significant improvement in the polynomial fit.

The last row in Fig.~\ref{fig:PT_Retrieval_Results} indicates that the \citet{Guillot:PT_Profile} model fails to fit the inversion in Earth's P-T profile. This further manifests itself in the corresponding $\mathrm{ln}(\mathcal{Z})$, which is an order of magnitude smaller than the $\mathrm{ln}(\mathcal{Z})$ associated with most of the other models considered (order $\geq$~4 for the polynomial model).

The \citet{2009ApJ...707...24M} profile accurately models the P-T structure of all considered atmospheres. However, when comparing $\mathrm{ln}(\mathcal{Z})$ to the other models via the Bayes' factor, we find that for Venus and Earth  $\mathrm{log_{10}}(K)$ exceeds 2. This suggests that both polynomials of order $\geq4$ and the \citet{2019AA...627A..67M} model provide a significantly better P-T fit (see Table~\ref{table:1}). For the Martian atmosphere, the \citet{2009ApJ...707...24M} outperforms the \citet{2019AA...627A..67M} model, but still underperforms compared to all polynomials of order $\geq2$.

The \citet{2019AA...627A..67M} accurately fits all three considered P-T structures. For Venus and Earth, the model provides a better fit than all polynomials of order $\leq6$. However, compared to the $4$\textsuperscript{th} order polynomial, one additional parameter is required. For Earth, the main difference between the \citet{2019AA...627A..67M} model and the 4\textsuperscript{th} polynomial occurs at the inversions at $10^{-1}$ and $10^{-3}$ bar, where both models struggle to describe the true P-T structure accurately. For Mars, all polynomials of order $\geq2$ outperform the modified Guillot model, as is indicated by the significantly larger $\mathrm{ln}(\mathcal{Z})$.

For our purposes, we choose the $4$\textsuperscript{th} order polynomial P-T model, since it adds fewer parameters to our retrievals. Despite relying on less parameter, it manages to yield a comparable fit and describes all three considered P-T structures satisfactorily. The saved parameter allows us to retrieve for one additional atmospheric parameter of interest.

\section{Posterior classification}\label{PostClass}

\begin{figure}[]
\centering
\subfloat[][]{\includegraphics[width=.23\textwidth]{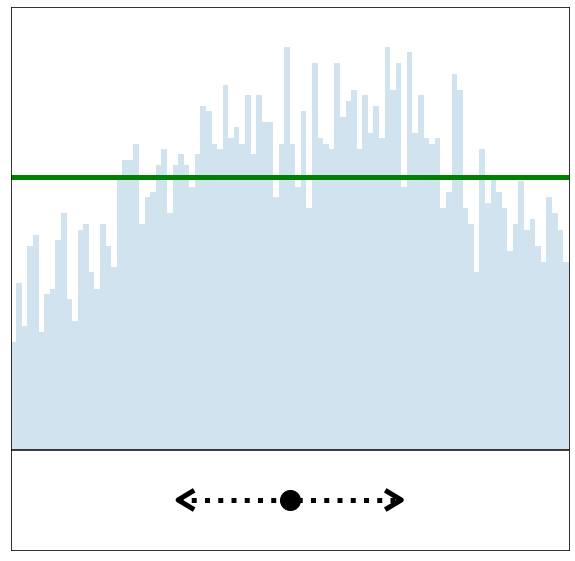}\label{fig:UC}}
\quad
\subfloat[][]{\includegraphics[width=.23\textwidth]{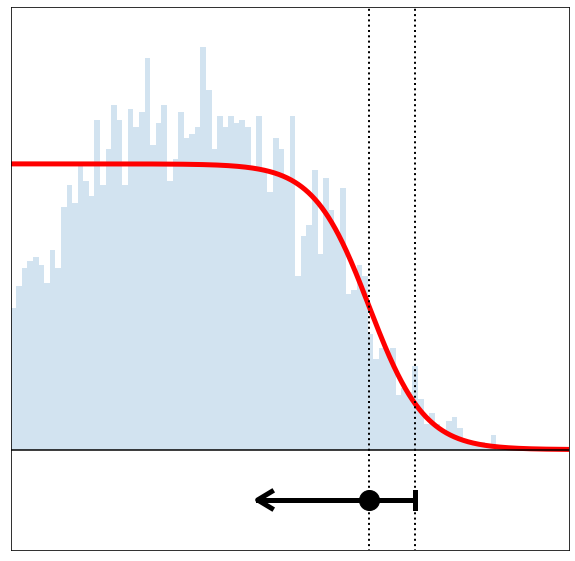}\label{fig:UL}}
\quad
\subfloat[][]{\includegraphics[width=.23\textwidth]{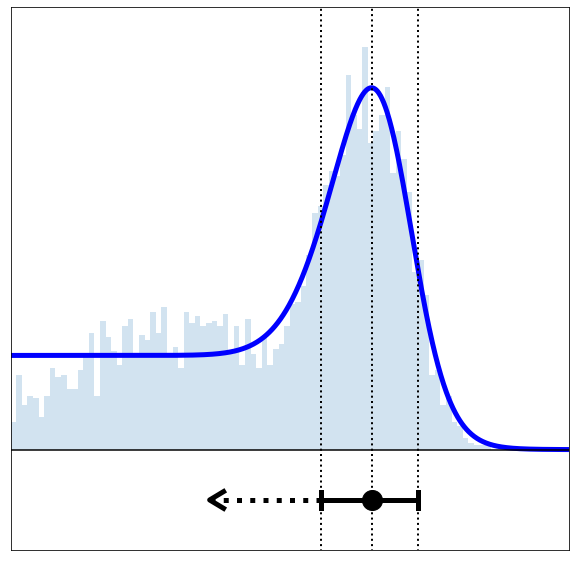}\label{fig:SL}}
\quad
\subfloat[][]{\includegraphics[width=.23\textwidth]{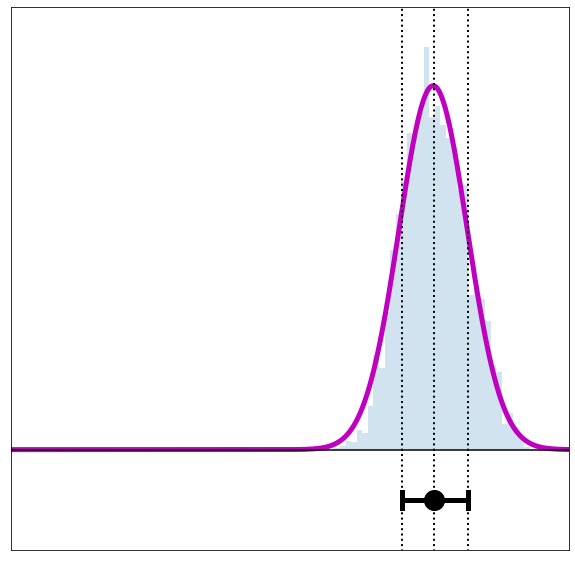}\label{fig:C}}
\caption{Classifications used for the retrieved posterior distributions. The histogram gives the posterior distribution as found by the retrieval framework. The colored lines represent the best fit model for the posterior distribution, the black symbols below show how such a posterior will be represented in our data analysis plots. (a): unconstrained posterior (UC), (b): upper limit posterior (UL), (c): sensitivity limit posterior (SL), (d): constrained posterior (C).}
\label{fig:posterior_types}
\end{figure}

For the presentation of the grid retrieval results, we chose to classify the retrieved posterior distributions for the different parameters into four main classes based on their visual appearance. Fig.~\ref{fig:posterior_types} gives illustrative examples for each of the four different posterior shapes. We define:
\begin{itemize}
    \item Unconstrained Posterior (UC): The retrieved posterior distribution does not pose a strong constraint on the parameter of interest for the assumed prior distribution. In this case, we describe the posterior distribution as a constant throughout the entire prior range (Fig.~\ref{fig:UC}).
    
    \item Upper Limit Posterior (UL): The retrieval yields the detection (upper) limit for the retrieved parameter. For the abundance of a molecule, this implies that the retrieval rules out any concentration above the found threshold value. Below this threshold, all abundances are equally likely and cannot be ruled out. We describe such a posterior via the Logistic function:
    \begin{equation}\label{equ:logistic_f_2}
        f(x) = \frac{c}{1+e^{a\cdot x+b}}.
    \end{equation}
    The constants a, b, and c are unique for each posterior. In the data analysis, we mark the half-maximum as well as the $16$\textsuperscript{th} percentile of the logistic function (Fig.~\ref{fig:UL}).
    
    \item Sensitivity Limit Posterior (SL): Similar to a UL posterior, the retrieval is capable of excluding the high but not the low molecular abundances. However, the posterior also shows a distinct peak at the boundary between these two regimes, which roughly corresponds to the true abundance. We convolve the Logistic function (Eq.~(\ref{equ:logistic_f_2})) with a Gaussian distribution to describe this posterior type:
    \begin{equation}\label{equ:SSG}
        f(x) = \frac{c+d\cdot e^{-(x+\mu)^2/\sigma^2}/\left(\sqrt{2\pi}\sigma\right)}{1+e^{a\cdot x+b}}.
    \end{equation}
    The constants a, b, c, d, $\mu$ and $\sigma$ are unique for each posterior. $\mu$ corresponds to the mean, $\sigma$ to the standard deviation of the Gaussian. In our analysis, we mark the maximum value of the distribution and the position of the half maximum to the left and the right of it (Fig.~\ref{fig:SL}).
    
    \item Constrained Posterior (C): Such a posterior distribution strongly constrains the prior range and can be approximated via a Gaussian distribution:
    \begin{equation}\label{Gauss}
        f(x) =  \frac{d\cdot e^{-(x+\mu)^2/\sigma^2}}{\sqrt{2\pi}\sigma}.
    \end{equation}
    The constants d, $\mu$ and $\sigma$ are unique for each posterior. $\mu$ corresponds to the mean and $\sigma$ to the standard deviation of the Gaussian. In the analysis we mark the $50$\textsuperscript{th}, $16$\textsuperscript{th} and $84$\textsuperscript{th} percentile which corresponds to the $1\,\sigma$ range (Fig.~\ref{fig:C}).
\end{itemize}
We determine the best fit model for each of the retrieved posterior distributions by fitting all four models. Thereafter, we use the log-likelihood function (Eq.~(\ref{equ:loglike}), Sect.~\ref{bayesiantechnique}) to determine which model best describes the retrieved posterior distribution. For the SL case, we require the maximum value of the fitted function to be at least $1.3$ times and maximally $10$ times larger than the continuum probability at low abundances. If the peak probability is less than $1.3$ times the continuum value, we assume an UL model for the posterior. If the peak abundance is at least $10$ times more probable than the low abundance continuum, we choose a C-type posterior distribution. These cutoff-values are chosen empirically to match what we would obtain in a manual classification.

\section{Impact of randomized noise and additional species on retrieval results}\label{RandNoise}

\begin{figure}[]
\centering
\subfloat{\includegraphics[width=0.49\textwidth]{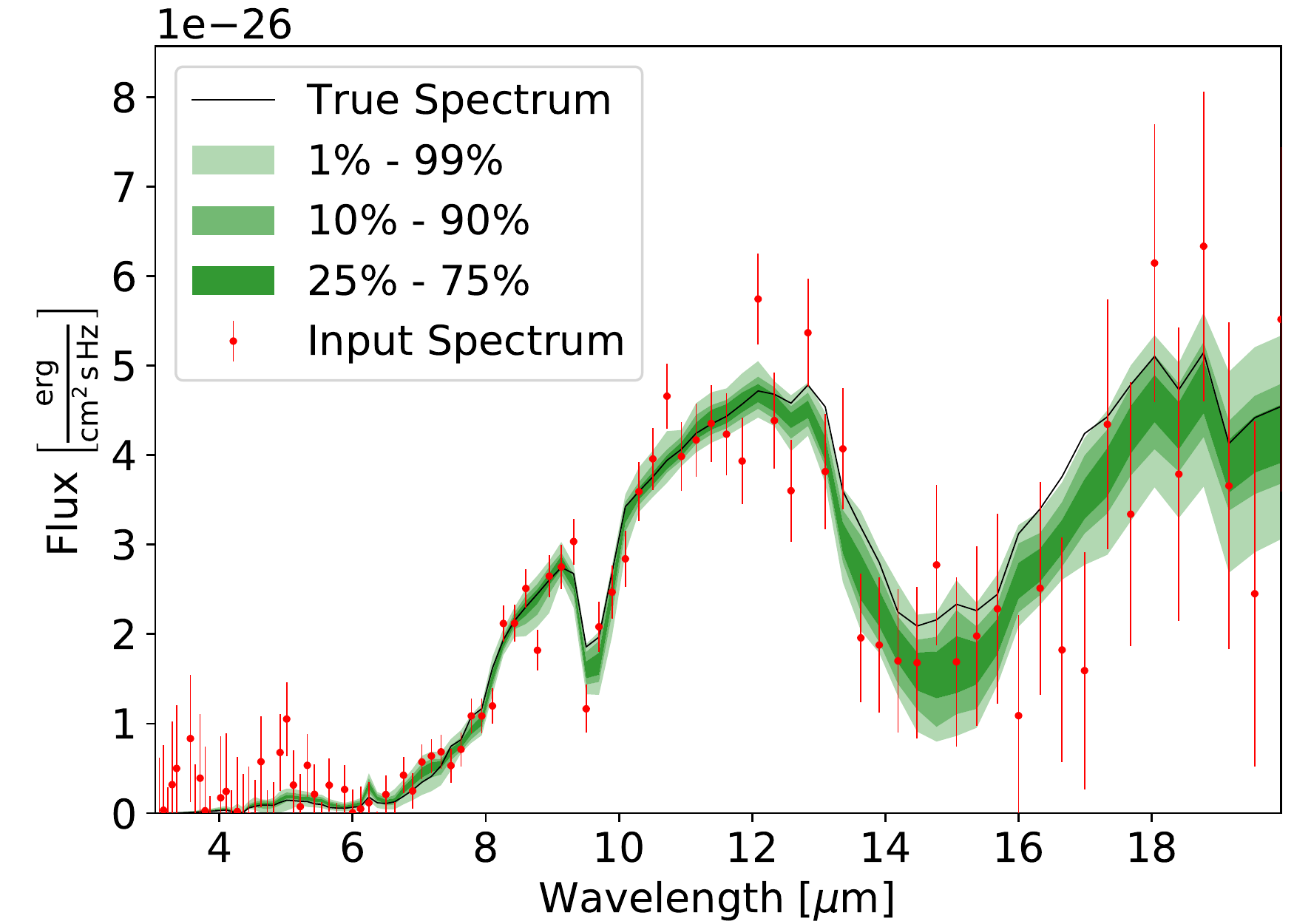}}
\caption{Fitted MIR emission spectra to a randomized input spectrum. The non-randomized spectrum is marked by the black line, the randomized input for the retrieval are represented by red data points and error bars. The green shaded regions mark percentiles of the MIR spectra corresponding to the retrieved posteriors of the model parameters.}
\label{fig:ret_flux_rand}
\end{figure}

\begin{figure*}[]
\centering
\subfloat{\includegraphics[width=\textwidth]{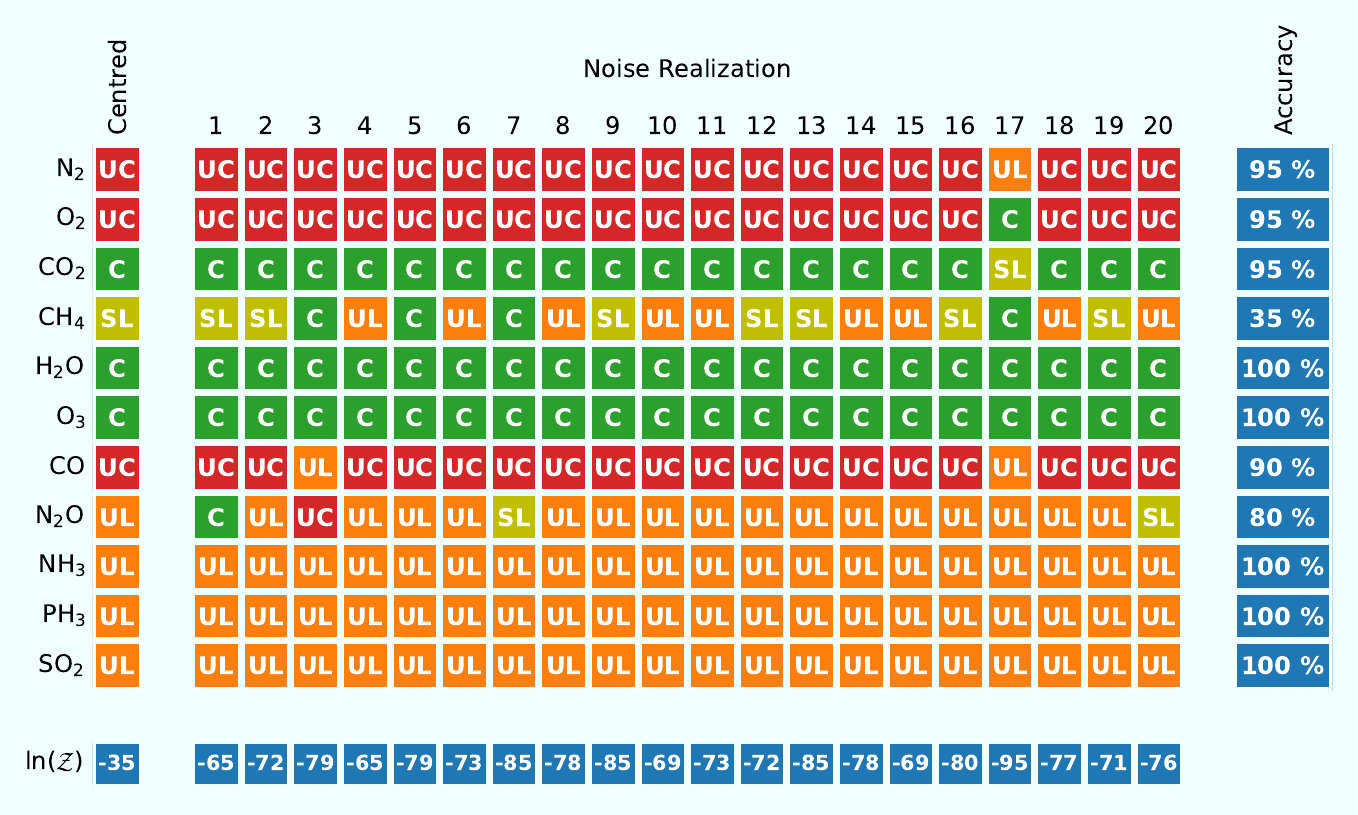}}\\
\subfloat{\includegraphics[width=.4\textwidth]{Figures/Posterior_Types_Horizontal.png}}
\caption{Retrieved posterior types for atmospheric gases for the non-randomized case (first column) and the 20 different noise realizations (columns 1 to 20). For posterior classification, we used the method outlined in Appendix \ref{PostClass}. The last row lists the Bayesian log-evidence $\ln(\mathcal{Z})$, the last column provides the retrieval accuracy (percentage of noise realizations resulting in the same posterior type as the centered case).}
\label{fig:noise_rand}
\end{figure*}

\begin{figure*}[]
\centering
\subfloat{\includegraphics[width=0.24\textwidth]{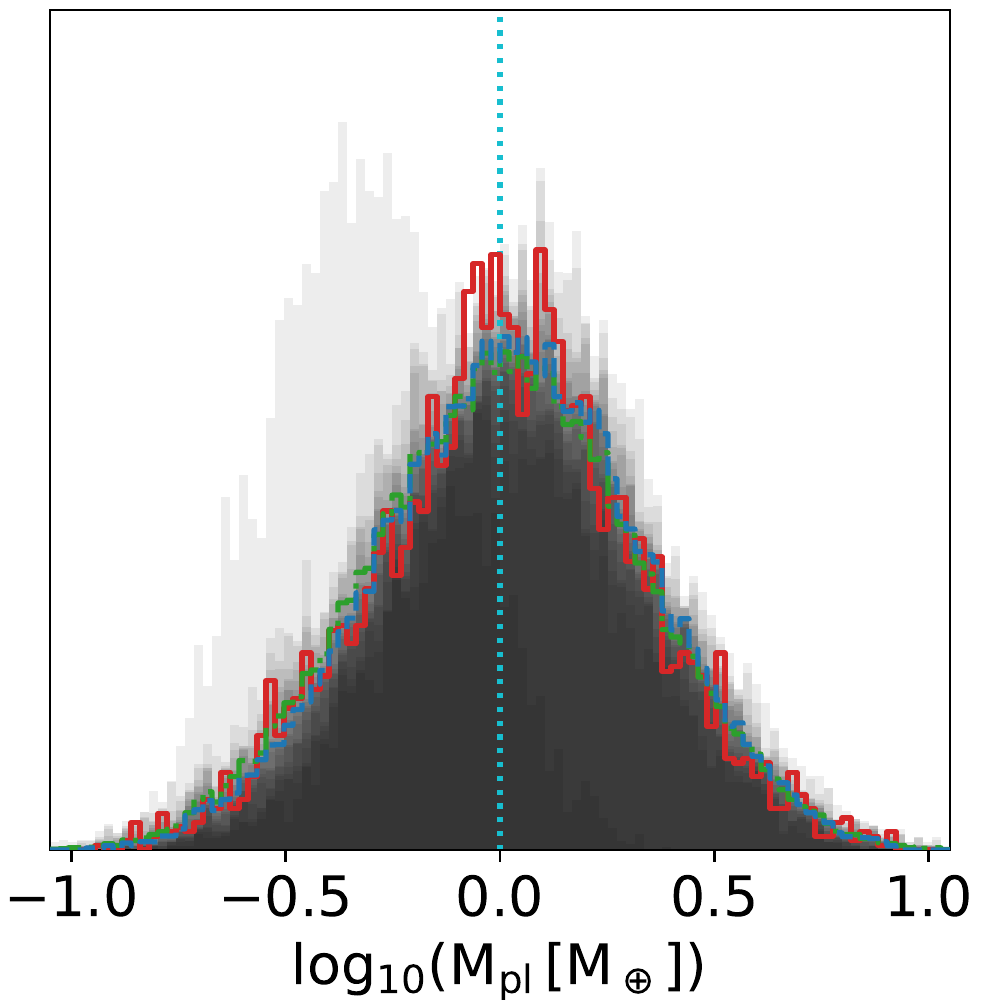}}\,\,
\subfloat{\includegraphics[width=0.24\textwidth]{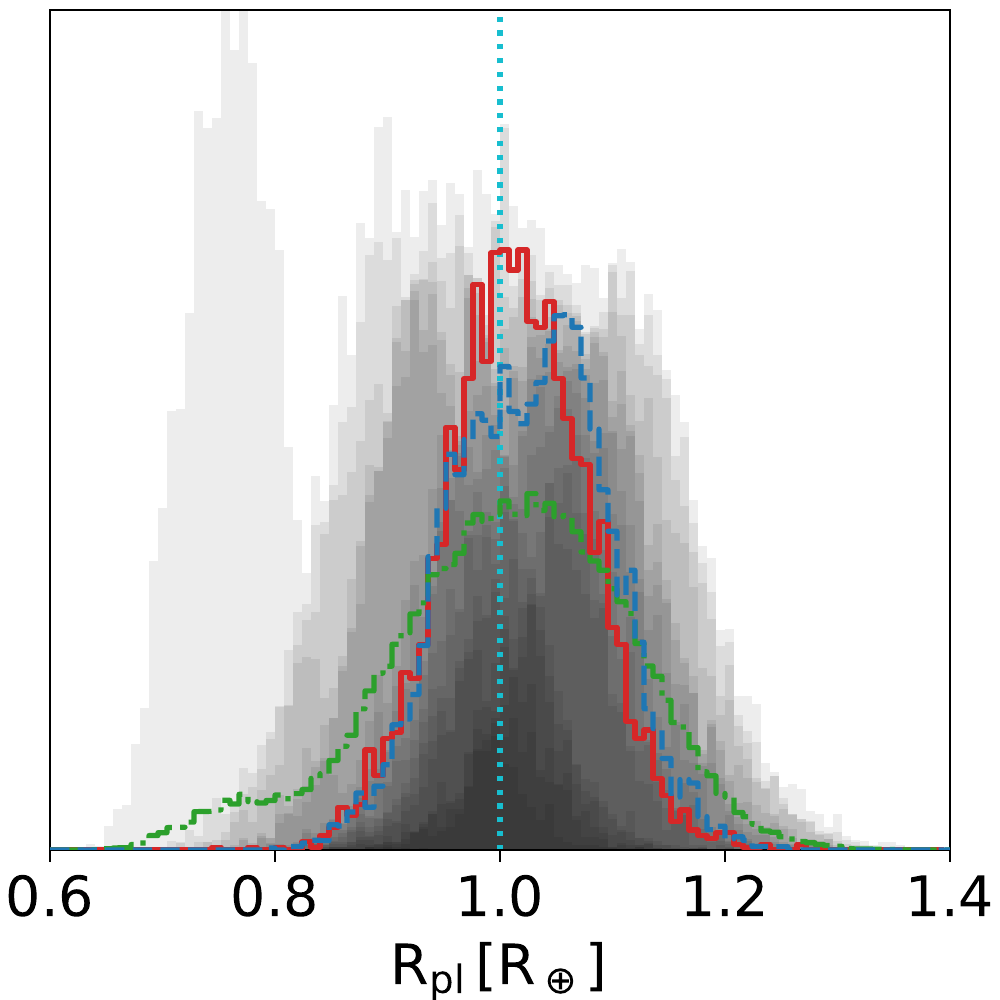}}\,\,
\subfloat{\includegraphics[width=0.24\textwidth]{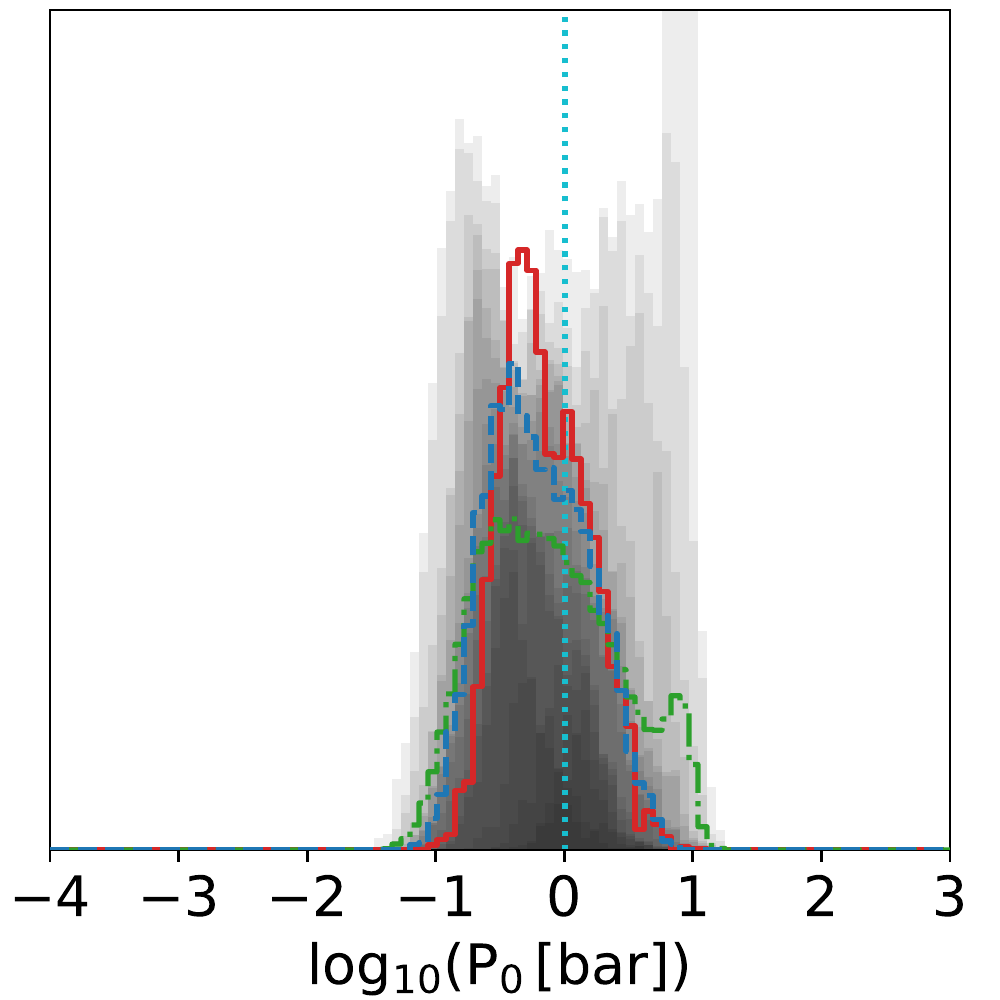}}\,\,
\subfloat{\includegraphics[width=0.24\textwidth]{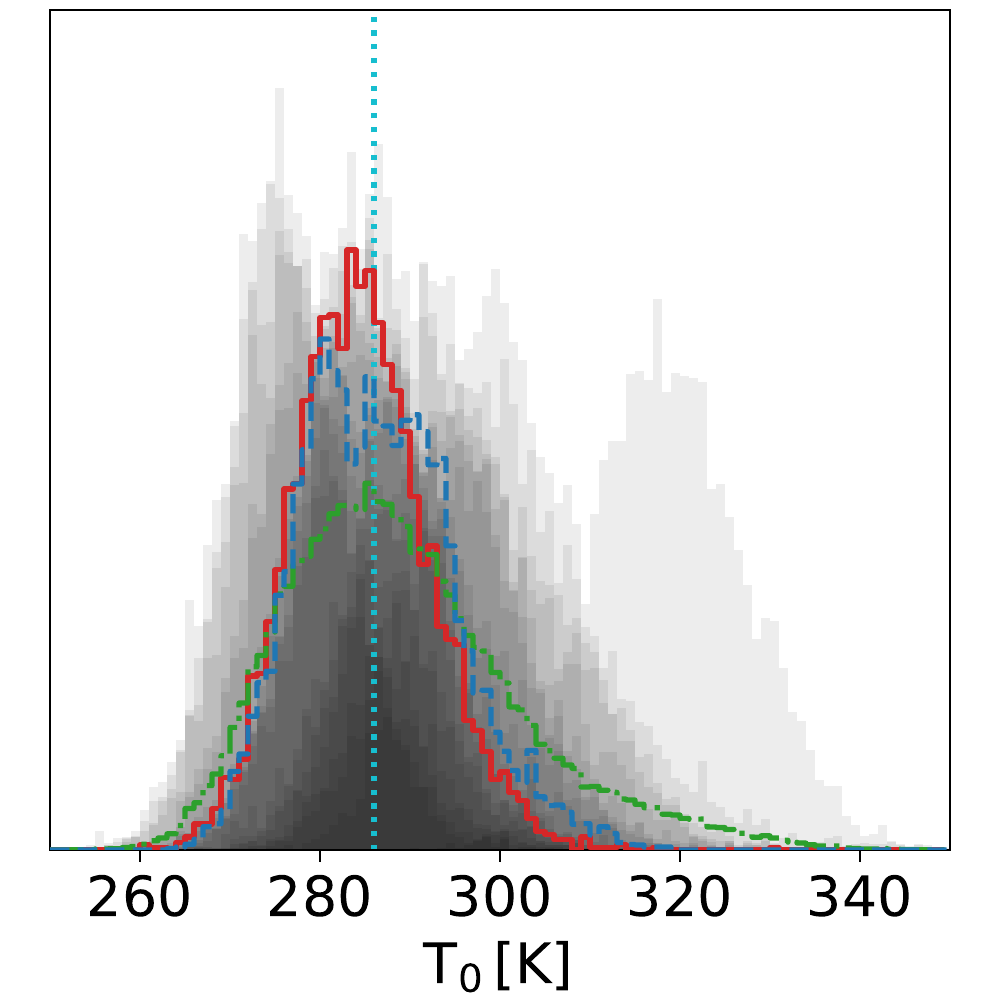}}\,\,
\subfloat{\includegraphics[width=0.24\textwidth]{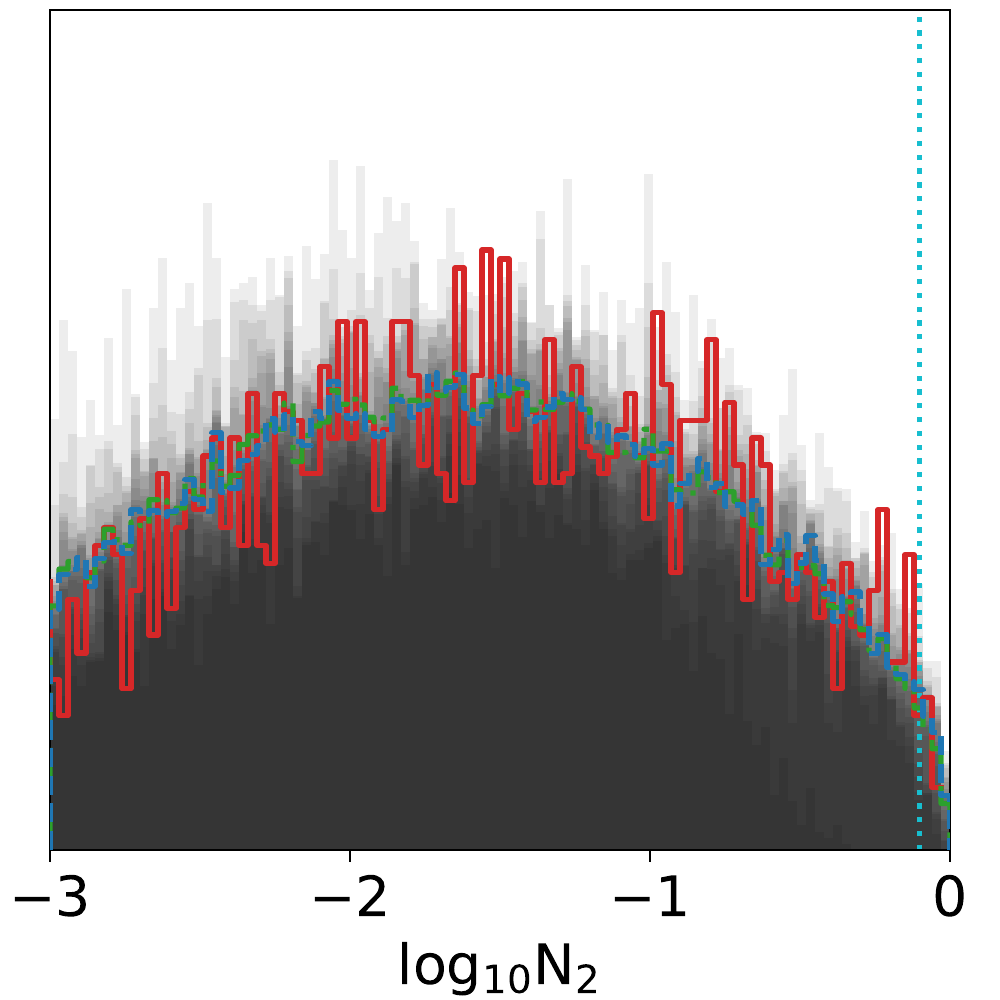}}\,\,
\subfloat{\includegraphics[width=0.24\textwidth]{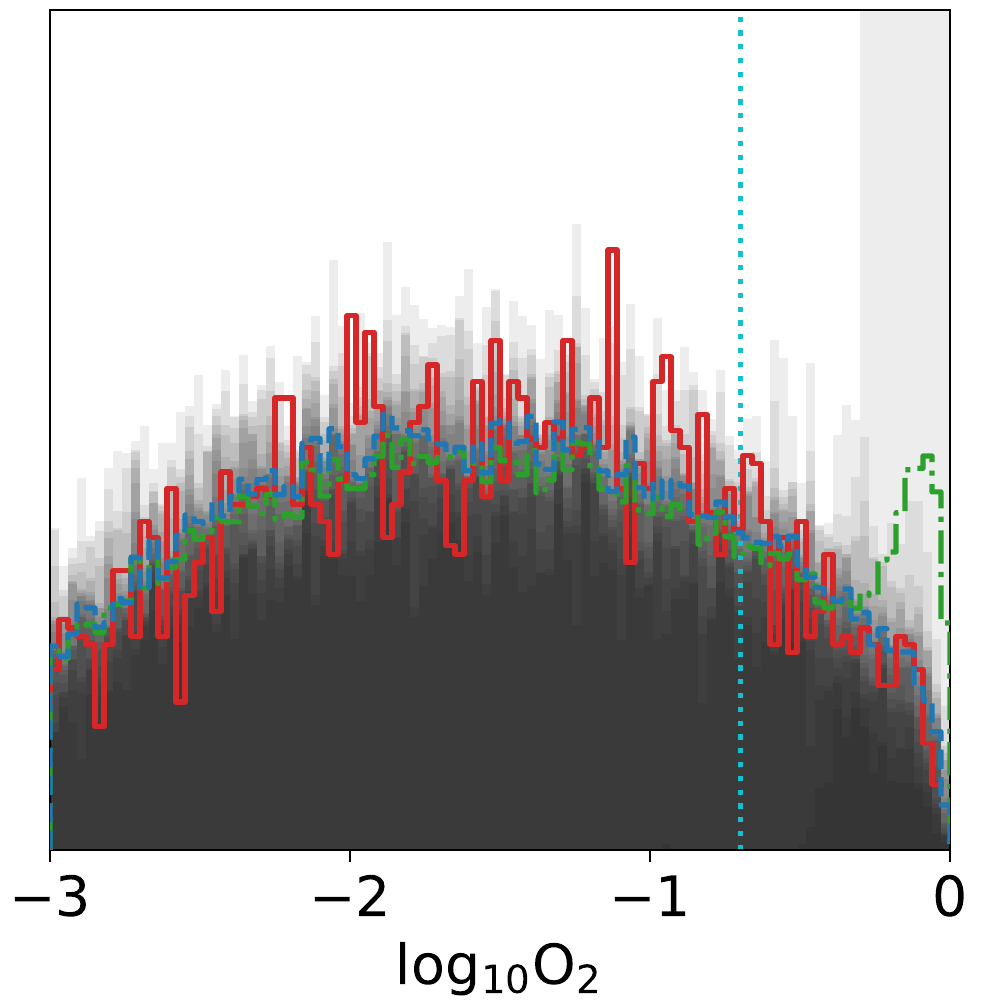}}\,\,
\subfloat{\includegraphics[width=0.24\textwidth]{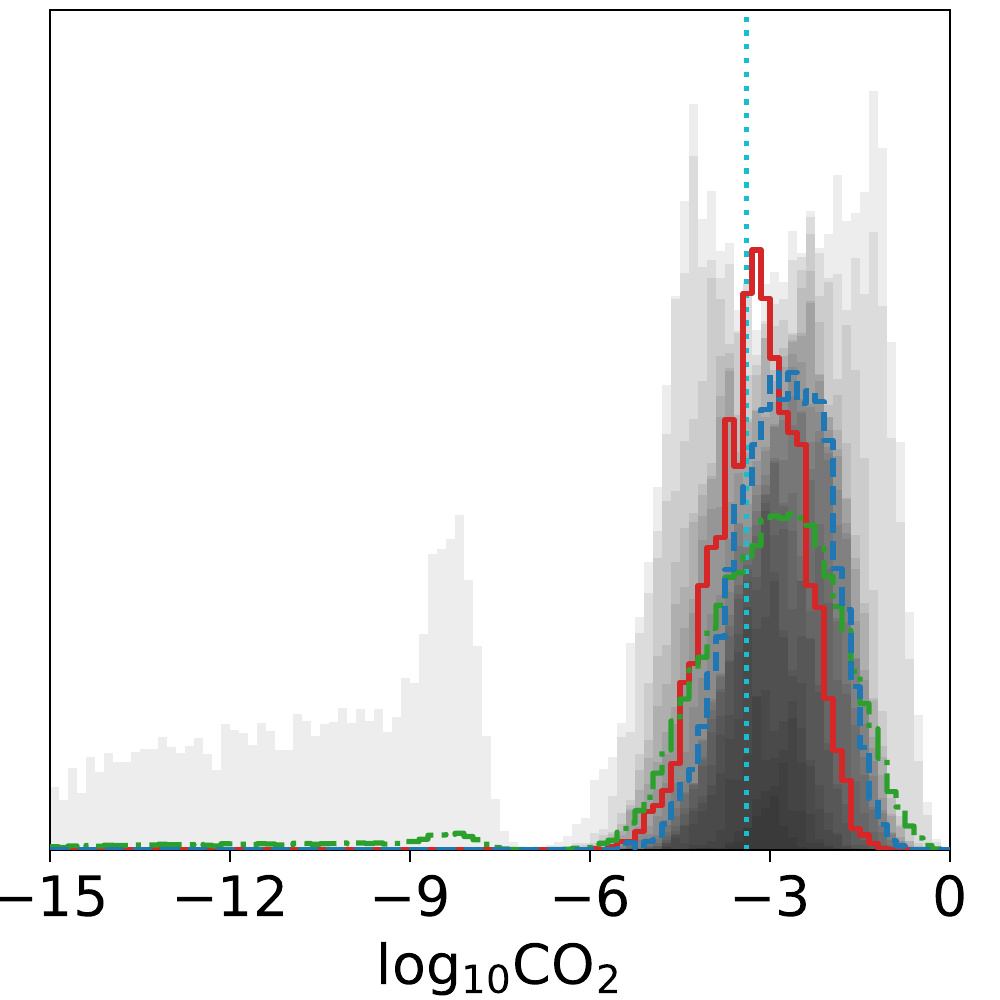}}\,\,
\subfloat{\includegraphics[width=0.24\textwidth]{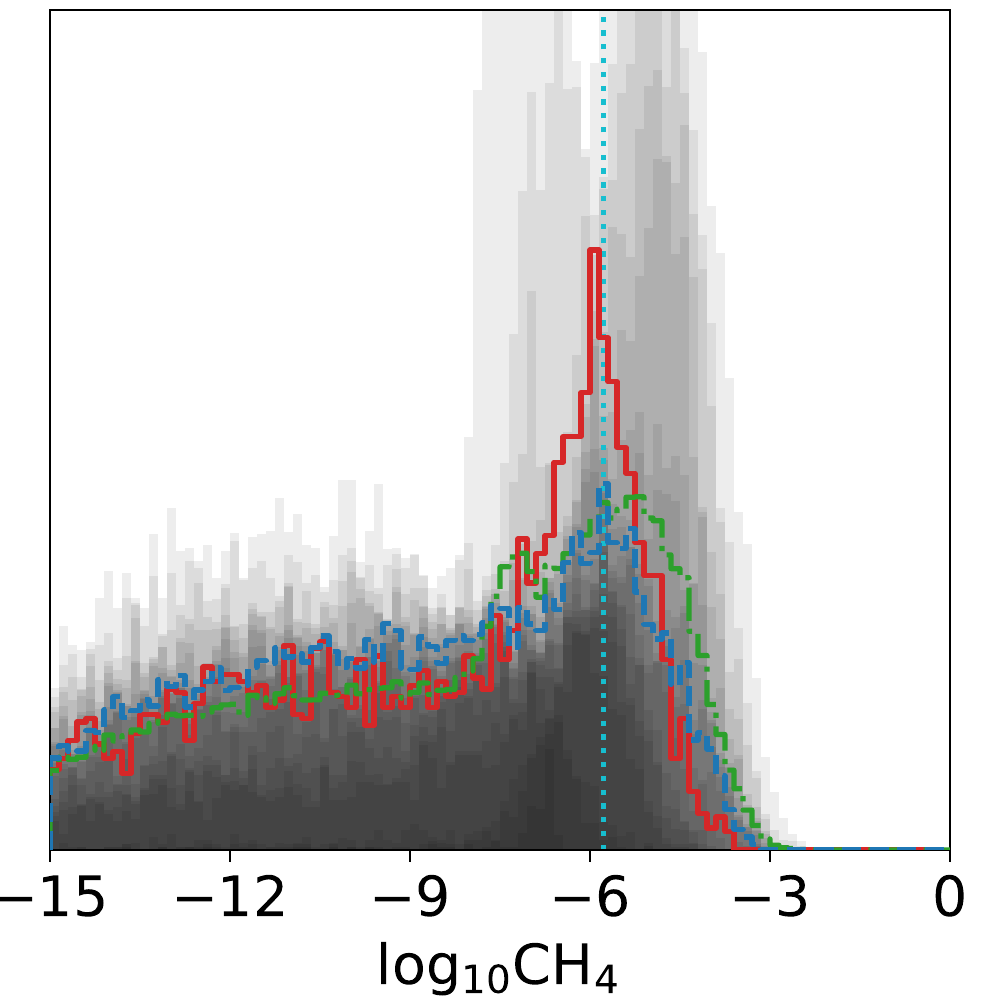}}\,\,
\subfloat{\includegraphics[width=0.24\textwidth]{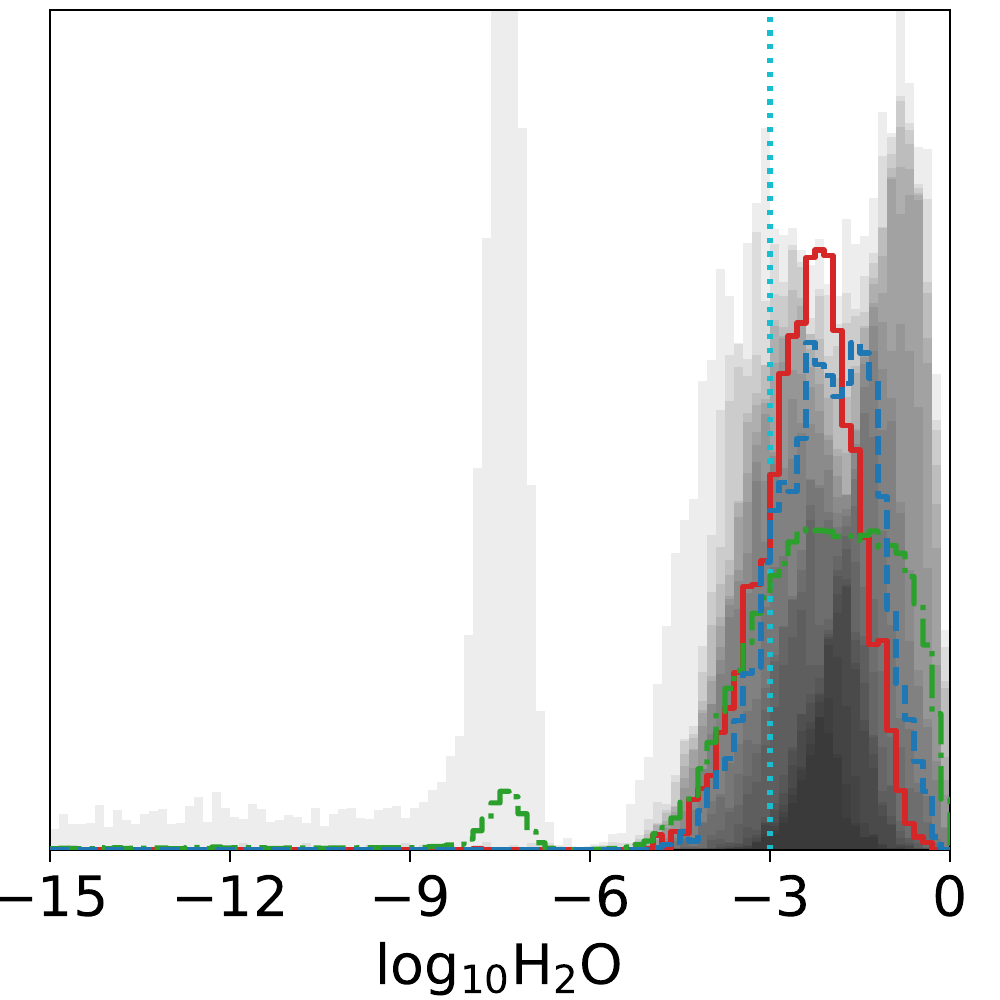}}\,\,
\subfloat{\includegraphics[width=0.24\textwidth]{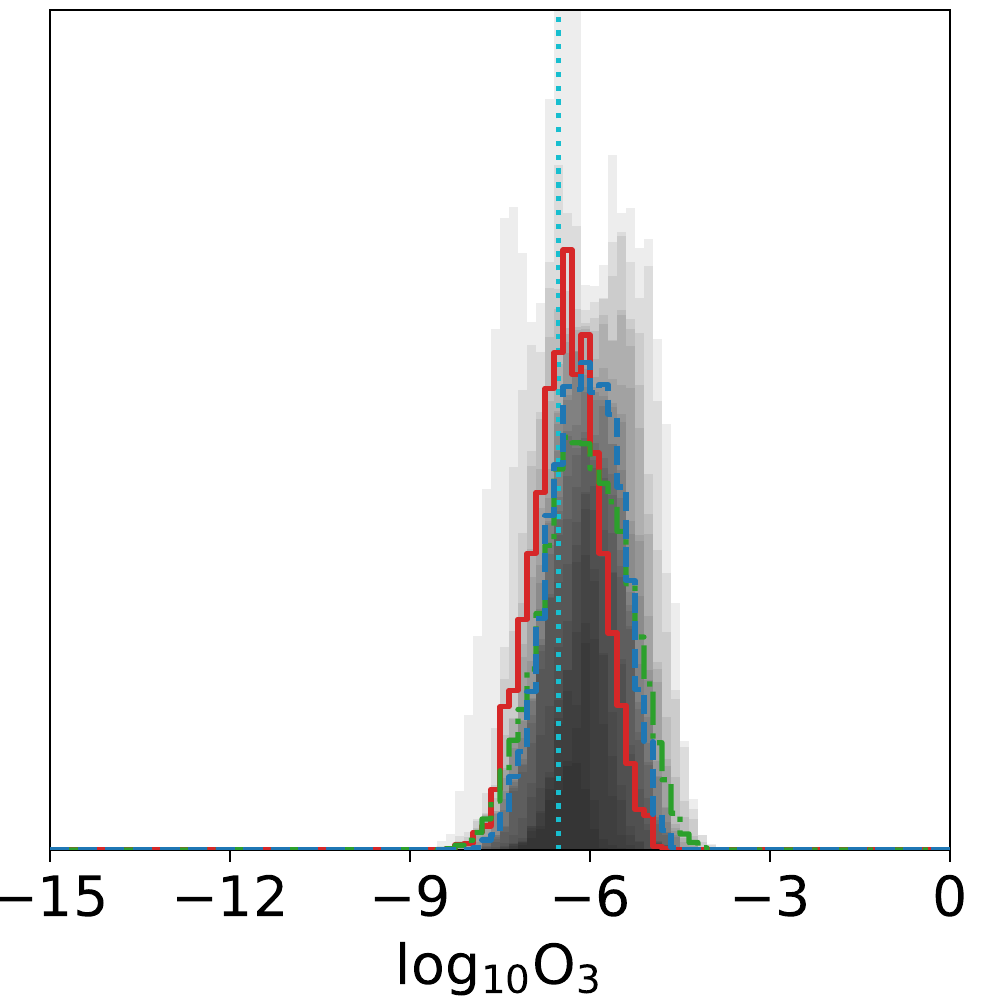}}\,\,
\subfloat{\includegraphics[width=0.24\textwidth]{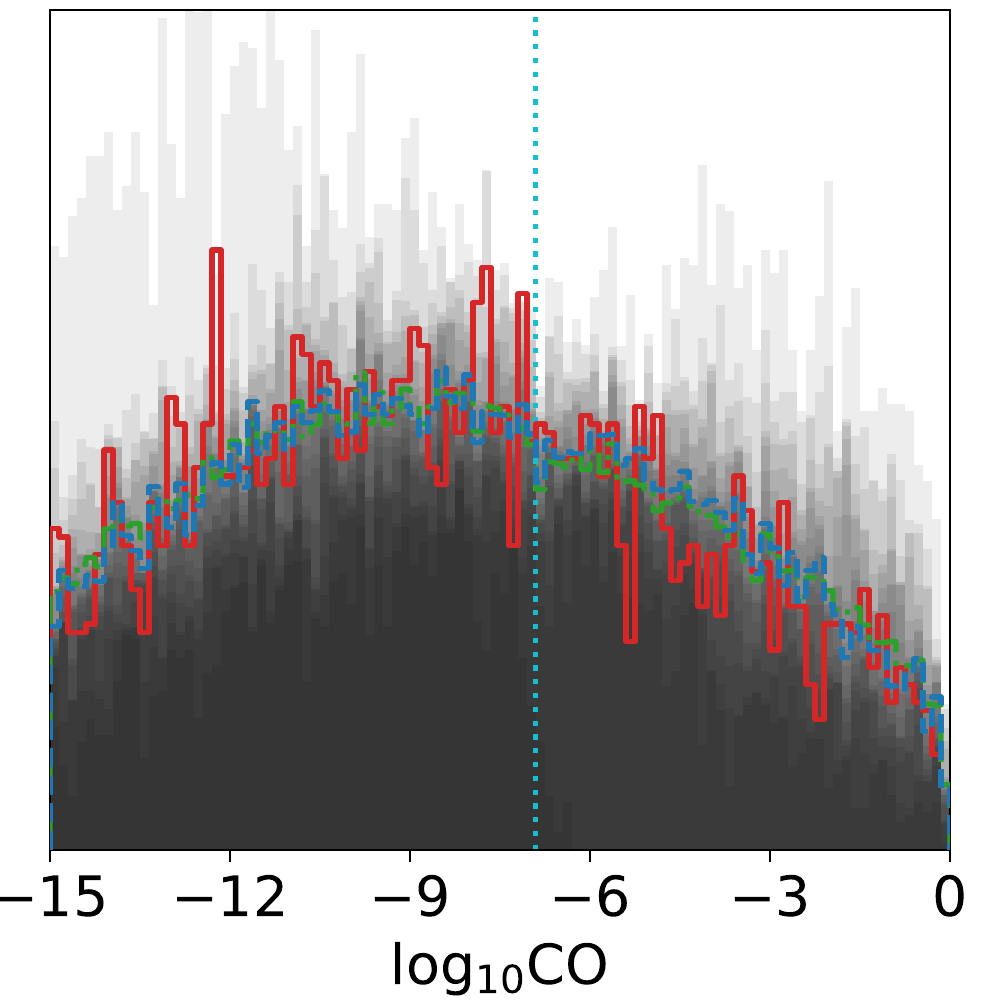}}\,\,
\subfloat{\includegraphics[width=0.24\textwidth]{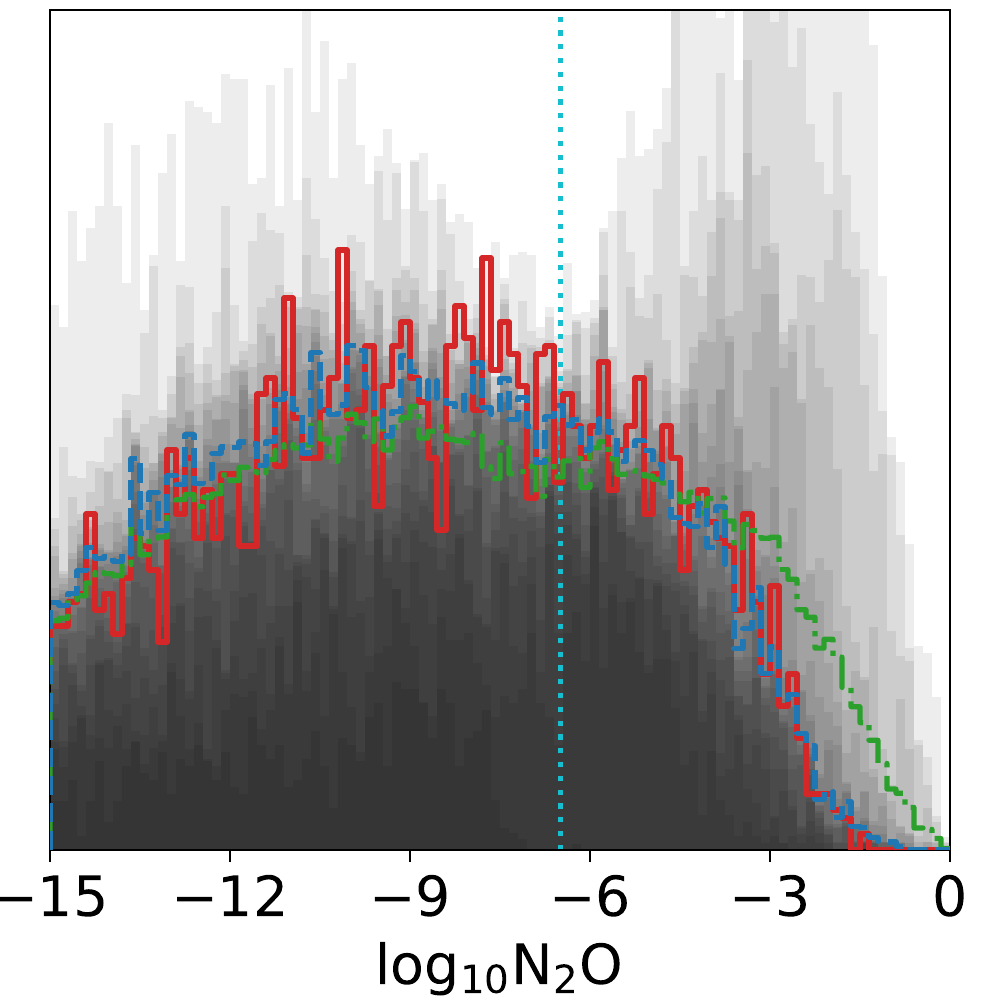}}\,\,
\subfloat{\includegraphics[width=0.24\textwidth]{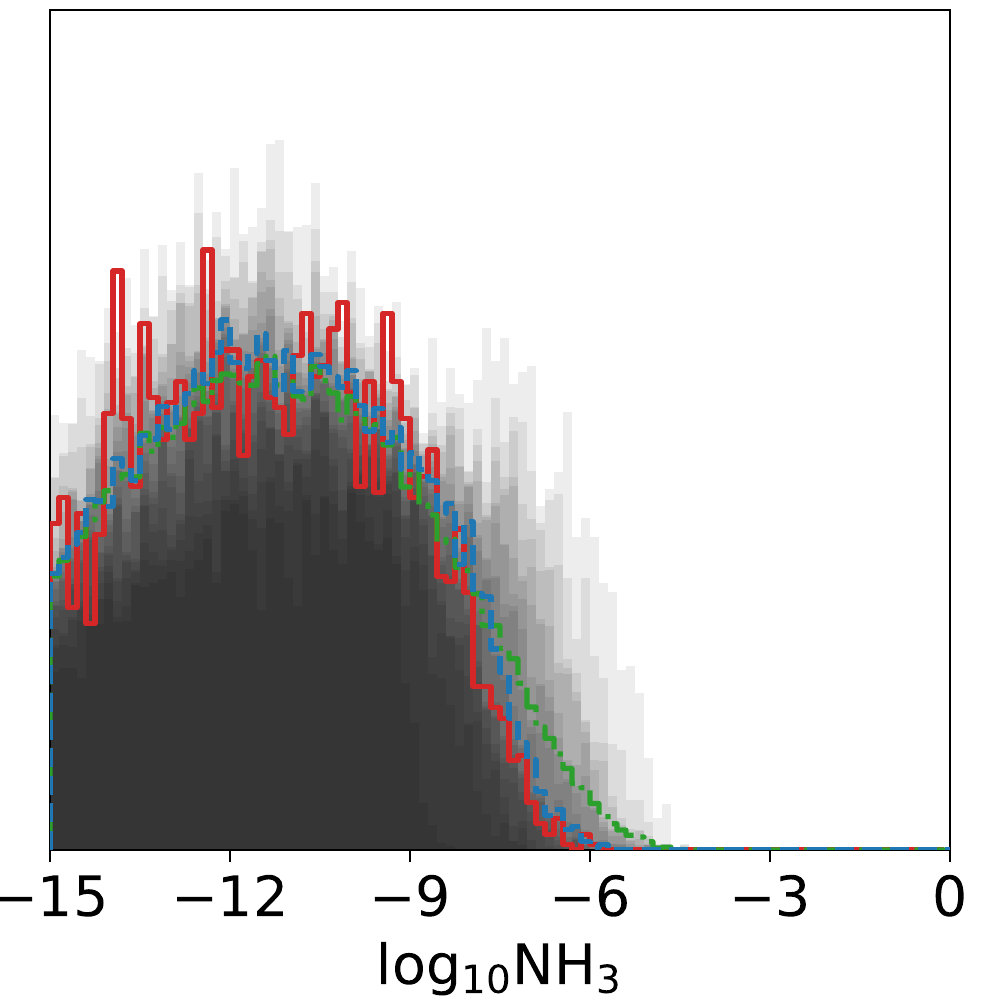}}\,\,
\subfloat{\includegraphics[width=0.24\textwidth]{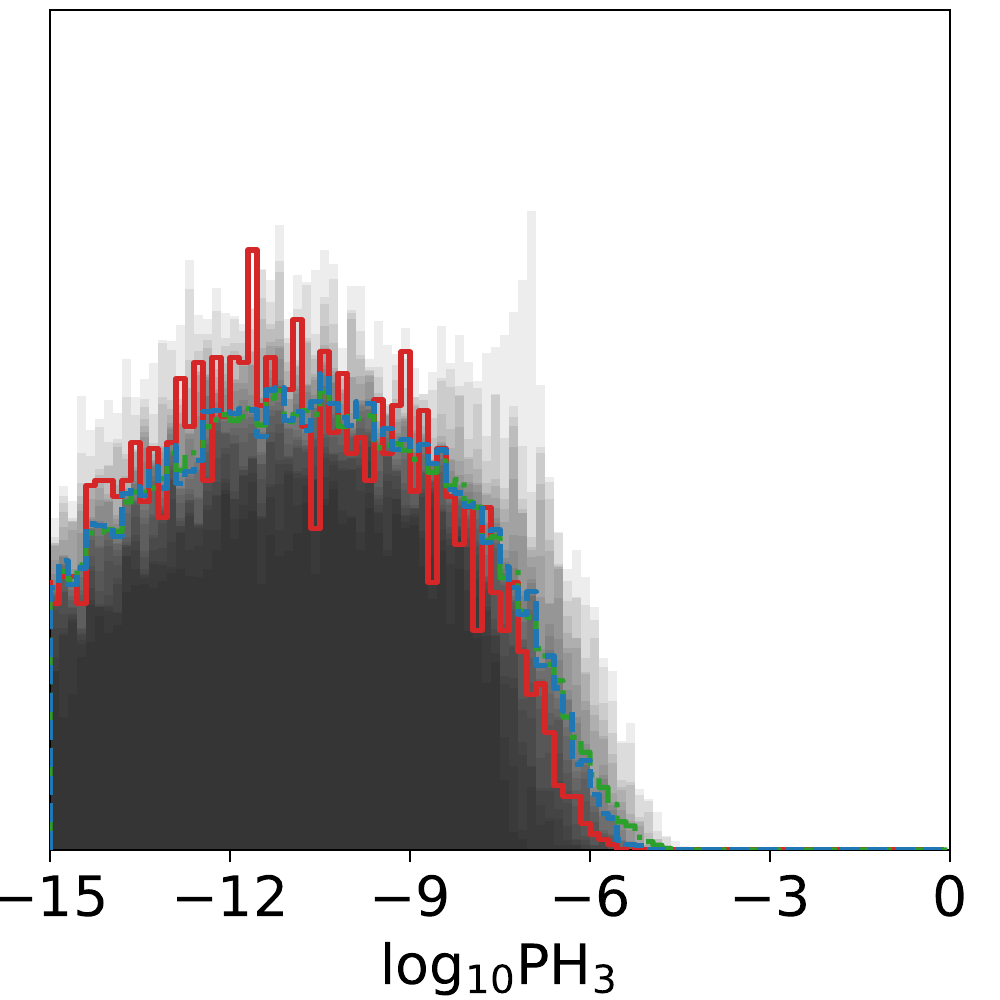}}\,\,
\subfloat{\includegraphics[width=0.24\textwidth]{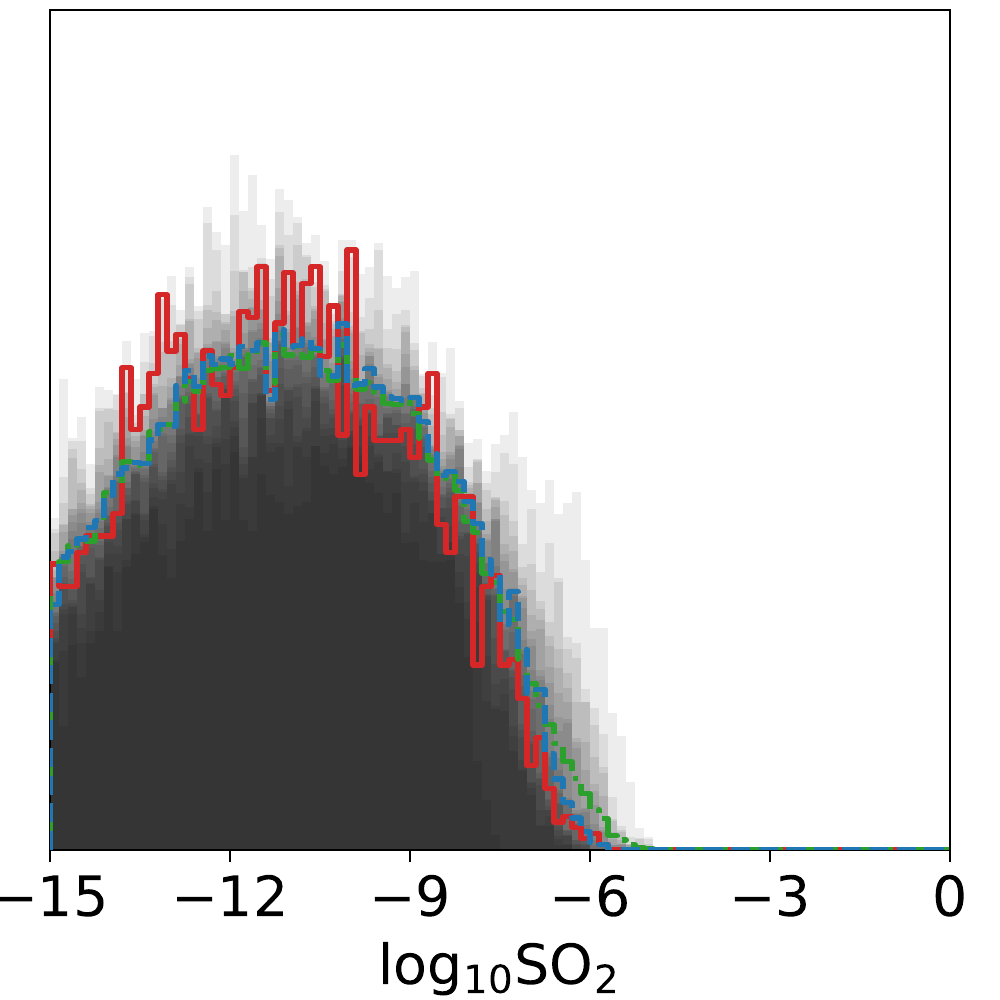}}\,\,
\subfloat{\includegraphics[width=0.24\textwidth]{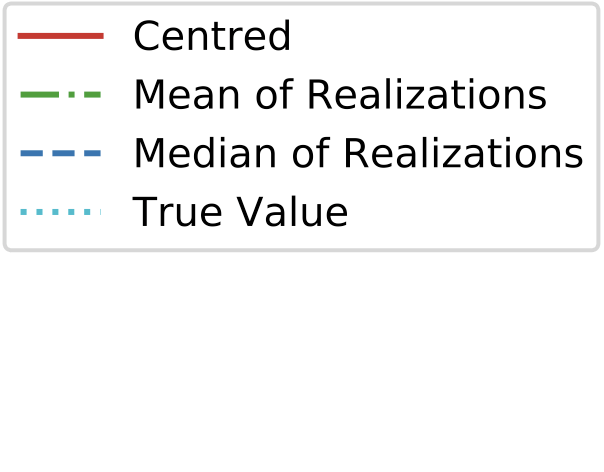}}\,

\caption{Retrieved normalized posterior distributions of some model parameters for the 20 noise realizations. The posteriors of the individual noise realizations are plotted in light gray. The more posteriors overlap, the darker the gray becomes. The green dash-dotted line marks the mean, the blue dashed line the median of the posteriors for the different noise realizations. The posteriors found in the retrieval on the non-randomized spectrum are plotted in solid red. The dotted line marks the true value for the Earth-twin atmosphere. Absence of the dotted line indicates that the species were not present in the atmosphere used to generate the input spectrum.}
\label{fig:post_rand}
\end{figure*}

As discussed in Sect.~\ref{pR_setup}, we rely on two major assumptions in order to render our retrieval study computationally feasible and bias-free. Firstly, we did not randomize the placement of the individual spectral points. Instead, we ran retrievals for the non-randomized spectral points and treated the LIFE\textsc{sim} noise as uncertainty on the spectrum. Secondly, our retrieval model did not allow for any additional gas species in the atmosphere. However, in real observations, we will not know what species are present. Therefore, it is important to investigate how robust these retrievals are with respect to false positive detections of molecules. There could be a finite probability of false positive detections, especially if combined with the randomized placement of the spectral points.

In this section, we investigate how randomizing spectral points and adding more species to the forward model impacts the performance of our retrieval routine at the proposed instrument requirements for \textit{LIFE} (R~$=$~50, S/N~$=$~10, optimized noise scenario). To this purpose, we generate 20 different noise realizations of the spectrum at these requirements. We randomize the placement of each individual spectral point by randomly sampling a Gaussian distribution, where the standard deviation is determined by the wavelength dependent LIFE\textsc{sim} noise. Next, we run a retrieval for every noise realization using the \texttt{pyMultiNest} settings outlined in Sect.~\ref{Validation}. In addition to the molecules present in the input spectrum, we also retrieve for traces of \ce{NH3}, \ce{PH3}, and \ce{SO2}. The resulting posteriors are then analyzed and classified using the method outlined in Appendix \ref{PostClass}.

We summarize our results in Figs.~\ref{fig:ret_flux_rand}, \ref{fig:noise_rand}, and \ref{fig:post_rand}. Figure~\ref{fig:ret_flux_rand} compares the MIR spectra corresponding to the posteriors retrieved from a single noise realization to the true spectrum. In Fig.~\ref{fig:noise_rand}, we summarize all retrieved posterior types for the considered atmospheric gases. The "Centered" column provides the retrieval findings for the non-randomized input. Columns 1 to 20 list the results for the randomized input spectra. The "Accuracy" column gives the percentage of retrievals on the randomized input that yield the same posterior type as the non-randomized retrieval. The last row provides the log-evidences $\ln(\mathcal{Z})$ for each retrieval run. In Fig.~\ref{fig:post_rand}, we show all retrieved posterior distributions as histograms and overlay them. Further, we plot the bin-wise mean and median of the posteriors obtained in the randomized retrievals and the posteriors found in the non-randomized retrieval.

\subsection{Truth versus spectra from noisy retrievals}\label{spec_fit}

As can be seen from Fig.~\ref{fig:ret_flux_rand}, our retrieval framework shows robust behavior even for randomized input spectra. The spectra corresponding to the retrieved posteriors roughly follow the true spectrum and generally do not significantly overfit to the noise. This observation holds for all noise realizations and is reinforced by the finding that the $\ln(\mathcal{Z})$ corresponding to the randomized retrievals are significantly smaller than in the non-randomized case (see Fig.~\ref{fig:noise_rand}). When overfitting, the retrieval models noise induced features accurately and thus the difference between the fitted spectrum and the input spectrum is small. This results in a $\ln(\mathcal{Z})$ value similar to the one obtained in retrievals of non-randomized spectra. In our case, we are not fitting to the noise induced features. Thus, there is a larger difference between the fitted and input spectrum than in the non-randomized case, which results in a smaller $\ln(\mathcal{Z})$.

Despite the good overall fit, a systematic offset between the true and the retrieved spectra starts to emerge above 10~$\mu$m. This offset becomes more pronounced at the longer wavelengths and results from the decreasing density of spectral points and increasing LIFE\textsc{sim} noise in this wavelength range. Between 10 and 12~$\mu$m, the molecules we consider have no strong absorption features and thus this range probes the planetary surface conditions. Offsets in this range will manifest themselves as shifts in the retrieved posteriors of $P_0$, $T_0$, and $R_{\mathrm{pl}}$. Above 12~$\mu$m the shape of the MIR spectrum is dominated first by \ce{CO2} (up to $\approx$~17~$\mu$m) and then by \ce{H2O} features. Offsets from the truth in this range correspond to shifts in the retrieved abundance posteriors of these molecules. The described offsets are unique to each of the considered noise realizations and are diminished when averaging the retrieval results obtained from multiple noise realizations (see Sect.~\ref{interpretation_unrandomized} for a detailed discussion).

\subsection{Robustness of the retrieved abundance posterior types}\label{abund_type}

In this section, we take a closer look at the retrieved abundance posterior types to understand how robust different types are with respect to the noise randomization of the input spectrum. There are two underlying questions. Firstly, how does randomization affect the results for molecules that are present in the atmosphere? Secondly, does randomization trigger false positive detections of molecules not present in the atmosphere?

From Fig.~\ref{fig:noise_rand}, we see that the C-type posterior is robust under randomization. This means that, if retrieving the non-randomized spectrum yields a C-type posterior, retrievals on noise realizations will mostly also yield C-type posteriors. This also shows in the high accuracy percentage. For our study, this observation implies that, at the determined \textit{LIFE} requirements, we can expect to be capable of detecting \ce{CO2}, \ce{H2O}, and \ce{O3} in an Earth-twin atmosphere. We observe a similarly strong robustness for the UC-type posterior. From this, we can conclude that retrieval behavior for \ce{N2}, \ce{O2}, and \ce{CO} is accurately predicted via the non-randomized retrievals presented in our study.

Also, the UL-Type posterior shows robust behavior with respect to noise randomization. Here we have to differentiate between two different scenarios:

\begin{itemize}
    \item \ce{N2O}, which is present in the input atmosphere,
    
    \item \ce{NH3}, \ce{PH3}, and \ce{SO2}, which are not present.
\end{itemize}

For \ce{N2O}, the high robustness of the UL-type posterior reassures our finding that \ce{N2O} is likely not detectable in an Earth-twin at the proposed \textit{LIFE} requirements. However, we do observe rare exceptions where an SL- or even C-Type posterior is retrieved. These exceptions correspond to cases where the randomization of the input spectrum results in a perceived amplification of the \ce{N2O} absorption feature. This results in a detection of \ce{N2O}. Naturally, the occurrence of a \ce{N2O} detection is characterized by a strong overestimation of the retrieved \ce{N2O} abundance. This can be seen from the retrieved posteriors of \ce{N2O} displayed in Fig.~\ref{interpretation_unrandomized}.

For \ce{NH3}, \ce{PH3}, and \ce{SO2}, we retrieve UL-type posteriors in all cases, which yields upper limits on the abundances of these molecules. We emphasize that retrieving a UL-type posterior is not a false positive detection. A UL-type posterior indicates that molecular abundances below the corresponding upper limit do not lead to an observable signature in the spectrum and therefore cannot be ruled out. The invariance of the retrieved UL-type posterior with respect to the different noise realizations indicates, that at \textit{LIFE} requirements, we are robust to false positive detections of these molecules. We are aware that this analysis does not generally rule out false positive detections of arbitrary  molecules. However, the fact that there was no false positive suggests that we are likely robust with respect to a large variety of false positives.

The most interesting case in this study is \ce{CH4}. When not randomizing the input spectrum, we retrieve an SL-type posterior, which indicates that \ce{CH_4} lies at the sensitivity limit (for the considered input data). When using the randomized spectra as input, the retrieved posterior types for \ce{CH4} vary between UL, SL and C. Therefore, SL-type posteriors are not robust with respect to spectrum randomization. The observed variance in retrieved posterior type originates because the randomization of input spectra can lead to an amplification or a reduction of the \ce{CH4} feature (similar as for \ce{N2O}). Since the true \ce{CH4} abundance is close to the sensitivity limit, already small differences in the input spectrum can lead to differences in the retrieved posterior type. However, in contrast to \ce{N2O}, the peaks of the retrieved SL- and U-type posteriors lie roughly within $\pm 1$~dex of the true \ce{CH4} abundance and do not have a tendency to overestimate it (see Fig.~\ref{interpretation_unrandomized}). Still, the fact that the posterior type for \ce{CH4} depends on the noise realization leads us to the question of how we should interpret an SL-type posterior in a non-randomized retrieval.

\subsection{Interpretation of retrievals results for unrandomized spectra}\label{interpretation_unrandomized}

As we discussed in the previous section, SL-type posteriors do not show robust behavior with respect to noise randomization. This observation raises the question of how to correctly interpret results obtained in retrievals of unrandomized input spectra. In Fig.~\ref{interpretation_unrandomized}, we illustrate one possible answer to this question, which we further motivate in this section.

When considering the posteriors from the individual retrieval runs on randomized input spectra (gray shaded areas), we observe that in some cases there are non-neglectable differences between the different results. For the UC-type posteriors (\ce{N2}, \ce{O2}, and \ce{CO}) we observe only small differences between the individual retrieval runs. This underlines the robustness of this posterior type with respect to noise randomization. Similarly, for \ce{NH3}, \ce{PH3}, and \ce{SO2}, we only observe little variance in the retrieved UL-type posteriors, which emphasizes the retrieval's robustness with respect to false positive detections. Small differences between the different posteriors correspond to small biases in the retrieved upper abundance limits of these molecules. Similar conclusions can be drawn for \ce{N2O}, where most of the retrieved posteriors are of the UL-type. The few deviances from the norm correspond to the previously described cases where an SL-type posterior that overestimates the abundance is retrieved.

The interpretation of posteriors is more subtle when considering C-Type ($\mathrm{M_{pl}}$, $\mathrm{R_{pl}}$, $\mathrm{P_0}$, $\mathrm{T_0}$, \ce{CO2}, \ce{H2O}, and \ce{O3}) or SL-type (\ce{CH4}) posteriors. Here we observe a considerable variance in both the shape and the position of the individual posteriors. This observed variance in the posteriors is a bias which is evoked by the noise randomization of the input spectra. We further find that the variation occurs around the true value and its magnitude is comparable to the standard deviation of the posteriors obtained in the non-randomized retrievals. Furthermore, we observe that the regions where many posteriors overlap (the dark regions) are similar in shape and position to the posteriors found in the unrandomized retrieval. In other words, the regions of parameter space that most retrievals agree on corresponds to the retrieval result obtained in the unrandomized retrieval.

Given the observations outlined above, the straightforward interpretation for retrieval results from unrandomized spectra is the following: the retrieved posterior distributions provide an estimate for the average retrieval behavior. Thus, if we take the average over the results from retrievals on different noise instances, it will converge toward the posterior distributions found in a retrieval on the unrandomized spectrum. To further motivate this interpretation, we plot the bin-wise mean and median of the posteriors retrieved from the 20 noise realizations in Fig.~\ref{interpretation_unrandomized}. We find that already for 20 noise realizations, both mean and median are similar in shape and position to the posteriors from the unrandomized retrieval. For all cases, the median resembles the unrandomized case more closely because it is less sensitive to outliers.

In conclusion, this study suggests that retrieving on an unrandomized input spectrum eliminates biases that arise from noise randomization and will provide reliable estimates for the average behavior of the retrieval on randomized spectra. Generally, UC, UL and C-type posteriors show robust behavior with respect to noise randomization. In contrast, SL-type posteriors are less robust and vary between C-, SL-, and UL-type in retrievals of randomized spectra. This variance in posterior type is closely linked to the fact that SL-type posteriors signify that an abundance is at the sensitivity limit of the retrieval for the studied spectral input.
\FloatBarrier

\section{Retrieved parameter values}\label{app:res}

The retrieved parameter values are listed in Tables:
\begin{itemize}
    \item Table \ref{table:3-20}: $3-20\,\mu$m, nominal case,
    \item Table \ref{table:3-20_opt}: $3-20\,\mu$m, optimized case,
    \item Table \ref{table:4-18.5}: $4-18.5\,\mu$m, nominal case,
    \item Table \ref{table:4-18.5_opt}: $4-18.5\,\mu$m, optimized case,
    \item Table \ref{table:6-17}: $6-17\,\mu$m, nominal case,
    \item Table \ref{table:6-17_opt}: $6-17\,\mu$m, optimized case.
\end{itemize}
\bgroup
\def\arraystretch{1.5}%
\begin{table*}[]
\caption{Retrieval results for the $3-20\,\mu$m wavelength range for the nominal case.}           
\label{table:3-20}      
\centering                          
\begin{tabular}{lc|cccc|cccc}        
\hline\hline                 
&&\multicolumn{4}{c|}{R=20}&\multicolumn{4}{c}{R=35}\\
Parameter&Input&S/N=5&S/N=10&S/N=15&S/N=20&S/N=5&S/N=10&S/N=15&S/N=20\\
\hline
$\sqrt[4]{a_4}$&1.14&$1.2_{-0.3}^{+0.3}$&$1.1_{-0.2}^{+0.2}$&$1.0_{-0.2}^{+0.2}$&$1.0_{-0.2}^{+0.2}$&$1.3_{-0.3}^{+0.3}$&$1.1_{-0.2}^{+0.2}$&$1.1_{-0.2}^{+0.2}$&$1.0_{-0.2}^{+0.2}$\\ 

$a_3$&23.12&$34.3_{-19.4}^{+29.5}$&$15.8_{-8.0}^{+13.0}$&$13.2_{-6.6}^{+9.3}$&$13.3_{-6.7}^{+11.2}$&$26.7_{-14.2}^{+21.3}$&$14.7_{-7.2}^{+11.2}$&$14.2_{-7.0}^{+11.6}$&$13.7_{-6.2}^{+10.1}$\\ 

$a_2$&99.70&$163.8_{-80.0}^{+114.4}$&$65.0_{-33.3}^{+49.7}$&$56.0_{-30.0}^{+38.9}$&$52.4_{-25.7}^{+37.2}$&$102.3_{-48.4}^{+63.0}$&$57.2_{-27.4}^{+35.5}$&$57.4_{-24.9}^{+33.5}$&$57.8_{-21.7}^{+29.9}$\\ 

$a_1$&146.63&$147.5_{-76.7}^{+125.6}$&$104.1_{-44.5}^{+71.1}$&$99.2_{-38.5}^{+68.4}$&$80.3_{-31.3}^{+62.4}$&$108.0_{-50.5}^{+68.6}$&$90.3_{-28.8}^{+47.1}$&$87.6_{-31.1}^{+46.5}$&$88.3_{-32.5}^{+56.0}$\\ 

$a_0$&285.22&$234.2_{-93.5}^{+94.6}$&$281.4_{-42.3}^{+45.3}$&$288.1_{-34.3}^{+45.7}$&$266.5_{-26.2}^{+34.6}$&$240.1_{-54.8}^{+42.8}$&$276.3_{-29.7}^{+34.1}$&$269.8_{-27.2}^{+33.0}$&$266.6_{-26.1}^{+43.3}$\\ 

\cdashline{1-10}$L\left(P_0\left[\mathrm{bar}\right]\right)$&0.006&$0.4_{-0.5}^{+0.3}$&$0.1_{-0.4}^{+0.4}$&$-0.0_{-0.4}^{+0.4}$&$0.2_{-0.3}^{+0.3}$&$0.4_{-0.3}^{+0.2}$&$0.1_{-0.3}^{+0.3}$&$0.2_{-0.3}^{+0.3}$&$0.2_{-0.4}^{+0.3}$\\ 

$R\,\left[R_\oplus\right]$&1.0&$0.82_{-0.1}^{+0.1}$&$0.96_{-0.09}^{+0.09}$&$1.01_{-0.08}^{+0.08}$&$1.02_{-0.07}^{+0.06}$&$0.88_{-0.10}^{+0.10}$&$0.99_{-0.08}^{+0.08}$&$1.01_{-0.06}^{+0.06}$&$1.03_{-0.05}^{+0.05}$\\ 

$L\left(M\,\left[M_\oplus\right]\right)$&0.0&$0.1_{-0.3}^{+0.3}$&$0.0_{-0.3}^{+0.3}$&$0.0_{-0.3}^{+0.3}$&$0.0_{-0.3}^{+0.3}$&$0.0_{-0.3}^{+0.3}$&$0.0_{-0.3}^{+0.3}$&$0.0_{-0.3}^{+0.3}$&$0.0_{-0.3}^{+0.3}$\\ 

\cdashline{1-10}$L(\mathrm{N_2})$&-0.107&$-7.3_{-4.4}^{+4.1}$&$-7.8_{-4.2}^{+4.6}$&$-7.6_{-4.5}^{+4.6}$&$-7.6_{-4.5}^{+4.5}$&$-7.5_{-4.3}^{+4.4}$&$-7.4_{-4.4}^{+4.4}$&$-7.7_{-4.4}^{+4.6}$&$-7.8_{-4.4}^{+4.7}$\\ 

$L(\mathrm{O_2})$&-0.679&$-7.0_{-4.6}^{+4.2}$&$-7.1_{-4.6}^{+4.4}$&$-7.4_{-4.5}^{+4.5}$&$-7.5_{-4.7}^{+4.6}$&$-7.7_{-4.3}^{+4.5}$&$-7.4_{-4.6}^{+4.3}$&$-7.6_{-4.5}^{+4.5}$&$-7.5_{-4.8}^{+4.7}$\\ 

$L(\mathrm{H_2O})$&-3.000&$-9.4_{-3.4}^{+3.8}$&$-3.2_{-1.1}^{+1.1}$&$-2.7_{-0.8}^{+0.9}$&$-3.1_{-0.7}^{+0.7}$&$-5.3_{-5.2}^{+1.7}$&$-2.9_{-0.8}^{+0.8}$&$-3.0_{-0.7}^{+0.7}$&$-3.0_{-0.7}^{+0.8}$\\ 

$L(\mathrm{CO_2})$&-3.387&$-6.6_{-4.8}^{+1.9}$&$-3.6_{-0.8}^{+0.9}$&$-3.2_{-0.7}^{+0.7}$&$-3.6_{-0.6}^{+0.7}$&$-4.6_{-1.0}^{+0.8}$&$-3.4_{-0.7}^{+0.8}$&$-3.5_{-0.6}^{+0.7}$&$-3.5_{-0.6}^{+0.7}$\\ 

$L(\mathrm{CH_4})$&-5.770&$-9.9_{-3.0}^{+3.1}$&$-8.7_{-3.8}^{+3.2}$&$-9.2_{-3.6}^{+3.4}$&$-9.6_{-3.2}^{+3.3}$&$-9.2_{-3.3}^{+3.4}$&$-9.7_{-3.2}^{+3.6}$&$-9.3_{-3.3}^{+3.2}$&$-9.0_{-3.8}^{+2.8}$\\ 

$L(\mathrm{O_3})$&-6.523&$-10.3_{-2.8}^{+2.7}$&$-6.8_{-2.2}^{+0.9}$&$-6.4_{-0.6}^{+0.6}$&$-6.7_{-0.5}^{+0.5}$&$-9.6_{-3.3}^{+2.5}$&$-6.5_{-0.6}^{+0.6}$&$-6.6_{-0.5}^{+0.5}$&$-6.7_{-0.5}^{+0.6}$\\ 

$L(\mathrm{CO})$&-6.903&$-9.8_{-3.0}^{+3.3}$&$-7.8_{-4.3}^{+4.3}$&$-7.5_{-4.5}^{+4.3}$&$-7.7_{-4.3}^{+4.5}$&$-8.7_{-3.7}^{+4.0}$&$-7.7_{-4.2}^{+4.3}$&$-7.7_{-4.4}^{+4.4}$&$-7.7_{-4.5}^{+4.3}$\\ 

$L(\mathrm{N_2O})$&-6.495&$-10.3_{-2.6}^{+2.8}$&$-8.5_{-3.9}^{+3.8}$&$-8.1_{-3.9}^{+3.7}$&$-8.5_{-4.0}^{+3.6}$&$-9.4_{-3.2}^{+3.5}$&$-8.3_{-3.9}^{+3.6}$&$-8.7_{-3.9}^{+3.6}$&$-8.9_{-3.7}^{+3.4}$\\ 
\hline
\multicolumn{10}{c}{} \\

\hline\hline                 
&&\multicolumn{4}{c|}{R=50}&\multicolumn{4}{c}{R=100}\\
Parameter&Input&S/N=5&S/N=10&S/N=15&S/N=20&S/N=5&S/N=10&S/N=15&S/N=20\\
\hline
$\sqrt[4]{a_4}$&1.14&$1.2_{-0.2}^{+0.3}$&$1.0_{-0.2}^{+0.2}$&$1.1_{-0.2}^{+0.3}$&$1.0_{-0.2}^{+0.2}$&$1.1_{-0.2}^{+0.2}$&$1.0_{-0.2}^{+0.2}$&$1.0_{-0.2}^{+0.2}$&$1.1_{-0.2}^{+0.2}$\\ 

$a_3$&23.12&$22.2_{-10.6}^{+29.6}$&$14.4_{-7.4}^{+11.8}$&$18.0_{-8.4}^{+20.9}$&$15.3_{-6.3}^{+11.2}$&$12.9_{-6.6}^{+10.0}$&$14.1_{-6.2}^{+11.4}$&$16.2_{-6.6}^{+12.1}$&$19.3_{-8.4}^{+16.8}$\\ 

$a_2$&99.70&$94.0_{-41.9}^{+69.0}$&$61.6_{-29.0}^{+43.5}$&$76.6_{-33.1}^{+69.7}$&$71.8_{-26.9}^{+34.7}$&$47.8_{-21.8}^{+26.9}$&$62.6_{-24.5}^{+35.2}$&$73.5_{-22.7}^{+34.7}$&$88.6_{-30.0}^{+56.3}$\\ 

$a_1$&146.63&$120.6_{-47.8}^{+72.6}$&$102.0_{-38.2}^{+70.2}$&$134.5_{-69.7}^{+83.4}$&$114.0_{-42.8}^{+81.2}$&$70.3_{-24.8}^{+33.5}$&$101.3_{-42.1}^{+64.2}$&$115.4_{-36.3}^{+68.2}$&$154.1_{-62.4}^{+68.0}$\\ 

$a_0$&285.22&$265.1_{-41.7}^{+47.0}$&$285.1_{-34.3}^{+39.9}$&$294.1_{-44.8}^{+49.0}$&$280.4_{-32.2}^{+61.1}$&$256.0_{-24.4}^{+30.6}$&$273.9_{-30.8}^{+45.1}$&$279.2_{-26.3}^{+47.7}$&$298.8_{-38.8}^{+41.7}$\\ 

\cdashline{1-10}$L\left(P_0\left[\mathrm{bar}\right]\right)$&0.006&$0.2_{-0.3}^{+0.3}$&$0.0_{-0.3}^{+0.4}$&$-0.1_{-0.3}^{+0.5}$&$0.0_{-0.4}^{+0.3}$&$0.3_{-0.3}^{+0.3}$&$0.1_{-0.4}^{+0.4}$&$0.0_{-0.3}^{+0.3}$&$-0.1_{-0.2}^{+0.3}$\\ 

$R\,\left[R_\oplus\right]$&1.0&$0.92_{-0.09}^{+0.09}$&$1.01_{-0.07}^{+0.07}$&$1.03_{-0.05}^{+0.06}$&$1.03_{-0.05}^{+0.05}$&$0.98_{-0.08}^{+0.08}$&$1.02_{-0.06}^{+0.06}$&$1.02_{-0.04}^{+0.04}$&$1.01_{-0.03}^{+0.04}$\\ 

$L\left(M\,\left[M_\oplus\right]\right)$&0.0&$0.0_{-0.3}^{+0.3}$&$0.0_{-0.3}^{+0.3}$&$0.0_{-0.3}^{+0.3}$&$0.0_{-0.3}^{+0.3}$&$-0.0_{-0.3}^{+0.3}$&$-0.0_{-0.3}^{+0.3}$&$-0.0_{-0.3}^{+0.3}$&$0.0_{-0.3}^{+0.3}$\\ 

\cdashline{1-10}$L(\mathrm{N_2})$&-0.107&$-6.5_{-4.6}^{+4.0}$&$-7.6_{-4.5}^{+4.5}$&$-7.3_{-4.7}^{+4.4}$&$-7.6_{-4.5}^{+4.7}$&$-7.9_{-4.1}^{+4.5}$&$-7.9_{-4.5}^{+4.8}$&$-7.4_{-4.6}^{+4.4}$&$-7.4_{-4.7}^{+4.6}$\\ 

$L(\mathrm{O_2})$&-0.679&$-7.8_{-4.3}^{+4.5}$&$-7.2_{-4.5}^{+4.4}$&$-7.1_{-4.7}^{+4.5}$&$-7.5_{-4.5}^{+4.7}$&$-7.4_{-4.4}^{+4.2}$&$-7.8_{-4.3}^{+5.0}$&$-7.3_{-4.7}^{+4.5}$&$-7.5_{-4.6}^{+4.6}$\\ 

$L(\mathrm{H_2O})$&-3.000&$-3.7_{-2.1}^{+1.1}$&$-2.8_{-0.8}^{+0.8}$&$-2.7_{-0.8}^{+0.7}$&$-2.8_{-0.7}^{+0.8}$&$-3.6_{-0.8}^{+0.9}$&$-3.0_{-0.8}^{+0.8}$&$-2.9_{-0.6}^{+0.6}$&$-2.7_{-0.6}^{+0.5}$\\ 

$L(\mathrm{CO_2})$&-3.387&$-4.0_{-0.8}^{+0.8}$&$-3.3_{-0.7}^{+0.7}$&$-3.2_{-0.7}^{+0.7}$&$-3.3_{-0.7}^{+0.7}$&$-3.9_{-0.7}^{+0.8}$&$-3.4_{-0.7}^{+0.7}$&$-3.3_{-0.5}^{+0.6}$&$-3.2_{-0.5}^{+0.5}$\\ 

$L(\mathrm{CH_4})$&-5.770&$-8.5_{-3.8}^{+3.1}$&$-9.0_{-3.6}^{+3.1}$&$-8.8_{-3.7}^{+2.9}$&$-7.7_{-4.4}^{+2.1}$&$-9.5_{-3.3}^{+3.2}$&$-8.7_{-3.7}^{+2.7}$&$-6.8_{-4.2}^{+1.3}$&$-5.9_{-1.4}^{+0.7}$\\ 

$L(\mathrm{O_3})$&-6.523&$-7.7_{-4.1}^{+1.4}$&$-6.5_{-0.6}^{+0.6}$&$-6.5_{-0.6}^{+0.6}$&$-6.5_{-0.5}^{+0.5}$&$-7.0_{-0.7}^{+0.7}$&$-6.6_{-0.5}^{+0.6}$&$-6.5_{-0.4}^{+0.4}$&$-6.4_{-0.4}^{+0.4}$\\ 

$L(\mathrm{CO})$&-6.903&$-7.8_{-4.0}^{+4.0}$&$-7.8_{-4.3}^{+4.5}$&$-7.7_{-4.5}^{+4.4}$&$-7.9_{-4.3}^{+4.4}$&$-8.1_{-4.0}^{+4.5}$&$-7.7_{-4.5}^{+4.7}$&$-7.8_{-4.4}^{+4.6}$&$-8.0_{-4.4}^{+4.6}$\\ 

$L(\mathrm{N_2O})$&-6.495&$-9.0_{-3.4}^{+3.8}$&$-8.5_{-3.8}^{+3.7}$&$-8.7_{-3.9}^{+3.5}$&$-8.9_{-3.7}^{+3.4}$&$-9.1_{-3.4}^{+3.5}$&$-9.0_{-3.7}^{+3.6}$&$-9.0_{-3.6}^{+3.4}$&$-9.3_{-3.7}^{+3.5}$\\ 
\hline
\end{tabular}
\tablefoot{Here, $L(\cdot)$ stands for $\log_{10}(\cdot)$.}
\end{table*}

\begin{table*}
\caption{Retrieval results for the $3-20\,\mu$m wavelength range for the optimized case.}            
\label{table:3-20_opt}      
\centering                          
\begin{tabular}{lc|cccc|cccc}        
\hline\hline                 
&&\multicolumn{4}{c|}{R=20}&\multicolumn{4}{c}{R=35}\\
Parameter&Input&S/N=5&S/N=10&S/N=15&S/N=20&S/N=5&S/N=10&S/N=15&S/N=20\\
\hline

$\sqrt[4]{a_4}$&1.14&$1.2_{-0.3}^{+0.2}$&$1.0_{-0.2}^{+0.2}$&$1.0_{-0.2}^{+0.2}$&$1.0_{-0.2}^{+0.2}$&$1.1_{-0.2}^{+0.2}$&$1.0_{-0.2}^{+0.2}$&$1.0_{-0.2}^{+0.2}$&$1.0_{-0.2}^{+0.2}$\\ 

$a_3$&23.12&$25.5_{-12.3}^{+20.0}$&$14.2_{-6.7}^{+9.4}$&$10.7_{-5.3}^{+8.0}$&$13.2_{-6.4}^{+11.7}$&$16.4_{-8.2}^{+14.3}$&$10.6_{-5.1}^{+7.1}$&$11.5_{-5.4}^{+10.3}$&$11.8_{-5.0}^{+8.2}$\\ 

$a_2$&99.70&$108.3_{-41.4}^{+59.0}$&$60.4_{-27.5}^{+36.3}$&$38.9_{-20.8}^{+30.2}$&$61.3_{-32.6}^{+51.6}$&$64.1_{-30.6}^{+41.2}$&$43.3_{-22.8}^{+33.8}$&$47.0_{-21.8}^{+36.5}$&$50.0_{-20.2}^{+26.1}$\\ 

$a_1$&146.63&$128.8_{-60.6}^{+67.5}$&$105.4_{-37.0}^{+51.8}$&$71.4_{-25.9}^{+42.2}$&$130.8_{-60.7}^{+97.0}$&$86.5_{-43.8}^{+56.6}$&$91.4_{-31.0}^{+53.7}$&$89.4_{-32.5}^{+59.0}$&$89.1_{-30.6}^{+42.1}$\\ 

$a_0$&285.22&$201.5_{-73.0}^{+73.6}$&$266.4_{-33.2}^{+42.4}$&$266.6_{-27.8}^{+33.0}$&$323.4_{-45.3}^{+66.8}$&$227.9_{-59.1}^{+46.2}$&$294.7_{-37.9}^{+40.7}$&$285.8_{-32.5}^{+43.6}$&$277.9_{-24.8}^{+29.5}$\\ 

\cdashline{1-10}$L\left(P_0\left[\mathrm{bar}\right]\right)$&0.006&$0.5_{-0.4}^{+0.3}$&$0.2_{-0.4}^{+0.3}$&$0.2_{-0.4}^{+0.4}$&$-0.3_{-0.3}^{+0.4}$&$0.6_{-0.4}^{+0.3}$&$-0.1_{-0.4}^{+0.4}$&$0.0_{-0.4}^{+0.4}$&$0.1_{-0.3}^{+0.3}$\\ 

$R\,\left[R_\oplus\right]$&1.0&$0.9_{-0.1}^{+0.1}$&$0.99_{-0.09}^{+0.09}$&$1.01_{-0.06}^{+0.07}$&$1.02_{-0.05}^{+0.06}$&$0.9_{-0.1}^{+0.1}$&$1.01_{-0.07}^{+0.07}$&$1.02_{-0.05}^{+0.05}$&$1.02_{-0.04}^{+0.04}$\\ 

$L\left(M\,\left[M_\oplus\right]\right)$&0.0&$0.0_{-0.3}^{+0.3}$&$0.0_{-0.3}^{+0.3}$&$0.0_{-0.3}^{+0.3}$&$0.0_{-0.3}^{+0.3}$&$0.0_{-0.3}^{+0.3}$&$0.0_{-0.3}^{+0.3}$&$0.0_{-0.3}^{+0.3}$&$0.0_{-0.3}^{+0.3}$\\ 

\cdashline{1-10}$L(\mathrm{N_2})$&-0.107&$-7.6_{-4.2}^{+4.3}$&$-7.9_{-4.2}^{+4.5}$&$-7.7_{-4.5}^{+4.4}$&$-7.8_{-4.3}^{+4.7}$&$-7.6_{-4.3}^{+4.2}$&$-7.2_{-4.6}^{+4.2}$&$-7.5_{-4.5}^{+4.6}$&$-7.5_{-4.4}^{+4.6}$\\ 

$L(\mathrm{O_2})$&-0.679&$-7.0_{-4.6}^{+4.8}$&$-7.3_{-4.6}^{+4.4}$&$-7.1_{-4.3}^{+4.3}$&$-7.9_{-4.3}^{+4.5}$&$-7.1_{-4.5}^{+4.2}$&$-7.8_{-4.4}^{+4.6}$&$-7.4_{-4.5}^{+4.4}$&$-7.5_{-4.5}^{+4.6}$\\ 

$L(\mathrm{H_2O})$&-3.000&$-4.2_{-3.0}^{+1.1}$&$-2.9_{-0.8}^{+0.9}$&$-2.9_{-0.8}^{+0.8}$&$-1.7_{-0.9}^{+0.7}$&$-3.8_{-0.9}^{+1.0}$&$-2.1_{-1.0}^{+0.9}$&$-2.4_{-0.9}^{+0.9}$&$-2.7_{-0.6}^{+0.7}$\\ 

$L(\mathrm{CO_2})$&-3.387&$-6.6_{-4.6}^{+2.0}$&$-3.9_{-0.9}^{+0.9}$&$-3.7_{-0.8}^{+0.8}$&$-2.6_{-0.8}^{+0.6}$&$-5.1_{-2.4}^{+1.2}$&$-3.1_{-0.9}^{+0.9}$&$-3.1_{-0.8}^{+0.8}$&$-3.3_{-0.6}^{+0.6}$\\ 

$L(\mathrm{CH_4})$&-5.770&$-9.8_{-3.0}^{+2.8}$&$-9.9_{-3.1}^{+3.1}$&$-9.5_{-3.4}^{+3.0}$&$-7.9_{-4.1}^{+2.6}$&$-10.2_{-2.8}^{+2.8}$&$-9.4_{-3.3}^{+3.1}$&$-8.0_{-4.2}^{+2.2}$&$-6.5_{-4.0}^{+0.9}$\\ 

$L(\mathrm{O_3})$&-6.523&$-9.7_{-3.0}^{+2.3}$&$-6.7_{-0.7}^{+0.7}$&$-6.6_{-0.6}^{+0.6}$&$-5.9_{-0.6}^{+0.6}$&$-7.5_{-2.7}^{+0.8}$&$-6.2_{-0.7}^{+0.7}$&$-6.3_{-0.6}^{+0.6}$&$-6.5_{-0.4}^{+0.5}$\\ 

$L(\mathrm{CO})$&-6.903&$-9.0_{-3.6}^{+3.9}$&$-7.5_{-4.5}^{+4.4}$&$-7.8_{-4.5}^{+4.5}$&$-7.9_{-4.3}^{+4.5}$&$-8.3_{-3.9}^{+3.9}$&$-7.5_{-4.5}^{+4.4}$&$-7.7_{-4.4}^{+4.4}$&$-8.1_{-4.4}^{+4.3}$\\ 

$L(\mathrm{N_2O})$&-6.495&$-9.8_{-3.0}^{+3.3}$&$-8.9_{-3.6}^{+3.7}$&$-9.0_{-3.6}^{+3.6}$&$-8.3_{-3.9}^{+3.4}$&$-8.8_{-3.5}^{+3.4}$&$-8.8_{-3.7}^{+3.8}$&$-9.4_{-3.3}^{+3.8}$&$-9.4_{-3.4}^{+3.3}$\\ 

\hline
\multicolumn{10}{c}{} \\

\hline\hline                 
&&\multicolumn{4}{c|}{R=50}&\multicolumn{4}{c}{R=100}\\
Parameter&Input&S/N=5&S/N=10&S/N=15&S/N=20&S/N=5&S/N=10&S/N=15&S/N=20\\
\hline
$\sqrt[4]{a_4}$&1.14&$1.3_{-0.2}^{+0.2}$&$1.0_{-0.2}^{+0.2}$&$1.0_{-0.2}^{+0.2}$&$1.1_{-0.2}^{+0.2}$&$1.0_{-0.2}^{+0.2}$&$1.0_{-0.2}^{+0.2}$&$1.1_{-0.2}^{+0.2}$&$1.0_{-0.2}^{+0.2}$\\ 

$a_3$&23.12&$30.4_{-13.0}^{+16.3}$&$14.1_{-6.7}^{+11.0}$&$11.7_{-5.1}^{+9.5}$&$13.9_{-6.2}^{+9.7}$&$10.7_{-5.1}^{+7.6}$&$14.7_{-6.4}^{+10.1}$&$19.1_{-8.4}^{+18.2}$&$17.0_{-6.7}^{+10.6}$\\ 

$a_2$&99.70&$112.7_{-41.3}^{+47.3}$&$64.4_{-30.6}^{+41.8}$&$43.8_{-18.1}^{+25.2}$&$53.2_{-19.9}^{+28.0}$&$40.3_{-21.0}^{+31.5}$&$67.9_{-26.6}^{+38.4}$&$86.6_{-37.8}^{+59.2}$&$78.0_{-26.9}^{+41.7}$\\ 

$a_1$&146.63&$141.1_{-54.4}^{+81.8}$&$126.8_{-44.3}^{+72.6}$&$69.1_{-20.7}^{+32.7}$&$79.3_{-27.2}^{+37.2}$&$79.7_{-33.1}^{+55.1}$&$128.1_{-46.1}^{+72.9}$&$148.4_{-72.5}^{+81.3}$&$132.2_{-56.2}^{+61.8}$\\ 

$a_0$&285.22&$255.5_{-41.8}^{+48.8}$&$311.1_{-31.6}^{+52.0}$&$260.9_{-19.7}^{+23.3}$&$260.8_{-19.2}^{+22.6}$&$273.8_{-32.1}^{+44.5}$&$306.9_{-39.7}^{+56.2}$&$298.4_{-42.5}^{+54.4}$&$286.7_{-34.3}^{+39.6}$\\ 

\cdashline{1-10}$L\left(P_0\left[\mathrm{bar}\right]\right)$&0.006&$0.2_{-0.3}^{+0.2}$&$-0.2_{-0.3}^{+0.3}$&$0.3_{-0.3}^{+0.2}$&$0.3_{-0.2}^{+0.3}$&$0.1_{-0.4}^{+0.4}$&$-0.2_{-0.3}^{+0.4}$&$-0.1_{-0.3}^{+0.4}$&$-0.0_{-0.2}^{+0.4}$\\ 

$R\,\left[R_\oplus\right]$&1.0&$0.98_{-0.09}^{+0.09}$&$1.02_{-0.06}^{+0.06}$&$1.01_{-0.04}^{+0.05}$&$1.01_{-0.03}^{+0.03}$&$1.01_{-0.08}^{+0.08}$&$1.01_{-0.04}^{+0.05}$&$1.0_{-0.03}^{+0.03}$&$1.0_{-0.02}^{+0.02}$\\ 

$L\left(M\,\left[M_\oplus\right]\right)$&0.0&$0.1_{-0.3}^{+0.2}$&$0.0_{-0.3}^{+0.3}$&$0.0_{-0.3}^{+0.3}$&$0.0_{-0.3}^{+0.3}$&$0.0_{-0.3}^{+0.3}$&$0.0_{-0.3}^{+0.3}$&$0.0_{-0.3}^{+0.3}$&$0.0_{-0.3}^{+0.3}$\\ 

\cdashline{1-10}$L(\mathrm{N_2})$&-0.107&$-7.8_{-4.0}^{+4.2}$&$-7.7_{-4.3}^{+4.7}$&$-7.8_{-4.4}^{+4.6}$&$-7.1_{-4.5}^{+4.2}$&$-7.3_{-4.8}^{+4.5}$&$-7.4_{-4.6}^{+4.6}$&$-7.5_{-4.7}^{+4.6}$&$-7.1_{-4.9}^{+4.5}$\\ 

$L(\mathrm{O_2})$&-0.679&$-7.1_{-4.2}^{+4.2}$&$-7.6_{-4.6}^{+4.5}$&$-7.5_{-4.6}^{+4.5}$&$-7.4_{-4.7}^{+4.6}$&$-7.2_{-4.6}^{+4.3}$&$-7.1_{-4.7}^{+4.4}$&$-7.7_{-4.5}^{+4.8}$&$-7.9_{-4.2}^{+5.0}$\\ 

$L(\mathrm{H_2O})$&-3.000&$-3.2_{-0.7}^{+0.8}$&$-2.0_{-0.7}^{+0.7}$&$-3.2_{-0.6}^{+0.7}$&$-3.3_{-0.6}^{+0.6}$&$-2.8_{-0.9}^{+1.1}$&$-2.4_{-0.8}^{+0.8}$&$-2.8_{-0.7}^{+0.6}$&$-2.9_{-0.6}^{+0.5}$\\ 

$L(\mathrm{CO_2})$&-3.387&$-4.2_{-1.0}^{+0.8}$&$-2.9_{-0.7}^{+0.7}$&$-3.7_{-0.5}^{+0.6}$&$-3.7_{-0.5}^{+0.5}$&$-3.5_{-0.9}^{+1.0}$&$-2.9_{-0.7}^{+0.7}$&$-3.2_{-0.6}^{+0.6}$&$-3.3_{-0.5}^{+0.5}$\\ 

$L(\mathrm{CH_4})$&-5.770&$-9.5_{-3.0}^{+2.9}$&$-8.0_{-4.4}^{+2.6}$&$-6.6_{-3.3}^{+0.7}$&$-6.2_{-0.6}^{+0.5}$&$-8.9_{-3.8}^{+2.8}$&$-5.7_{-1.1}^{+0.9}$&$-5.7_{-0.7}^{+0.6}$&$-5.8_{-0.6}^{+0.5}$\\ 

$L(\mathrm{O_3})$&-6.523&$-7.0_{-0.9}^{+0.6}$&$-6.1_{-0.5}^{+0.5}$&$-6.7_{-0.4}^{+0.4}$&$-6.7_{-0.4}^{+0.4}$&$-6.6_{-0.6}^{+0.8}$&$-6.2_{-0.5}^{+0.6}$&$-6.4_{-0.5}^{+0.5}$&$-6.5_{-0.4}^{+0.4}$\\ 

$L(\mathrm{CO})$&-6.903&$-7.8_{-4.1}^{+4.3}$&$-8.0_{-4.4}^{+4.6}$&$-8.0_{-4.2}^{+4.5}$&$-8.2_{-4.0}^{+4.5}$&$-7.9_{-4.6}^{+4.6}$&$-7.8_{-4.4}^{+4.3}$&$-8.6_{-4.0}^{+4.4}$&$-8.5_{-4.1}^{+4.1}$\\ 

$L(\mathrm{N_2O})$&-6.495&$-9.8_{-3.0}^{+3.6}$&$-8.3_{-3.9}^{+3.5}$&$-9.3_{-3.4}^{+3.2}$&$-9.5_{-3.5}^{+3.2}$&$-8.5_{-3.9}^{+3.7}$&$-9.0_{-3.6}^{+3.6}$&$-9.4_{-3.5}^{+3.2}$&$-9.7_{-3.2}^{+3.2}$\\ 
\hline
\end{tabular}
\tablefoot{Here, $L(\cdot)$ stands for $\log_{10}(\cdot)$.}
\end{table*}

\begin{table*}
\caption{Retrieval results for the $4-18.5\,\mu$m wavelength range for the nominal case.}       
\label{table:4-18.5}      
\centering                          
\begin{tabular}{lc|cccc|cccc}        
\hline\hline                 
&&\multicolumn{4}{c|}{R=20}&\multicolumn{4}{c}{R=35}\\
Parameter&Input&S/N=5&S/N=10&S/N=15&S/N=20&S/N=5&S/N=10&S/N=15&S/N=20\\
\hline
$\sqrt[4]{a_4}$&1.14&$1.3_{-0.3}^{+0.3}$&$1.1_{-0.2}^{+0.2}$&$1.1_{-0.2}^{+0.2}$&$1.0_{-0.2}^{+0.2}$&$1.1_{-0.2}^{+0.2}$&$1.0_{-0.2}^{+0.2}$&$1.0_{-0.2}^{+0.2}$&$1.0_{-0.2}^{+0.2}$\\ 

$a_3$&23.12&$39.3_{-22.0}^{+28.4}$&$15.1_{-7.1}^{+10.1}$&$14.9_{-7.4}^{+13.0}$&$11.7_{-5.7}^{+10.5}$&$18.9_{-9.3}^{+17.0}$&$12.1_{-5.9}^{+10.1}$&$12.6_{-5.8}^{+9.7}$&$13.5_{-6.5}^{+11.1}$\\ 

$a_2$&99.70&$192.8_{-90.8}^{+137.4}$&$60.8_{-28.7}^{+34.4}$&$59.1_{-29.8}^{+46.5}$&$45.3_{-22.1}^{+37.7}$&$85.3_{-41.0}^{+58.3}$&$53.4_{-26.3}^{+36.6}$&$52.4_{-24.2}^{+28.2}$&$57.6_{-24.2}^{+35.7}$\\ 

$a_1$&146.63&$176.0_{-90.0}^{+176.3}$&$98.3_{-31.5}^{+38.2}$&$99.4_{-35.8}^{+60.6}$&$80.5_{-31.1}^{+56.6}$&$127.6_{-60.3}^{+89.4}$&$103.3_{-36.5}^{+53.6}$&$84.0_{-32.6}^{+47.5}$&$93.0_{-33.2}^{+55.1}$\\ 

$a_0$&285.22&$272.1_{-107.3}^{+93.4}$&$267.3_{-30.9}^{+32.8}$&$280.4_{-31.4}^{+40.9}$&$275.1_{-33.0}^{+41.1}$&$265.1_{-53.7}^{+59.5}$&$291.7_{-33.7}^{+44.0}$&$270.0_{-28.4}^{+33.3}$&$273.6_{-26.8}^{+36.6}$\\ 

\cdashline{1-10}$L\left(P_0\left[\mathrm{bar}\right]\right)$&0.006&$0.1_{-0.5}^{+0.5}$&$0.2_{-0.3}^{+0.2}$&$0.0_{-0.3}^{+0.3}$&$0.1_{-0.4}^{+0.4}$&$0.2_{-0.4}^{+0.4}$&$-0.1_{-0.4}^{+0.4}$&$0.2_{-0.3}^{+0.4}$&$0.1_{-0.3}^{+0.3}$\\ 

$R\,\left[R_\oplus\right]$&1.0&$0.86_{-0.09}^{+0.1}$&$0.99_{-0.1}^{+0.1}$&$1.01_{-0.08}^{+0.07}$&$1.03_{-0.06}^{+0.06}$&$0.9_{-0.1}^{+0.1}$&$1.01_{-0.08}^{+0.08}$&$1.02_{-0.06}^{+0.06}$&$1.03_{-0.05}^{+0.05}$\\ 

$L\left(M\,\left[M_\oplus\right]\right)$&0.0&$0.1_{-0.3}^{+0.3}$&$0.0_{-0.3}^{+0.3}$&$0.0_{-0.3}^{+0.3}$&$0.0_{-0.3}^{+0.3}$&$-0.0_{-0.3}^{+0.3}$&$0.0_{-0.3}^{+0.3}$&$0.0_{-0.3}^{+0.3}$&$0.0_{-0.3}^{+0.3}$\\ 

\cdashline{1-10}$L(\mathrm{N_2})$&-0.107&$-7.6_{-4.2}^{+4.2}$&$-7.4_{-4.4}^{+4.4}$&$-7.6_{-4.5}^{+4.7}$&$-7.1_{-4.7}^{+4.2}$&$-7.0_{-4.3}^{+4.0}$&$-7.2_{-4.7}^{+4.5}$&$-7.2_{-4.9}^{+4.5}$&$-7.4_{-4.7}^{+4.6}$\\ 

$L(\mathrm{O_2})$&-0.679&$-7.4_{-4.5}^{+4.5}$&$-7.6_{-4.5}^{+4.5}$&$-7.7_{-4.3}^{+4.5}$&$-7.2_{-4.5}^{+4.5}$&$-8.0_{-4.2}^{+4.6}$&$-8.0_{-4.4}^{+4.4}$&$-8.0_{-4.4}^{+4.8}$&$-7.5_{-4.8}^{+4.5}$\\ 

$L(\mathrm{H_2O})$&-3.000&$-9.5_{-3.2}^{+4.1}$&$-3.2_{-0.8}^{+0.9}$&$-2.7_{-0.8}^{+0.8}$&$-2.7_{-0.9}^{+0.9}$&$-3.8_{-1.6}^{+1.4}$&$-2.5_{-0.9}^{+0.9}$&$-2.9_{-0.8}^{+0.8}$&$-2.8_{-0.7}^{+0.7}$\\ 

$L(\mathrm{CO_2})$&-3.387&$-9.2_{-3.5}^{+3.7}$&$-3.8_{-0.8}^{+0.8}$&$-3.4_{-0.7}^{+0.8}$&$-3.4_{-0.7}^{+0.8}$&$-4.3_{-1.3}^{+1.1}$&$-3.2_{-0.8}^{+0.9}$&$-3.5_{-0.7}^{+0.7}$&$-3.4_{-0.6}^{+0.7}$\\ 

$L(\mathrm{CH_4})$&-5.770&$-10.2_{-2.8}^{+2.9}$&$-9.5_{-3.4}^{+3.3}$&$-9.2_{-3.5}^{+3.1}$&$-9.2_{-3.5}^{+3.1}$&$-9.2_{-3.3}^{+3.1}$&$-9.5_{-3.4}^{+3.4}$&$-9.0_{-3.5}^{+2.9}$&$-8.0_{-4.3}^{+2.1}$\\ 

$L(\mathrm{O_3})$&-6.523&$-10.5_{-2.5}^{+2.6}$&$-7.0_{-3.5}^{+0.8}$&$-6.4_{-0.6}^{+0.6}$&$-6.5_{-0.6}^{+0.6}$&$-8.5_{-3.7}^{+2.0}$&$-6.3_{-0.6}^{+0.7}$&$-6.6_{-0.6}^{+0.5}$&$-6.6_{-0.5}^{+0.5}$\\ 

$L(\mathrm{CO})$&-6.903&$-10.5_{-2.7}^{+3.1}$&$-7.8_{-4.4}^{+4.4}$&$-8.0_{-4.3}^{+4.5}$&$-8.2_{-4.2}^{+4.7}$&$-8.1_{-3.9}^{+4.0}$&$-7.5_{-4.5}^{+4.3}$&$-7.9_{-4.5}^{+4.7}$&$-7.9_{-4.4}^{+4.6}$\\ 

$L(\mathrm{N_2O})$&-6.495&$-10.5_{-2.6}^{+2.7}$&$-8.8_{-3.6}^{+3.8}$&$-8.5_{-3.8}^{+3.7}$&$-8.9_{-3.8}^{+3.5}$&$-9.6_{-3.1}^{+3.5}$&$-8.9_{-3.9}^{+4.0}$&$-9.2_{-3.6}^{+3.8}$&$-9.3_{-3.5}^{+3.6}$\\ 
\hline
\multicolumn{10}{c}{} \\

\hline\hline                 
&&\multicolumn{4}{c|}{R=50}&\multicolumn{4}{c}{R=100}\\
Parameter&Input&S/N=5&S/N=10&S/N=15&S/N=20&S/N=5&S/N=10&S/N=15&S/N=20\\
\hline
$\sqrt[4]{a_4}$&1.14&$1.1_{-0.2}^{+0.2}$&$1.1_{-0.2}^{+0.2}$&$1.0_{-0.2}^{+0.2}$&$1.0_{-0.2}^{+0.2}$&$1.0_{-0.2}^{+0.2}$&$1.0_{-0.2}^{+0.2}$&$1.0_{-0.2}^{+0.2}$&$1.1_{-0.2}^{+0.2}$\\ 

$a_3$&23.12&$17.4_{-8.8}^{+12.5}$&$16.5_{-8.9}^{+14.5}$&$14.4_{-6.6}^{+9.1}$&$14.4_{-6.5}^{+12.1}$&$14.5_{-7.4}^{+12.9}$&$13.5_{-6.0}^{+9.5}$&$14.9_{-6.1}^{+10.7}$&$19.8_{-8.1}^{+13.1}$\\ 

$a_2$&99.70&$77.0_{-39.5}^{+54.3}$&$62.6_{-29.8}^{+56.0}$&$63.4_{-26.0}^{+39.7}$&$60.2_{-21.5}^{+32.0}$&$61.2_{-29.6}^{+44.3}$&$59.6_{-23.1}^{+37.7}$&$67.2_{-24.4}^{+32.6}$&$86.9_{-27.1}^{+35.4}$\\ 

$a_1$&146.63&$127.5_{-56.3}^{+93.3}$&$100.0_{-38.0}^{+78.3}$&$109.7_{-44.5}^{+85.1}$&$90.2_{-29.3}^{+46.2}$&$105.6_{-40.4}^{+68.9}$&$109.0_{-51.3}^{+61.0}$&$107.0_{-43.6}^{+60.5}$&$135.3_{-43.5}^{+47.2}$\\ 

$a_0$&285.22&$286.7_{-47.0}^{+57.0}$&$280.9_{-30.5}^{+39.3}$&$286.4_{-35.9}^{+58.7}$&$264.2_{-20.0}^{+30.3}$&$287.3_{-37.2}^{+46.9}$&$286.0_{-41.4}^{+41.3}$&$276.4_{-30.8}^{+38.4}$&$284.5_{-26.5}^{+31.4}$\\ 

\cdashline{1-10}$L\left(P_0\left[\mathrm{bar}\right]\right)$&0.006&$0.0_{-0.3}^{+0.4}$&$0.0_{-0.3}^{+0.3}$&$-0.0_{-0.4}^{+0.4}$&$0.2_{-0.3}^{+0.2}$&$-0.0_{-0.3}^{+0.4}$&$-0.0_{-0.3}^{+0.5}$&$0.1_{-0.3}^{+0.4}$&$0.0_{-0.2}^{+0.2}$\\ 

$R\,\left[R_\oplus\right]$&1.0&$1.0_{-0.1}^{+0.1}$&$1.02_{-0.07}^{+0.07}$&$1.03_{-0.05}^{+0.05}$&$1.02_{-0.04}^{+0.04}$&$1.00_{-0.08}^{+0.08}$&$1.03_{-0.05}^{+0.05}$&$1.01_{-0.04}^{+0.04}$&$1.00_{-0.03}^{+0.03}$\\ 

$L\left(M\,\left[M_\oplus\right]\right)$&0.0&$-0.0_{-0.3}^{+0.3}$&$0.0_{-0.3}^{+0.3}$&$0.0_{-0.3}^{+0.3}$&$0.0_{-0.3}^{+0.3}$&$0.0_{-0.3}^{+0.3}$&$0.0_{-0.3}^{+0.3}$&$0.0_{-0.3}^{+0.3}$&$0.0_{-0.3}^{+0.3}$\\ 

\cdashline{1-10}$L(\mathrm{N_2})$&-0.107&$-7.1_{-4.5}^{+4.3}$&$-7.5_{-4.7}^{+4.6}$&$-7.4_{-4.6}^{+4.5}$&$-7.5_{-4.7}^{+4.6}$&$-7.5_{-4.7}^{+4.7}$&$-7.4_{-4.6}^{+4.5}$&$-7.5_{-4.7}^{+4.5}$&$-7.6_{-4.5}^{+4.8}$\\ 

$L(\mathrm{O_2})$&-0.679&$-7.6_{-4.3}^{+4.4}$&$-7.8_{-4.2}^{+4.6}$&$-7.5_{-4.6}^{+4.5}$&$-7.4_{-4.9}^{+4.7}$&$-7.5_{-4.5}^{+4.6}$&$-7.2_{-4.7}^{+4.4}$&$-7.5_{-4.5}^{+4.7}$&$-7.9_{-4.6}^{+4.9}$\\ 

$L(\mathrm{H_2O})$&-3.000&$-3.1_{-1.1}^{+1.0}$&$-2.8_{-0.8}^{+0.7}$&$-2.6_{-0.8}^{+0.8}$&$-3.1_{-0.6}^{+0.6}$&$-2.8_{-0.9}^{+0.9}$&$-2.7_{-0.9}^{+0.7}$&$-3.0_{-0.7}^{+0.6}$&$-2.9_{-0.5}^{+0.5}$\\ 

$L(\mathrm{CO_2})$&-3.387&$-3.7_{-0.9}^{+0.9}$&$-3.3_{-0.7}^{+0.7}$&$-3.1_{-0.7}^{+0.8}$&$-3.6_{-0.5}^{+0.6}$&$-3.3_{-0.8}^{+0.8}$&$-3.2_{-0.8}^{+0.6}$&$-3.4_{-0.6}^{+0.5}$&$-3.3_{-0.5}^{+0.4}$\\ 

$L(\mathrm{CH_4})$&-5.770&$-8.8_{-3.5}^{+3.0}$&$-9.3_{-3.5}^{+3.1}$&$-7.9_{-4.2}^{+2.3}$&$-6.9_{-4.3}^{+1.1}$&$-9.2_{-3.4}^{+3.2}$&$-7.9_{-4.2}^{+2.3}$&$-6.3_{-2.0}^{+0.9}$&$-5.8_{-0.6}^{+0.6}$\\ 

$L(\mathrm{O_3})$&-6.523&$-7.2_{-4.3}^{+1.1}$&$-6.5_{-0.5}^{+0.5}$&$-6.4_{-0.6}^{+0.6}$&$-6.7_{-0.4}^{+0.4}$&$-6.5_{-0.7}^{+0.7}$&$-6.5_{-0.6}^{+0.5}$&$-6.6_{-0.5}^{+0.4}$&$-6.5_{-0.4}^{+0.4}$\\ 

$L(\mathrm{CO})$&-6.903&$-7.6_{-4.2}^{+4.4}$&$-7.5_{-4.6}^{+4.5}$&$-8.0_{-4.4}^{+4.8}$&$-7.9_{-4.3}^{+4.6}$&$-7.7_{-4.3}^{+4.5}$&$-7.8_{-4.5}^{+4.4}$&$-8.0_{-4.3}^{+4.4}$&$-8.3_{-4.2}^{+4.3}$\\ 

$L(\mathrm{N_2O})$&-6.495&$-8.3_{-3.7}^{+3.8}$&$-8.9_{-3.6}^{+3.9}$&$-8.6_{-3.8}^{+3.5}$&$-9.0_{-3.7}^{+3.2}$&$-8.5_{-4.0}^{+4.0}$&$-8.8_{-3.6}^{+3.6}$&$-9.2_{-3.6}^{+3.3}$&$-9.4_{-3.4}^{+3.3}$\\ 
\hline
\end{tabular}
\tablefoot{Here, $L(\cdot)$ stands for $\log_{10}(\cdot)$.}
\end{table*}

\begin{table*}
\caption{Retrieval results for the $4-18.5\,\mu$m wavelength range for the optimized case.}             
\label{table:4-18.5_opt}      
\centering                          
\begin{tabular}{lc|cccc|cccc}        
\hline\hline                 
&&\multicolumn{4}{c|}{R=20}&\multicolumn{4}{c}{R=35}\\
Parameter&Input&S/N=5&S/N=10&S/N=15&S/N=20&S/N=5&S/N=10&S/N=15&S/N=20\\
\hline
$\sqrt[4]{a_4}$&1.14&$1.2_{-0.2}^{+0.3}$&$1.0_{-0.2}^{+0.2}$&$1.0_{-0.2}^{+0.2}$&$1.0_{-0.2}^{+0.2}$&$1.1_{-0.2}^{+0.2}$&$1.0_{-0.2}^{+0.2}$&$1.0_{-0.2}^{+0.2}$&$1.0_{-0.2}^{+0.2}$\\ 

$a_3$&23.12&$22.6_{-11.5}^{+18.3}$&$13.1_{-6.3}^{+9.8}$&$13.0_{-6.1}^{+7.8}$&$11.7_{-5.6}^{+8.5}$&$16.4_{-8.0}^{+15.1}$&$10.8_{-5.4}^{+9.2}$&$12.2_{-5.8}^{+10.8}$&$12.1_{-5.4}^{+11.2}$\\ 

$a_2$&99.70&$101.3_{-42.6}^{+51.4}$&$61.5_{-31.9}^{+42.6}$&$51.7_{-26.4}^{+34.6}$&$49.0_{-25.0}^{+32.9}$&$71.5_{-34.7}^{+49.6}$&$39.5_{-20.3}^{+29.5}$&$55.6_{-27.4}^{+43.9}$&$47.8_{-20.3}^{+39.1}$\\ 

$a_1$&146.63&$135.8_{-56.7}^{+57.9}$&$132.1_{-47.6}^{+62.6}$&$94.7_{-39.4}^{+71.5}$&$94.7_{-32.5}^{+61.3}$&$113.8_{-44.9}^{+53.7}$&$65.7_{-23.4}^{+36.2}$&$112.4_{-45.4}^{+81.2}$&$89.8_{-40.3}^{+47.5}$\\ 

$a_0$&285.22&$211.1_{-79.7}^{+60.6}$&$304.6_{-36.4}^{+45.6}$&$283.3_{-37.6}^{+55.6}$&$290.9_{-28.4}^{+48.1}$&$246.6_{-67.9}^{+56.3}$&$259.4_{-28.2}^{+30.6}$&$310.0_{-44.4}^{+55.4}$&$275.8_{-28.1}^{+29.8}$\\ 

\cdashline{1-10}$L\left(P_0\left[\mathrm{bar}\right]\right)$&0.006&$0.5_{-0.3}^{+0.3}$&$-0.1_{-0.3}^{+0.3}$&$0.1_{-0.4}^{+0.5}$&$-0.0_{-0.4}^{+0.3}$&$0.4_{-0.4}^{+0.3}$&$0.4_{-0.4}^{+0.3}$&$-0.2_{-0.4}^{+0.5}$&$0.1_{-0.3}^{+0.4}$\\ 

$R\,\left[R_\oplus\right]$&1.0&$0.9_{-0.1}^{+0.1}$&$0.98_{-0.09}^{+0.09}$&$0.99_{-0.07}^{+0.07}$&$1.01_{-0.05}^{+0.06}$&$0.9_{-0.1}^{+0.1}$&$1.00_{-0.08}^{+0.07}$&$1.01_{-0.05}^{+0.05}$&$1.01_{-0.04}^{+0.04}$\\ 

$L\left(M\,\left[M_\oplus\right]\right)$&0.0&$0.0_{-0.3}^{+0.3}$&$0.0_{-0.3}^{+0.3}$&$0.0_{-0.3}^{+0.3}$&$0.0_{-0.3}^{+0.3}$&$0.0_{-0.3}^{+0.3}$&$-0.0_{-0.3}^{+0.3}$&$0.0_{-0.3}^{+0.3}$&$0.0_{-0.3}^{+0.3}$\\ 

\cdashline{1-10}$L(\mathrm{N_2})$&-0.107&$-6.6_{-4.4}^{+4.0}$&$-7.2_{-4.6}^{+4.1}$&$-7.0_{-4.6}^{+4.1}$&$-7.1_{-4.7}^{+4.3}$&$-8.0_{-4.1}^{+4.3}$&$-7.8_{-4.4}^{+4.6}$&$-7.8_{-4.4}^{+4.7}$&$-7.6_{-4.6}^{+4.6}$\\ 

$L(\mathrm{O_2})$&-0.679&$-7.3_{-4.4}^{+4.4}$&$-7.8_{-4.4}^{+4.4}$&$-7.9_{-4.2}^{+4.5}$&$-7.3_{-4.5}^{+4.3}$&$-7.7_{-4.5}^{+4.6}$&$-7.3_{-4.5}^{+4.2}$&$-7.5_{-4.5}^{+4.5}$&$-7.3_{-4.8}^{+4.6}$\\ 

$L(\mathrm{H_2O})$&-3.000&$-3.9_{-1.4}^{+0.9}$&$-2.1_{-0.9}^{+0.9}$&$-2.5_{-0.9}^{+1.1}$&$-2.2_{-0.8}^{+0.8}$&$-3.3_{-0.8}^{+1.1}$&$-3.1_{-0.8}^{+0.9}$&$-2.0_{-0.9}^{+0.9}$&$-2.7_{-0.8}^{+0.7}$\\

$L(\mathrm{CO_2})$&-3.387&$-6.3_{-4.3}^{+1.9}$&$-3.5_{-1.1}^{+1.0}$&$-3.5_{-0.8}^{+1.0}$&$-3.1_{-0.8}^{+0.7}$&$-5.0_{-3.1}^{+1.4}$&$-3.9_{-0.8}^{+0.9}$&$-2.8_{-0.9}^{+0.8}$&$-3.4_{-0.7}^{+0.7}$\\ 

$L(\mathrm{CH_4})$&-5.770&$-9.6_{-3.1}^{+2.8}$&$-9.5_{-3.2}^{+3.3}$&$-9.4_{-3.4}^{+3.0}$&$-8.8_{-3.7}^{+2.9}$&$-10.1_{-2.9}^{+2.9}$&$-9.9_{-3.1}^{+3.0}$&$-7.8_{-4.4}^{+2.5}$&$-6.7_{-3.9}^{+1.1}$\\ 

$L(\mathrm{O_3})$&-6.523&$-9.7_{-3.1}^{+2.4}$&$-6.2_{-0.7}^{+0.7}$&$-6.5_{-0.6}^{+0.7}$&$-6.2_{-0.6}^{+0.6}$&$-7.2_{-2.1}^{+0.9}$&$-6.8_{-0.6}^{+0.6}$&$-6.0_{-0.7}^{+0.7}$&$-6.5_{-0.5}^{+0.5}$\\ 

$L(\mathrm{CO})$&-6.903&$-8.9_{-3.6}^{+3.9}$&$-7.8_{-4.5}^{+4.6}$&$-7.2_{-4.4}^{+4.3}$&$-7.7_{-4.5}^{+4.3}$&$-8.2_{-4.1}^{+4.5}$&$-8.1_{-4.3}^{+4.7}$&$-7.8_{-4.5}^{+4.6}$&$-8.0_{-4.3}^{+4.3}$\\ 

$L(\mathrm{N_2O})$&-6.495&$-9.7_{-3.1}^{+3.5}$&$-8.2_{-4.0}^{+3.8}$&$-8.4_{-3.8}^{+3.7}$&$-8.5_{-3.8}^{+3.6}$&$-8.7_{-3.5}^{+3.6}$&$-8.9_{-3.6}^{+3.4}$&$-9.3_{-3.5}^{+3.7}$&$-9.2_{-3.4}^{+3.2}$\\ 
\hline
\multicolumn{10}{c}{} \\

\hline\hline                 
&&\multicolumn{4}{c|}{R=50}&\multicolumn{4}{c}{R=100}\\
Parameter&Input&S/N=5&S/N=10&S/N=15&S/N=20&S/N=5&S/N=10&S/N=15&S/N=20\\
\hline
$\sqrt[4]{a_4}$&1.14&$1.0_{-0.2}^{+0.2}$&$1.0_{-0.2}^{+0.2}$&$1.0_{-0.2}^{+0.2}$&$1.0_{-0.2}^{+0.2}$&$1.0_{-0.2}^{+0.2}$&$1.1_{-0.2}^{+0.2}$&$1.1_{-0.2}^{+0.2}$&$1.1_{-0.2}^{+0.2}$\\ 

$a_3$&23.12&$13.1_{-6.5}^{+10.4}$&$11.5_{-5.5}^{+7.8}$&$13.9_{-5.8}^{+10.5}$&$15.7_{-6.7}^{+13.0}$&$13.2_{-6.3}^{+9.8}$&$17.0_{-7.5}^{+13.7}$&$16.3_{-7.2}^{+13.5}$&$19.7_{-8.7}^{+14.4}$\\ 

$a_2$&99.70&$59.2_{-31.6}^{+44.2}$&$36.8_{-17.5}^{+22.5}$&$62.0_{-25.8}^{+42.8}$&$74.7_{-28.0}^{+44.0}$&$53.6_{-26.3}^{+36.4}$&$81.4_{-35.6}^{+57.7}$&$73.6_{-27.1}^{+41.0}$&$84.6_{-27.7}^{+40.5}$\\ 

$a_1$&146.63&$125.9_{-51.9}^{+69.8}$&$60.8_{-20.1}^{+29.0}$&$113.3_{-41.9}^{+92.3}$&$130.1_{-48.3}^{+68.4}$&$97.0_{-33.6}^{+55.4}$&$156.6_{-70.0}^{+101.0}$&$118.9_{-43.0}^{+62.2}$&$132.7_{-47.1}^{+51.0}$\\ 

$a_0$&285.22&$298.7_{-49.8}^{+56.7}$&$257.1_{-20.7}^{+24.7}$&$299.9_{-37.3}^{+63.9}$&$298.7_{-37.0}^{+40.7}$&$280.5_{-31.4}^{+41.7}$&$325.9_{-53.1}^{+71.9}$&$282.2_{-28.6}^{+43.2}$&$283.1_{-27.3}^{+31.3}$\\ 

\cdashline{1-10}$L\left(P_0\left[\mathrm{bar}\right]\right)$&0.006&$-0.1_{-0.4}^{+0.4}$&$0.4_{-0.3}^{+0.3}$&$-0.1_{-0.4}^{+0.4}$&$-0.1_{-0.2}^{+0.3}$&$0.1_{-0.3}^{+0.3}$&$-0.3_{-0.3}^{+0.4}$&$0.0_{-0.3}^{+0.3}$&$0.0_{-0.2}^{+0.3}$\\ 

$R\,\left[R_\oplus\right]$&1.0&$1.0_{-0.1}^{+0.1}$&$1.01_{-0.06}^{+0.07}$&$1.01_{-0.05}^{+0.05}$&$1.00_{-0.03}^{+0.04}$&$1.00_{-0.08}^{+0.08}$&$1.01_{-0.05}^{+0.05}$&$1.00_{-0.03}^{+0.03}$&$1.00_{-0.02}^{+0.02}$\\ 

$L\left(M\,\left[M_\oplus\right]\right)$&0.0&$0.0_{-0.3}^{+0.3}$&$0.0_{-0.3}^{+0.3}$&$0.0_{-0.3}^{+0.3}$&$0.0_{-0.3}^{+0.3}$&$0.0_{-0.3}^{+0.3}$&$0.0_{-0.3}^{+0.3}$&$0.0_{-0.3}^{+0.3}$&$0.0_{-0.3}^{+0.3}$\\ 

\cdashline{1-10}$L(\mathrm{N_2})$&-0.107&$-7.6_{-4.2}^{+4.3}$&$-7.6_{-4.3}^{+4.5}$&$-7.3_{-4.3}^{+4.3}$&$-7.5_{-4.6}^{+4.6}$&$-7.9_{-4.4}^{+4.7}$&$-7.6_{-4.8}^{+4.6}$&$-7.5_{-4.4}^{+4.6}$&$-7.4_{-4.6}^{+4.7}$\\ 

$L(\mathrm{O_2})$&-0.679&$-7.7_{-4.3}^{+4.5}$&$-7.4_{-4.6}^{+4.4}$&$-7.4_{-4.6}^{+4.5}$&$-7.2_{-4.7}^{+4.4}$&$-7.1_{-4.7}^{+4.3}$&$-8.1_{-4.3}^{+4.9}$&$-7.4_{-4.7}^{+4.5}$&$-7.8_{-4.7}^{+4.9}$\\ 

$L(\mathrm{H_2O})$&-3.000&$-2.4_{-1.1}^{+1.0}$&$-3.1_{-0.7}^{+0.8}$&$-2.3_{-0.9}^{+0.8}$&$-2.7_{-0.7}^{+0.6}$&$-2.6_{-0.8}^{+0.8}$&$-2.2_{-0.8}^{+0.8}$&$-2.9_{-0.6}^{+0.6}$&$-3.0_{-0.5}^{+0.5}$\\ 

$L(\mathrm{CO_2})$&-3.387&$-3.6_{-1.3}^{+1.1}$&$-3.8_{-0.6}^{+0.7}$&$-3.0_{-0.7}^{+0.8}$&$-3.1_{-0.6}^{+0.5}$&$-3.5_{-0.8}^{+0.8}$&$-2.8_{-0.7}^{+0.7}$&$-3.3_{-0.6}^{+0.6}$&$-3.4_{-0.5}^{+0.5}$\\ 

$L(\mathrm{CH_4})$&-5.770&$-9.9_{-3.1}^{+3.2}$&$-9.1_{-3.7}^{+2.6}$&$-6.2_{-3.9}^{+1.2}$&$-5.7_{-0.7}^{+0.6}$&$-8.4_{-3.8}^{+2.5}$&$-5.5_{-1.1}^{+0.9}$&$-5.8_{-0.6}^{+0.6}$&$-5.8_{-0.5}^{+0.5}$\\ 

$L(\mathrm{O_3})$&-6.523&$-6.4_{-1.0}^{+0.8}$&$-6.8_{-0.5}^{+0.5}$&$-6.2_{-0.6}^{+0.6}$&$-6.3_{-0.5}^{+0.4}$&$-6.5_{-0.6}^{+0.6}$&$-6.1_{-0.5}^{+0.6}$&$-6.5_{-0.4}^{+0.5}$&$-6.5_{-0.4}^{+0.4}$\\ 

$L(\mathrm{CO})$&-6.903&$-8.2_{-3.9}^{+4.5}$&$-8.0_{-4.2}^{+4.6}$&$-8.0_{-4.4}^{+4.5}$&$-8.3_{-4.1}^{+4.6}$&$-7.4_{-4.4}^{+4.6}$&$-7.8_{-4.5}^{+4.3}$&$-7.9_{-4.4}^{+4.3}$&$-8.7_{-4.0}^{+4.2}$\\ 

$L(\mathrm{N_2O})$&-6.495&$-8.5_{-3.9}^{+3.9}$&$-9.2_{-3.3}^{+3.4}$&$-9.0_{-3.5}^{+3.5}$&$-9.3_{-3.5}^{+3.2}$&$-8.7_{-3.7}^{+3.7}$&$-9.0_{-3.7}^{+3.7}$&$-9.5_{-3.4}^{+3.2}$&$-9.9_{-3.2}^{+3.1}$\\ 
\hline
\end{tabular}
\tablefoot{Here, $L(\cdot)$ stands for $\log_{10}(\cdot)$.}
\end{table*}

\begin{table*}
\caption{Retrieval results for the $6-17\,\mu$m wavelength range for the nominal case.}           
\label{table:6-17}      
\centering                          
\begin{tabular}{lc|cccc|cccc}        
\hline\hline                 
&&\multicolumn{4}{c|}{R=20}&\multicolumn{4}{c}{R=35}\\
Parameter&Input&S/N=5&S/N=10&S/N=15&S/N=20&S/N=5&S/N=10&S/N=15&S/N=20\\
\hline
$\sqrt[4]{a_4}$&1.14&$1.2_{-0.3}^{+0.3}$&$1.2_{-0.3}^{+0.3}$&$1.1_{-0.2}^{+0.2}$&$1.1_{-0.2}^{+0.2}$&$1.2_{-0.3}^{+0.3}$&$1.1_{-0.2}^{+0.2}$&$1.1_{-0.2}^{+0.2}$&$1.1_{-0.2}^{+0.2}$\\ 

$a_3$&23.12&$43.0_{-23.8}^{+29.6}$&$30.0_{-17.0}^{+29.9}$&$17.9_{-9.1}^{+17.6}$&$15.4_{-7.3}^{+11.3}$&$35.5_{-18.3}^{+26.6}$&$15.5_{-7.5}^{+11.6}$&$16.8_{-8.7}^{+16.3}$&$16.8_{-7.8}^{+14.4}$\\ 

$a_2$&99.70&$167.4_{-77.4}^{+96.9}$&$120.6_{-62.6}^{+96.3}$&$71.6_{-33.6}^{+50.3}$&$59.6_{-25.6}^{+31.5}$&$130.7_{-62.0}^{+97.7}$&$63.4_{-27.9}^{+34.7}$&$61.4_{-28.2}^{+57.6}$&$63.7_{-26.0}^{+41.9}$\\ 

$a_1$&146.63&$148.2_{-79.8}^{+122.3}$&$118.5_{-47.4}^{+71.1}$&$101.1_{-37.0}^{+59.8}$&$80.6_{-28.1}^{+41.1}$&$103.7_{-54.7}^{+117.6}$&$96.6_{-32.6}^{+40.7}$&$84.1_{-34.6}^{+81.8}$&$86.8_{-38.3}^{+55.1}$\\ 

$a_0$&285.22&$217.6_{-88.8}^{+92.3}$&$235.7_{-70.0}^{+47.4}$&$269.6_{-34.3}^{+37.1}$&$258.0_{-26.5}^{+26.7}$&$213.8_{-61.3}^{+53.4}$&$269.6_{-32.5}^{+35.4}$&$261.2_{-27.0}^{+45.5}$&$258.2_{-22.8}^{+29.4}$\\ 

\cdashline{1-10}$L\left(P_0\left[\mathrm{bar}\right]\right)$&0.006&$0.4_{-0.4}^{+0.3}$&$0.3_{-0.3}^{+0.2}$&$0.2_{-0.3}^{+0.3}$&$0.3_{-0.3}^{+0.2}$&$0.5_{-0.3}^{+0.2}$&$0.2_{-0.3}^{+0.3}$&$0.2_{-0.4}^{+0.3}$&$0.3_{-0.3}^{+0.3}$\\ 

$R\,\left[R_\oplus\right]$&1.0&$0.9_{-0.1}^{+0.1}$&$1.01_{-0.09}^{+0.1}$&$1.00_{-0.08}^{+0.09}$&$1.01_{-0.07}^{+0.07}$&$0.9_{-0.1}^{+0.1}$&$0.99_{-0.08}^{+0.09}$&$1.01_{-0.07}^{+0.07}$&$1.01_{-0.06}^{+0.06}$\\ 

$L\left(M\,\left[M_\oplus\right]\right)$&0.0&$0.0_{-0.3}^{+0.3}$&$-0.0_{-0.3}^{+0.3}$&$0.0_{-0.3}^{+0.3}$&$0.0_{-0.3}^{+0.3}$&$0.0_{-0.3}^{+0.3}$&$0.0_{-0.3}^{+0.3}$&$0.0_{-0.3}^{+0.3}$&$0.0_{-0.3}^{+0.3}$\\ 

\cdashline{1-10}$L(\mathrm{N_2})$&-0.107&$-7.5_{-4.3}^{+4.2}$&$-7.4_{-4.4}^{+4.3}$&$-7.4_{-4.5}^{+4.4}$&$-7.5_{-4.6}^{+4.8}$&$-8.1_{-4.1}^{+4.4}$&$-7.6_{-4.3}^{+4.7}$&$-7.1_{-4.8}^{+4.4}$&$-7.6_{-4.8}^{+4.7}$\\ 

$L(\mathrm{O_2})$&-0.679&$-7.7_{-4.1}^{+4.5}$&$-7.5_{-4.1}^{+4.4}$&$-7.2_{-4.5}^{+4.4}$&$-7.8_{-4.7}^{+4.8}$&$-6.8_{-4.9}^{+4.2}$&$-7.3_{-4.5}^{+4.4}$&$-7.7_{-4.6}^{+4.7}$&$-7.6_{-4.6}^{+4.7}$\\ 

$L(\mathrm{H_2O})$&-3.000&$-9.5_{-3.1}^{+3.3}$&$-5.5_{-5.1}^{+2.2}$&$-3.2_{-1.1}^{+0.9}$&$-3.2_{-0.8}^{+0.8}$&$-8.6_{-3.6}^{+3.4}$&$-3.2_{-1.2}^{+1.0}$&$-3.2_{-0.8}^{+0.9}$&$-3.2_{-0.7}^{+0.7}$\\ 

$L(\mathrm{CO_2})$&-3.387&$-5.1_{-2.2}^{+1.0}$&$-4.0_{-0.5}^{+0.6}$&$-3.6_{-0.6}^{+0.7}$&$-3.8_{-0.6}^{+0.6}$&$-4.5_{-0.7}^{+0.6}$&$-3.6_{-0.7}^{+0.8}$&$-3.7_{-0.7}^{+0.7}$&$-3.8_{-0.6}^{+0.6}$\\ 

$L(\mathrm{CH_4})$&-5.770&$-9.6_{-3.1}^{+2.8}$&$-9.1_{-3.5}^{+3.1}$&$-8.4_{-3.6}^{+2.8}$&$-9.3_{-3.5}^{+3.0}$&$-9.6_{-3.2}^{+3.1}$&$-8.9_{-3.9}^{+3.2}$&$-9.2_{-3.5}^{+3.0}$&$-9.2_{-3.5}^{+2.9}$\\ 

$L(\mathrm{O_3})$&-6.523&$-10.4_{-2.7}^{+2.7}$&$-7.8_{-3.5}^{+1.1}$&$-6.7_{-0.6}^{+0.6}$&$-6.8_{-0.5}^{+0.5}$&$-9.6_{-3.1}^{+2.3}$&$-6.7_{-0.7}^{+0.7}$&$-6.8_{-0.5}^{+0.6}$&$-6.8_{-0.4}^{+0.5}$\\ 

$L(\mathrm{CO})$&-6.903&$-7.8_{-4.1}^{+4.4}$&$-7.1_{-4.4}^{+4.5}$&$-7.8_{-4.3}^{+4.4}$&$-7.4_{-4.8}^{+4.7}$&$-7.7_{-4.2}^{+4.4}$&$-7.4_{-4.4}^{+4.4}$&$-7.6_{-4.8}^{+4.7}$&$-7.4_{-4.7}^{+4.7}$\\ 

$L(\mathrm{N_2O})$&-6.495&$-9.7_{-3.0}^{+3.0}$&$-9.4_{-3.3}^{+3.3}$&$-8.9_{-3.7}^{+3.9}$&$-8.5_{-4.0}^{+3.6}$&$-10.0_{-2.9}^{+2.9}$&$-8.7_{-3.7}^{+3.8}$&$-9.0_{-3.8}^{+3.7}$&$-8.9_{-3.7}^{+3.6}$\\
\hline
\multicolumn{10}{c}{} \\

\hline\hline                 
&&\multicolumn{4}{c|}{R=50}&\multicolumn{4}{c}{R=100}\\
Parameter&Input&S/N=5&S/N=10&S/N=15&S/N=20&S/N=5&S/N=10&S/N=15&S/N=20\\
\hline
$\sqrt[4]{a_4}$&1.14&$1.2_{-0.3}^{+0.3}$&$1.1_{-0.2}^{+0.2}$&$1.1_{-0.2}^{+0.2}$&$1.1_{-0.2}^{+0.2}$&$1.1_{-0.2}^{+0.3}$&$1.1_{-0.2}^{+0.2}$&$1.1_{-0.2}^{+0.2}$&$1.1_{-0.2}^{+0.2}$\\ 

$a_3$&23.12&$33.5_{-17.3}^{+22.8}$&$15.0_{-7.6}^{+14.3}$&$16.3_{-7.8}^{+13.6}$&$15.5_{-6.6}^{+12.2}$&$17.8_{-9.1}^{+18.9}$&$16.2_{-7.7}^{+13.6}$&$19.0_{-8.5}^{+16.5}$&$19.4_{-8.7}^{+15.5}$\\ 

$a_2$&99.70&$127.0_{-57.5}^{+75.9}$&$62.9_{-28.5}^{+39.9}$&$63.1_{-28.5}^{+48.4}$&$66.3_{-24.3}^{+30.7}$&$70.1_{-31.0}^{+43.5}$&$67.0_{-26.8}^{+35.9}$&$78.3_{-26.8}^{+52.3}$&$81.7_{-30.0}^{+48.2}$\\ 

$a_1$&146.63&$118.8_{-58.6}^{+109.9}$&$94.5_{-31.0}^{+47.3}$&$92.7_{-48.3}^{+82.6}$&$92.8_{-33.7}^{+50.9}$&$102.3_{-35.8}^{+45.1}$&$98.4_{-39.0}^{+56.9}$&$126.9_{-54.5}^{+60.6}$&$119.9_{-50.9}^{+72.1}$\\ 

$a_0$&285.22&$232.6_{-50.4}^{+49.7}$&$271.5_{-30.5}^{+34.5}$&$266.7_{-32.9}^{+48.3}$&$261.4_{-22.0}^{+32.4}$&$269.0_{-32.6}^{+37.2}$&$271.0_{-29.7}^{+36.1}$&$281.0_{-32.5}^{+35.3}$&$275.9_{-30.8}^{+43.7}$\\ 

\cdashline{1-10}$L\left(P_0\left[\mathrm{bar}\right]\right)$&0.006&$0.3_{-0.3}^{+0.2}$&$0.1_{-0.3}^{+0.3}$&$0.2_{-0.4}^{+0.5}$&$0.2_{-0.3}^{+0.3}$&$0.1_{-0.3}^{+0.3}$&$0.1_{-0.3}^{+0.3}$&$0.0_{-0.2}^{+0.3}$&$0.1_{-0.3}^{+0.3}$\\ 

$R\,\left[R_\oplus\right]$&1.0&$1.0_{-0.1}^{+0.1}$&$1.01_{-0.08}^{+0.08}$&$1.01_{-0.06}^{+0.06}$&$1.02_{-0.05}^{+0.05}$&$1.0_{-0.1}^{+0.1}$&$1.01_{-0.06}^{+0.07}$&$1.02_{-0.05}^{+0.05}$&$1.01_{-0.04}^{+0.04}$\\ 

$L\left(M\,\left[M_\oplus\right]\right)$&0.0&$0.0_{-0.3}^{+0.3}$&$0.0_{-0.3}^{+0.3}$&$0.0_{-0.3}^{+0.3}$&$-0.0_{-0.3}^{+0.3}$&$-0.0_{-0.3}^{+0.3}$&$0.0_{-0.3}^{+0.3}$&$-0.0_{-0.3}^{+0.3}$&$0.0_{-0.3}^{+0.3}$\\ 

\cdashline{1-10}$L(\mathrm{N_2})$&-0.107&$-8.0_{-4.1}^{+4.5}$&$-7.4_{-4.8}^{+4.6}$&$-7.6_{-4.5}^{+4.7}$&$-7.5_{-4.6}^{+4.6}$&$-7.4_{-4.4}^{+4.3}$&$-7.3_{-4.7}^{+4.5}$&$-7.7_{-4.7}^{+4.9}$&$-7.4_{-4.7}^{+4.7}$\\ 

$L(\mathrm{O_2})$&-0.679&$-7.3_{-4.4}^{+4.4}$&$-7.4_{-4.5}^{+4.4}$&$-7.6_{-4.6}^{+4.6}$&$-7.7_{-4.7}^{+4.9}$&$-7.4_{-4.6}^{+4.5}$&$-7.7_{-4.5}^{+4.7}$&$-7.3_{-4.6}^{+4.5}$&$-7.5_{-4.6}^{+4.8}$\\ 

$L(\mathrm{H_2O})$&-3.000&$-7.5_{-4.5}^{+3.2}$&$-3.0_{-0.9}^{+0.9}$&$-3.1_{-0.9}^{+0.8}$&$-3.2_{-0.6}^{+0.7}$&$-3.5_{-1.6}^{+1.1}$&$-3.0_{-0.7}^{+0.7}$&$-2.9_{-0.6}^{+0.6}$&$-3.0_{-0.6}^{+0.6}$\\ 

$L(\mathrm{CO_2})$&-3.387&$-4.1_{-0.6}^{+0.6}$&$-3.5_{-0.7}^{+0.7}$&$-3.6_{-0.8}^{+0.7}$&$-3.6_{-0.6}^{+0.6}$&$-3.6_{-0.6}^{+0.7}$&$-3.5_{-0.6}^{+0.7}$&$-3.4_{-0.6}^{+0.5}$&$-3.5_{-0.6}^{+0.6}$\\ 

$L(\mathrm{CH_4})$&-5.770&$-8.8_{-3.5}^{+2.9}$&$-9.1_{-3.6}^{+3.1}$&$-9.3_{-3.5}^{+3.1}$&$-8.2_{-3.9}^{+2.2}$&$-9.0_{-3.5}^{+3.1}$&$-8.4_{-4.1}^{+2.5}$&$-7.1_{-4.5}^{+1.4}$&$-6.3_{-2.6}^{+0.9}$\\ 

$L(\mathrm{O_3})$&-6.523&$-8.4_{-3.7}^{+1.5}$&$-6.6_{-0.5}^{+0.6}$&$-6.7_{-0.6}^{+0.6}$&$-6.7_{-0.5}^{+0.5}$&$-6.8_{-1.3}^{+0.7}$&$-6.6_{-0.5}^{+0.5}$&$-6.6_{-0.4}^{+0.5}$&$-6.6_{-0.5}^{+0.4}$\\ 

$L(\mathrm{CO})$&-6.903&$-8.0_{-4.1}^{+4.7}$&$-7.4_{-4.7}^{+4.5}$&$-7.4_{-4.7}^{+4.5}$&$-7.9_{-4.4}^{+4.6}$&$-7.2_{-4.5}^{+4.4}$&$-7.7_{-4.6}^{+4.8}$&$-7.2_{-4.6}^{+4.4}$&$-7.2_{-4.8}^{+4.5}$\\ 

$L(\mathrm{N_2O})$&-6.495&$-9.5_{-3.1}^{+3.0}$&$-8.9_{-3.8}^{+3.9}$&$-8.8_{-3.6}^{+3.8}$&$-9.3_{-3.5}^{+3.6}$&$-8.5_{-3.8}^{+3.5}$&$-8.5_{-3.9}^{+3.6}$&$-9.2_{-3.6}^{+3.8}$&$-9.3_{-3.6}^{+3.6}$\\ 
\hline
\end{tabular}
\tablefoot{Here, $L(\cdot)$ stands for $\log_{10}(\cdot)$.}
\end{table*}

\begin{table*}
\caption{Retrieval results for the $6-17\,\mu$m wavelength range for the optimized case.}           
\label{table:6-17_opt}      
\centering                          
\begin{tabular}{lc|cccc|cccc}        
\hline\hline                 
&&\multicolumn{4}{c|}{R=20}&\multicolumn{4}{c}{R=35}\\
Parameter&Input&S/N=5&S/N=10&S/N=15&S/N=20&S/N=5&S/N=10&S/N=15&S/N=20\\
\hline
$\sqrt[4]{a_4}$&1.14&$1.3_{-0.3}^{+0.3}$&$1.1_{-0.2}^{+0.2}$&$1.0_{-0.2}^{+0.2}$&$1.0_{-0.2}^{+0.2}$&$1.1_{-0.2}^{+0.2}$&$1.0_{-0.2}^{+0.2}$&$1.0_{-0.2}^{+0.2}$&$1.1_{-0.2}^{+0.2}$\\ 

$a_3$&23.12&$32.1_{-16.6}^{+27.0}$&$14.7_{-7.7}^{+11.3}$&$11.4_{-5.5}^{+8.5}$&$12.6_{-5.9}^{+9.3}$&$18.4_{-9.5}^{+12.0}$&$13.3_{-6.7}^{+10.0}$&$12.2_{-5.9}^{+10.3}$&$14.3_{-6.9}^{+10.4}$\\ 

$a_2$&99.70&$113.7_{-48.6}^{+76.8}$&$64.6_{-35.0}^{+43.4}$&$46.9_{-23.7}^{+33.7}$&$54.6_{-26.9}^{+38.5}$&$80.2_{-36.5}^{+41.0}$&$55.2_{-28.9}^{+32.8}$&$49.5_{-22.9}^{+32.5}$&$54.5_{-23.4}^{+37.8}$\\ 

$a_1$&146.63&$114.4_{-57.1}^{+79.8}$&$114.1_{-44.2}^{+71.8}$&$87.6_{-31.7}^{+55.1}$&$110.4_{-37.0}^{+63.1}$&$125.6_{-51.4}^{+64.4}$&$97.7_{-36.3}^{+55.9}$&$91.4_{-36.8}^{+48.6}$&$88.7_{-31.1}^{+65.9}$\\ 

$a_0$&285.22&$194.6_{-72.8}^{+68.0}$&$277.5_{-44.3}^{+53.6}$&$276.4_{-30.8}^{+46.8}$&$310.2_{-35.3}^{+46.0}$&$247.1_{-59.5}^{+47.8}$&$281.9_{-34.2}^{+44.8}$&$283.4_{-31.7}^{+38.2}$&$277.0_{-26.1}^{+42.8}$\\ 

\cdashline{1-10}$L\left(P_0\left[\mathrm{bar}\right]\right)$&0.006&$0.5_{-0.4}^{+0.3}$&$0.1_{-0.4}^{+0.3}$&$0.2_{-0.4}^{+0.3}$&$-0.2_{-0.3}^{+0.4}$&$0.3_{-0.3}^{+0.3}$&$0.1_{-0.4}^{+0.3}$&$0.1_{-0.3}^{+0.4}$&$0.1_{-0.4}^{+0.3}$\\ 

$R\,\left[R_\oplus\right]$&1.0&$1.0_{-0.1}^{+0.1}$&$0.96_{-0.09}^{+0.1}$&$0.99_{-0.08}^{+0.07}$&$0.99_{-0.06}^{+0.06}$&$0.9_{-0.1}^{+0.1}$&$0.98_{-0.08}^{+0.08}$&$1.00_{-0.06}^{+0.06}$&$1.00_{-0.05}^{+0.04}$\\ 

$L\left(M\,\left[M_\oplus\right]\right)$&0.0&$0.0_{-0.3}^{+0.3}$&$-0.0_{-0.3}^{+0.3}$&$0.0_{-0.3}^{+0.3}$&$0.0_{-0.3}^{+0.3}$&$0.0_{-0.3}^{+0.3}$&$0.1_{-0.3}^{+0.3}$&$0.0_{-0.3}^{+0.3}$&$0.0_{-0.3}^{+0.3}$\\ 

\cdashline{1-10}$L(\mathrm{N_2})$&-0.107&$-7.7_{-4.2}^{+4.2}$&$-7.4_{-4.4}^{+4.5}$&$-7.4_{-4.6}^{+4.4}$&$-7.5_{-4.7}^{+4.4}$&$-7.4_{-4.4}^{+4.5}$&$-7.2_{-4.6}^{+4.4}$&$-7.8_{-4.7}^{+4.7}$&$-7.4_{-4.6}^{+4.7}$\\ 

$L(\mathrm{O_2})$&-0.679&$-7.0_{-4.6}^{+4.8}$&$-7.5_{-4.4}^{+4.4}$&$-7.4_{-4.5}^{+4.4}$&$-7.9_{-4.3}^{+4.5}$&$-6.7_{-4.8}^{+4.3}$&$-7.4_{-4.5}^{+4.6}$&$-7.8_{-4.5}^{+4.7}$&$-7.6_{-4.5}^{+4.5}$\\ 

$L(\mathrm{H_2O})$&-3.000&$-5.1_{-5.5}^{+1.5}$&$-2.7_{-0.9}^{+1.0}$&$-2.6_{-0.8}^{+1.0}$&$-1.9_{-0.8}^{+0.8}$&$-3.3_{-1.0}^{+0.9}$&$-2.4_{-0.9}^{+0.8}$&$-2.5_{-0.8}^{+0.8}$&$-2.6_{-0.7}^{+0.8}$\\ 

$L(\mathrm{CO_2})$&-3.387&$-5.6_{-3.0}^{+1.5}$&$-4.0_{-1.1}^{+1.0}$&$-3.7_{-0.8}^{+0.9}$&$-3.0_{-0.8}^{+0.7}$&$-4.8_{-2.0}^{+1.2}$&$-3.5_{-0.8}^{+0.8}$&$-3.4_{-0.7}^{+0.8}$&$-3.4_{-0.7}^{+0.7}$\\ 

$L(\mathrm{CH_4})$&-5.770&$-9.9_{-2.9}^{+2.7}$&$-9.9_{-2.9}^{+3.1}$&$-9.5_{-3.4}^{+2.9}$&$-9.0_{-3.6}^{+3.1}$&$-9.6_{-3.0}^{+2.8}$&$-9.5_{-3.4}^{+3.0}$&$-9.1_{-3.5}^{+2.8}$&$-7.2_{-4.2}^{+1.4}$\\ 

$L(\mathrm{O_3})$&-6.523&$-10.6_{-2.6}^{+2.7}$&$-6.6_{-0.7}^{+0.7}$&$-6.5_{-0.6}^{+0.6}$&$-6.0_{-0.6}^{+0.6}$&$-7.2_{-3.1}^{+0.8}$&$-6.4_{-0.6}^{+0.6}$&$-6.4_{-0.6}^{+0.6}$&$-6.5_{-0.5}^{+0.5}$\\ 

$L(\mathrm{CO})$&-6.903&$-7.5_{-4.3}^{+4.3}$&$-7.2_{-4.5}^{+4.2}$&$-7.2_{-4.7}^{+4.3}$&$-7.5_{-4.5}^{+4.3}$&$-6.7_{-4.6}^{+4.0}$&$-7.5_{-4.6}^{+4.6}$&$-7.5_{-4.7}^{+4.4}$&$-7.8_{-4.5}^{+4.7}$\\ 

$L(\mathrm{N_2O})$&-6.495&$-10.1_{-2.8}^{+3.1}$&$-8.7_{-3.8}^{+4.0}$&$-8.9_{-3.7}^{+3.8}$&$-8.7_{-3.7}^{+3.7}$&$-8.7_{-3.6}^{+3.6}$&$-8.9_{-3.8}^{+3.8}$&$-9.1_{-3.6}^{+3.5}$&$-9.4_{-3.4}^{+3.4}$\\ 
\hline
\multicolumn{10}{c}{} \\

\hline\hline                 
&&\multicolumn{4}{c|}{R=50}&\multicolumn{4}{c}{R=100}\\
Parameter&Input&S/N=5&S/N=10&S/N=15&S/N=20&S/N=5&S/N=10&S/N=15&S/N=20\\
\hline
$\sqrt[4]{a_4}$&1.14&$1.1_{-0.2}^{+0.2}$&$1.0_{-0.2}^{+0.2}$&$1.1_{-0.2}^{+0.2}$&$1.0_{-0.2}^{+0.2}$&$1.0_{-0.2}^{+0.2}$&$1.0_{-0.2}^{+0.2}$&$1.1_{-0.2}^{+0.2}$&$1.1_{-0.2}^{+0.2}$\\ 

$a_3$&23.12&$16.4_{-8.1}^{+13.3}$&$14.0_{-7.1}^{+10.3}$&$13.0_{-6.5}^{+12.7}$&$14.9_{-6.8}^{+11.3}$&$12.5_{-6.0}^{+10.9}$&$15.0_{-7.5}^{+11.7}$&$19.4_{-8.9}^{+18.3}$&$21.8_{-9.8}^{+16.8}$\\ 

$a_2$&99.70&$76.1_{-37.6}^{+42.9}$&$57.8_{-31.1}^{+43.0}$&$49.6_{-23.1}^{+40.7}$&$67.8_{-30.5}^{+36.7}$&$56.3_{-28.2}^{+38.3}$&$68.5_{-33.7}^{+53.1}$&$88.9_{-37.4}^{+66.8}$&$93.1_{-32.7}^{+56.3}$\\ 

$a_1$&146.63&$130.4_{-49.4}^{+65.2}$&$105.9_{-43.8}^{+77.0}$&$84.5_{-32.7}^{+56.4}$&$119.0_{-59.8}^{+60.0}$&$99.9_{-37.2}^{+66.8}$&$126.8_{-59.3}^{+106.8}$&$161.8_{-83.2}^{+94.6}$&$149.2_{-59.5}^{+80.3}$\\ 

$a_0$&285.22&$268.3_{-45.0}^{+44.3}$&$293.2_{-34.0}^{+53.4}$&$274.2_{-27.1}^{+35.7}$&$286.0_{-37.7}^{+47.4}$&$277.1_{-33.1}^{+47.8}$&$308.5_{-51.4}^{+68.5}$&$307.7_{-51.1}^{+59.3}$&$295.9_{-37.8}^{+49.0}$\\ 

\cdashline{1-10}$L\left(P_0\left[\mathrm{bar}\right]\right)$&0.006&$0.2_{-0.3}^{+0.3}$&$-0.0_{-0.4}^{+0.4}$&$0.1_{-0.3}^{+0.4}$&$0.0_{-0.3}^{+0.4}$&$0.1_{-0.4}^{+0.3}$&$-0.2_{-0.4}^{+0.5}$&$-0.1_{-0.3}^{+0.4}$&$-0.1_{-0.3}^{+0.3}$\\ 

$R\,\left[R_\oplus\right]$&1.0&$0.9_{-0.1}^{+0.1}$&$0.99_{-0.07}^{+0.07}$&$1.00_{-0.05}^{+0.05}$&$1.00_{-0.04}^{+0.04}$&$0.99_{-0.09}^{+0.09}$&$1.00_{-0.05}^{+0.06}$&$1.00_{-0.03}^{+0.03}$&$1.00_{-0.03}^{+0.03}$\\ 

$L\left(M\,\left[M_\oplus\right]\right)$&0.0&$0.0_{-0.3}^{+0.3}$&$-0.0_{-0.3}^{+0.3}$&$0.0_{-0.3}^{+0.3}$&$0.0_{-0.3}^{+0.3}$&$-0.0_{-0.3}^{+0.3}$&$0.0_{-0.3}^{+0.3}$&$0.0_{-0.3}^{+0.3}$&$0.0_{-0.3}^{+0.3}$\\ 

\cdashline{1-10}$L(\mathrm{N_2})$&-0.107&$-7.6_{-4.2}^{+4.5}$&$-7.2_{-4.7}^{+4.4}$&$-7.6_{-4.7}^{+4.7}$&$-7.6_{-4.6}^{+4.6}$&$-7.8_{-4.1}^{+4.5}$&$-7.4_{-4.6}^{+4.6}$&$-7.7_{-4.5}^{+4.9}$&$-7.5_{-4.5}^{+4.7}$\\ 

$L(\mathrm{O_2})$&-0.679&$-7.7_{-4.3}^{+4.6}$&$-7.3_{-4.5}^{+4.4}$&$-7.4_{-4.8}^{+4.6}$&$-7.3_{-4.6}^{+4.5}$&$-7.5_{-4.5}^{+4.5}$&$-7.2_{-4.7}^{+4.5}$&$-7.5_{-4.5}^{+4.6}$&$-7.9_{-4.4}^{+4.7}$\\ 

$L(\mathrm{H_2O})$&-3.000&$-2.9_{-0.7}^{+0.8}$&$-2.4_{-0.8}^{+0.9}$&$-2.8_{-0.8}^{+0.8}$&$-2.8_{-0.8}^{+0.7}$&$-2.8_{-0.9}^{+0.9}$&$-2.4_{-0.9}^{+0.9}$&$-2.7_{-0.7}^{+0.6}$&$-2.8_{-0.6}^{+0.6}$\\ 

$L(\mathrm{CO_2})$&-3.387&$-4.3_{-1.9}^{+1.1}$&$-3.3_{-0.8}^{+0.8}$&$-3.4_{-0.7}^{+0.7}$&$-3.3_{-0.7}^{+0.7}$&$-3.7_{-0.8}^{+0.9}$&$-3.0_{-0.8}^{+0.8}$&$-3.1_{-0.6}^{+0.6}$&$-3.2_{-0.6}^{+0.6}$\\ 

$L(\mathrm{CH_4})$&-5.770&$-10.0_{-2.9}^{+3.1}$&$-8.7_{-3.9}^{+2.9}$&$-6.5_{-3.7}^{+1.0}$&$-5.9_{-0.9}^{+0.8}$&$-8.6_{-3.7}^{+2.6}$&$-5.8_{-1.6}^{+1.0}$&$-5.6_{-0.7}^{+0.6}$&$-5.6_{-0.6}^{+0.6}$\\ 

$L(\mathrm{O_3})$&-6.523&$-6.8_{-0.6}^{+0.6}$&$-6.3_{-0.6}^{+0.6}$&$-6.5_{-0.5}^{+0.5}$&$-6.5_{-0.6}^{+0.6}$&$-6.6_{-0.6}^{+0.6}$&$-6.2_{-0.6}^{+0.7}$&$-6.3_{-0.5}^{+0.5}$&$-6.4_{-0.4}^{+0.4}$\\ 

$L(\mathrm{CO})$&-6.903&$-7.1_{-4.5}^{+4.3}$&$-7.4_{-4.4}^{+4.6}$&$-7.5_{-4.7}^{+4.7}$&$-7.4_{-4.6}^{+4.5}$&$-7.4_{-4.6}^{+4.4}$&$-7.7_{-4.5}^{+4.8}$&$-7.1_{-4.7}^{+4.3}$&$-7.6_{-4.6}^{+4.7}$\\ 

$L(\mathrm{N_2O})$&-6.495&$-8.8_{-3.7}^{+3.6}$&$-8.4_{-3.8}^{+3.5}$&$-9.2_{-3.4}^{+3.4}$&$-9.4_{-3.5}^{+3.2}$&$-8.7_{-3.7}^{+3.8}$&$-8.8_{-3.7}^{+3.5}$&$-9.2_{-3.6}^{+3.3}$&$-9.6_{-3.3}^{+3.3}$\\
\hline
\end{tabular}
\tablefoot{Here, $L(\cdot)$ stands for $\log_{10}(\cdot)$.}
\end{table*}

\egroup
\FloatBarrier
\newpage

\section{Observation time estimates}\label{app:integration}


\begin{table}[h]
\centering

\caption{Required observation time in days.}

\subfloat[Nominal case]{
\begin{tabular}{clc:cccc}
\hline\hline
&&&\multicolumn{4}{c}{Observation Time [days]}\\
\multicolumn{3}{c:}{R}   & 20       & 35        & 50        & 100       \\ \hline
\multirow{12}{*}{\rotatebox[origin=c]{90}{S/N}} & \multirow{4}{*}{\rotatebox[origin=c]{90}{\begin{tabular}{cc}1 m\\Mirrors\end{tabular}}}   &5  & 48.6   & 85.7    & 122.5   & 249.1   \\ 
                                        &                                     &10 & 194.6  & 342.7   & 490.1   & 996.2  \\
                                        &                                     &15 & 437.8  & 771.0   & 1102.7  & 2241.5  \\ 
                                        &                                     &20 & 778.3  & 1370.6  & 1960.4  & 3984.8  \\ \cline{2-7} 
                                        & \multirow{4}{*}{\rotatebox[origin=c]{90}{\begin{tabular}{cc}2 m\\Mirrors\end{tabular}}}   & 5                       & 4.7    & 8.2    & 11.7    & 23.9    \\
                                        &                                     & 10                      & 18.6   & 32.7    & 46.7    & 95.4   \\
                                        &                                     & 15                      & 41.9   & 73.5    & 105.0   & 214.7   \\
                                        &                                     & 20                      & 74.5   & 130.7  & 186.6   & 381.7   \\ \cline{2-7} 
                                        & \multirow{4}{*}{\rotatebox[origin=c]{90}{\begin{tabular}{cc}3.5 m\\Mirrors\end{tabular}}} & 5                       & 1.0    & 1.7     & 2.4     & 4.9     \\
                                        &                                     & 10                      & 3.8    & 6.7    & 9.5    & 19.6    \\
                                        &                                     & 15                      & 8.6    & 15.1    & 21.5    & 44.2    \\
                                        &                                     & 20                      & 15.3   & 26.8    & 38.2    & 78.5    \\ \hline
\end{tabular}}

\subfloat[Optimized case]{
\begin{tabular}{clc:cccc}
\hline\hline
&&&\multicolumn{4}{c}{Observation Time [days]}\\
\multicolumn{3}{c:}{R}   & 20       & 35        & 50        & 100       \\ \hline
\multirow{12}{*}{\rotatebox[origin=c]{90}{S/N}} & \multirow{4}{*}{\rotatebox[origin=c]{90}{\begin{tabular}{cc}1m\\Mirrors\end{tabular}}}   &5  & 57.7   & 102.5    & 147.1   & 294.7   \\ 
                                        &                                     &10 & 230.6  & 410.0   & 588.5   & 1178.7  \\
                                        &                                     &15 & 518.9  & 922.2   & 1324.2  & 2652.2  \\ 
                                        &                                     &20 & 922.5  & 1639.5  & 2354.1  & 4714.9  \\ \cline{2-7}
                                        & \multirow{4}{*}{\rotatebox[origin=c]{90}{\begin{tabular}{cc}2m\\Mirrors\end{tabular}}}   & 5                       & 4.0    & 7.1     & 10.2    & 20.6    \\
                                        &                                     & 10                      & 16.0   & 28.5    & 41.0    & 82.3   \\ 
                                        &                                     & 15                      & 36.2   & 64.2   & 92.1   & 185.2   \\
                                        &                                     & 20                      & 64.4   & 114.2  & 163.8   & 329.2  \\\cline{2-7} 
                                        & \multirow{4}{*}{\rotatebox[origin=c]{90}{\begin{tabular}{cc}3.5m\\Mirrors\end{tabular}}} & 5                       & 0.6    & 1.0     & 1.4     & 2.9     \\ 
                                        &                                     & 10                      & 2.2    & 3.9    & 5.7   & 11.4   \\
                                        &                                     & 15                      & 5.0   & 8.9    & 12.7    & 25.7   \\
                                        &                                     & 20                      & 8.9   & 15.8    & 22.6    & 45.8    \\ \hline
\end{tabular}}
\tablefoot{Observation times necessary for \textit{LIFE} to investigate an Earth-twin at 10 pc in orbit around a G$2$ star for the grid of R and S/N discussed in this publication. The S/N in the Table corresponds to the S/N at $11.2\,\mu\mathrm{m}$. We consider three different aperture diameters (1m, 2m, and 3.5m) and conservatively assume a total instrument throughput of 5\% \citep[cf.][]{dannert2022large}.  (a): Nominal case; (b): Optimized case.}
\label{tab:all_integration_times}
\end{table}

\end{appendix}

\end{document}